\documentclass[preprint,authoryear,1p,times]{elsarticle}
\usepackage{graphicx}
\usepackage{epsfig}

\usepackage{amssymb}
\usepackage{amsthm}
\usepackage{amsbsy}
\usepackage{amsmath}
\usepackage{amsfonts}
\usepackage{dsfont}
\usepackage{rotating}
\usepackage{color}

\journal{Computer Physics Communications}

\begin{document}

\begin{frontmatter}



%\title{}
\title{A high order special relativistic hydrodynamic and magnetohydrodynamic
  code with space-time adaptive mesh refinement}

\author[UNITN]{Olindo Zanotti\corref{cor1}}
\ead{olindo.zanotti@unitn.it}
\author[UNITN]{Michael Dumbser}
\ead{michael.dumbser@unitn.it}

\address[UNITN]{Laboratory of Applied Mathematics, Department of Civil, Environmental and Mechanical Engineering, University of Trento, Via Mesiano 77, I-38123 Trento, Italy}

\cortext[cor1]{Corresponding author}

\begin{abstract}
We present a high order one-step ADER-WENO finite volume 
scheme with space-time adaptive mesh refinement (AMR) for the solution 
of the special relativistic hydrodynamic and magnetohydrodynamic equations. 
By adopting a local discontinuous Galerkin predictor
method, a high order one-step time discretization is
obtained, with no need for Runge--Kutta sub-steps. This
turns out to be particularly advantageous in combination
with space-time adaptive mesh refinement, which has been 
implemented following a ``cell-by-cell'' approach. As in 
existing second order AMR methods, also the present higher order AMR 
algorithm features time-accurate local time stepping (LTS), where 
grids on different spatial refinement levels are allowed to use 
different time steps. 

We also compare two different Riemann solvers for the computation of the 
numerical fluxes at the cell interfaces. The new scheme has been validated 
over a sample of numerical test problems in one, two and three spatial 
dimensions, exploring its ability in resolving the propagation of 
relativistic hydrodynamical and magnetohydrodynamical waves in different physical regimes. 
The astrophysical relevance of the new code for the study of the 
Richtmyer--Meshkov instability is briefly discussed in view of future applications.
\end{abstract}


\end{frontmatter}

% \linenumbers

\newcommand{\be}{\begin{equation}}
\newcommand{\ee}{\end{equation}}
\newcommand{\bdm}{\begin{displaymath}}
\newcommand{\edm}{\end{displaymath}}
\newcommand{\bea}{\begin{eqnarray}}
\newcommand{\eea}{\end{eqnarray}}
\newcommand{\PNM}{P_NP_M}
\newcommand{\halb}{\frac{1}{2}}
\newcommand{\FQi}{\tens{\mathbf{F}}\left(\Qi\right)}
\newcommand{\FQj}{\tens{\mathbf{F}}\left(\Qj\right)}
\newcommand{\FQjj}{\tens{\mathbf{F}}\left(\Qjj\right)}
\newcommand{\nj}{\vec n_j}
\newcommand{\FORCE}{\textnormal{FORCE}}
\newcommand{\GFORCE}{\textnormal{GFORCEN}}
\newcommand{\LF}{\textnormal{LF}'}
\newcommand{\LW}{\textnormal{LW}'}
\newcommand{\WL}{\mathcal{W}_h^-}
\newcommand{\WR}{\mathcal{W}_h^+}
\newcommand{\nur}{\boldsymbol{\nu}^\textbf{r} }
\newcommand{\nuf}{\boldsymbol{\nu}^{\boldsymbol{\phi}} }
\newcommand{\nut}{\boldsymbol{\nu}^{\boldsymbol{\theta}} }
\newcommand{\ar}{\phi_1\rho_1}
\newcommand{\arr}{\phi_2\rho_2}
\newcommand{\ur}{u_1^r}
\newcommand{\uf}{u_1^{\phi}}
\newcommand{\ut}{u_1^{\theta}}
\newcommand{\urr}{u_2^r}
\newcommand{\uff}{u_2^{\phi}}
\newcommand{\utt}{u_2^{\theta}}
\newcommand{\ub}{\textbf{u}_\textbf{1}}
\newcommand{\ubb}{\textbf{u}_\textbf{2}}
\newcommand{\RoeMat}{{\tilde A}_{\Path}^G} 
\renewcommand{\u}{\mathbf{u}}
\newcommand{\q}{\mathbf{q}}
\newcommand{\w}{\mathbf{w}}



\section{Introduction}
\label{introduction}

The numerical modeling of complex astrophysical flows
that involve relativistic processes requires the development
of more and more sophisticated codes. Relevant examples
of relativistic phenomena whose
understanding can greatly benefit from hydrodynamic and
magnetohydrodynamic numerical
simulations include extragalactic jets,
gamma-ray-bursts, accretion onto compact objects,
binary mergers of neutron stars (or black holes),
relativistic heavy-ion collisions and so on.

Until few years ago, most of the applications in
numerical relativistic hydrodynamic (RHD) and
magnetohydrodynamic (RMHD)
used second-order accurate, typically TVD, numerical codes. The
scientific progress that has been made possible by these
implementations is rather significant 
and not always appreciated enough,  
see the living reviews by
  \citet{Marti03} and by
  \citet{Font08} plus references therein.
However, the necessity of improving the accuracy of the
computations, especially in the presence of complex
phenomena such as
instabilities or turbulence, combined
with computational resources which are inevitably always
limited, 
has motivated a strong research effort along two different
directions. The first direction is represented by the 
development of high order schemes (better than second order in 
space and time), while the second direction consists 
in the implementation of efficient adaptive mesh 
refinement (AMR) algorithms. Taken separately, high order
numerical schemes and AMR techniques have a long history,
which is embarrassing to summarize in few words. The 
first high order special relativistic numerical scheme
is due to \citet{Dolezal1995}, who, in the context of
ultra-relativistic nuclear collision experiments,
implemented a conservative  
finite difference scheme using ENO reconstruction in space 
and Runge--Kutta time integration, but without Riemann solvers. 
The first transposition of this approach to the
astrophysical context is due to \citet{DelZanna2002}, who, in addition,   
used local Riemann problems to guarantee the upwind character
of the scheme.
Since then, several high order schemes
have been proposed and applied to a variety of different
astrophysical problems, with and without magnetic fields,
both in the special and in the
general relativistic regime [see \citet{DelZanna2003a}, \citet{DelZanna2007},
\citet{Tchekhovskoy2007}, 
\citet{Bucciantini2011}, \citet{Radice2011},
\citet{Radice2012a}]. 
Though different under many respects,
a common feature of all these
approaches is the use of a multi-step Runge--Kutta 
time integrator. 
A few years ago, \cite{DET2008}
proposed an alternative idea for obtaining a high order integration in time,
that avoids Runge--Kutta schemes altogether 
and originates from the ADER philosophy of 
\citet{Toro2002}. According to this idea, which is 
followed also in this work, an arbitrary high order 
numerical scheme with just one step for the time update can be 
obtained, 
provided a high order time evolution 
is performed locally (namely within each cell), 
for the reconstructed polynomials.
The first implementation of such ADER schemes in the  
context of ideal relativistic magnetohydrodynamics 
can be found in \citet{DBTM2008} and has later been 
successfully extended also to the non-ideal relativistic
MHD equations in \citet{Dumbser2009}. 

The implementation of AMR techniques has also a rich
tradition in relativistic hydrodynamics and
magnetohydrodynamics.\footnote{We are not mentioning here
the whole family of 
AMR implementations in vacuum space-times, namely without matter,
which were initiated by \citet{Choptuik89}.} 
The first occurrence is documented
in \citet{Winkler1984} 
%and \citet{Mihalas1984b}, 
followed by \citet{Donmez2004},
\citet{Evans2005a}, \citet{vanderHolst2008}, \citet{Tao2008},
\citet{Etienne:2010ui}, \citet{DeColle2011}, \citet{liebling_2010_emr}, \citet{Lehner2011}, \citet{East2012b}.

The combination of high order relativistic codes with AMR 
has a much more recent history. Relevant examples are given by 
the works of \citet{Mignone2012}, 
who combined the AMR library CHOMBO, which originated from the original 
work of \citet{Berger84}, \citet{Berger86} and \citet{Berger89}, with the versatile PLUTO code,
%\footnote{http://plutocode.ph.unito.it/}  
using a Corner-Transport-Upwind scheme together with a third-order WENO 
reconstruction in space; 
\citet{Anderson2006a}, 
who solved the equations of general relativistic magnetohydrodynamics using
a conservative finite difference scheme (with reconstruction of primitive variables 
plus Riemann solvers); 
\citet{Zhang2006}, who implemented both a conservative finite difference scheme 
(with reconstruction of fluxes and no need for Riemann solvers) and a finite volume 
method, within the block-structured AMR package PARAMESH of \citet{MacNeice00}.

Contrary to the above mentioned approaches,
all of them sharing
a TVD Runge--Kutta for the time integration,
in this paper we present an ADER-WENO finite volume
scheme for solving the special relativistic magnetohydrodynamics
equations, with adaptive mesh refinement. 
In \citet{Dumbser2012b} we have proposed the first
ADER-AMR finite volume numerical scheme for the Newtonian Euler
equations and here we propose  the
relativistic extension of our new approach. 
The use of a one-step scheme in time allows the implementation of time-accurate local time 
stepping (LTS) in a very natural and straight forward manner and has already been successfully 
applied in the context of high order Discontinuous Galerkin schemes with LTS (see 
\citet{Dumbser2007c,Taube2009,stedg1,stedg2}). 

The outline of this paper is the following. In 
Sect.~\ref{srh} we briefly recall the conservative
formulation of special relativistic
hydrodynamics. Sect.~\ref{Numerical_method} is devoted to
the description of the numerical method, while
Sect.~\ref{Results} contains the results of the new
scheme. Finally, Sect.~\ref{Conclusions} concludes our
work, with a discussion about future astrophysical applications.
In the following we will assume a signature $\{-,+,+,+\}$
for the space-time metric and we will use Greek letters
$\mu,\nu,\lambda,\ldots$ (running from 0 to 3) for
four-dimensional space-time tensor components, while
Latin letters $i,j,k,\ldots$ (running from 1 to 3) will
be employed for three-dimensional spatial tensor
components. Moreover, we set the speed of light $c=1$ and we adopt the Lorentz-Heaviside notation for the electromagnetic quantities, such that
all $\sqrt{4\pi}$ factors disappears.

%------------------------------------
\section{Special relativistic magnetohydrodynamics}
\label{srh}
%------------------------------------

In the following we consider a perfect magneto-fluid,
under the assumption of infinite conductivity (ideal RMHD), in a 
Minkowski space-time
with Cartesian coordinates, for which the metric is given by
\be
\mathrm{d}s^2=g_{\mu\nu}dx^\mu dx^\nu=-\mathrm{d}t^2+\mathrm{d}x^2+\mathrm{d}y^2+\mathrm{d}z^2\,.
\ee
The fluid is described by an energy momentum tensor $T^{\alpha\beta}$

\be
T^{\alpha\beta}=(\rho h+b^2) u^\alpha
u^\beta+(p + b^2/2) g^{\alpha\beta}- b^\alpha b^\beta \,,
%T^{\alpha\beta}=\rho h\,u^{\,\alpha}u^{\beta}+pg^{\,\alpha\beta}\,,
\label{eq:T_matter}
\ee
where $u^\alpha$ is the four-velocity of the fluid,
$b^\alpha$ is the four-vector of the magnetic field, $b^2=b_\alpha b^\alpha$, while $\rho$, $h=1+\epsilon + p/\rho$,
$\epsilon$ and $p$ are the rest-mass density, the specific enthalpy,
the specific internal energy, and the thermal pressure,
respectively. 
All these quantities are measured in the co-moving
frame of the fluid. We assume the pressure is related to $\rho$ and
$\epsilon$ through the 
ideal-gas equation of state (EOS), i.e.
\be
\label{eq:EOS}
p=\rho\epsilon(\gamma-1) \,,
\ee
where $\gamma$ is the (constant) adiabatic index of the gas. 
The equations of special relativistic magneto-hydrodynamics,
can be written in covariant form simply as
\bea
\label{0eq:mass}
&&\nabla_{\alpha} (\rho u^{\,\alpha})=0, \\
\label{0eq:momentum}
&&\nabla_{\alpha}T^{\alpha\beta}=0,\\
\label{0eq:maxwell}
&&\nabla_{\alpha}F^{\ast\alpha\beta}=0\,,
\eea
where $F^{\ast\alpha\beta}$ is the dual of the electromagnetic tensor~\citep{Anile_book}.
However, for numerical purposes it is convenient to recast them  
in conservative form as \citep{Marti91,Komissarov1999,Balsara2001} 
\be
\partial_t{\bf u} + \partial_i{\bf f}^i=0\,,
\label{eq:UFS}
\ee
where the conserved variables and the corresponding
fluxes in the $i$ direction are given by
\be
{\bf u}=\left[\begin{array}{c}
D \\ S_j \\ U \\ B^j 
\end{array}\right],~~~
{\bf f}^i=\left[\begin{array}{c}
 v^i D \\
 W^i_j \\
 S^i \\
\epsilon^{jik}E^k 
\end{array}\right]\,.
\label{eq:fluxes}
\ee
The conserved
variables $(D,S_j,U,B_j)$ 
are related to 
the rest-mass density $\rho$, to the thermal
pressure $p$, to the fluid velocity $v_i$ and to the magnetic field $B_i$ by\footnote{We note that, since the spacetime is flat and we are using Cartesian coordinates, the covariant and the contravariant components of spatial vectors can be used interchangeably, namely $A_i=A^i$, for the generic vector $\vec A$.}
\bea
\label{eq:cons1}
&&D   = \rho W ,\\
\label{eq:cons2}
&&S_i = \rho h W^2 v_i + \epsilon_{ijk}E_j B_k, \\
\label{eq:cons3}
&&U   = \rho h W^2 - p + \frac{1}{2}(E^2 + B^2)\,,
\eea
where $\epsilon_{ijk}$ is the Levi--Civita tensor and $\delta_{ij}$ is the Kronecker symbol.        
We have used $W=(1-v^2)^{-1/2}$ to denote the Lorentz factor of the fluid
with respect to the Eulerian observer at rest in the 
Cartesian coordinate system, while $E_i$ are the components of the electric field, which, in ideal magnetohydrodynamics, is simply given by
$\vec E = -\vec v \times \vec B$.
The tensor
\be
W_{ij} \equiv \rho h W^2 v_i v_j - E_i E_j - B_i B_j + \left[p +\frac{1}{2}(E^2+B^2)\right]\delta_{ij} \\
\label{eq:W} 
\ee
is the fully spatial projection of the energy-momentum
tensor of the fluid [see also \citet{DelZanna2007}].
In order to preserve the divergence-free property of the magnetic field, we augment the system (\ref{eq:UFS}) with an additional equation for a scalar 
field $\Phi$, whose role is to propagate away the deviations from $\vec \nabla\cdot\vec B=0$. In practice, we need to solve 
\be
\label{eq:divB}
\partial_t \Phi + \partial_i B^i = -\kappa \Phi\,,
\ee
while the fluxes for the evolution of the magnetic field are also modified, namely ${\bf f}^i(B^j)\rightarrow \epsilon^{jik}E^k + \Phi \delta^{ij}$.
This approach has been introduced by \citet{Dedner:2002} for the classical MHD equations, and it has been later extended to the relativistic regime by \citet{Palenzuela:2008sf}.
We emphasize that the damping coefficient $\kappa$ in Eq.~\eqref{eq:divB} drives the solution towards  $\vec \nabla\cdot\vec B=0$ over a timescale $1/\kappa$.
In our simulations with a non-zero magnetic field we have used $\kappa\in[1;10]$ (see also \citet{Dionysopoulou:2012pp}).
As a hyperbolic system in conservative form, the
equations (\ref{eq:UFS}) admit a well defined Jacobian
matrix with a corresponding set of eigenvectors and
eigenvalues. All these properties have been investigated
deeply over the years. 
We address the interested
reader to \citet{Aloy1999c,Marti03,Rezzolla_book:2013} 
and to \citet{Balsara99,Komissarov1999,DelZanna2007,Anton2010} 
for the  mathematical aspects
of the RHD and of the RMHD equations, respectively.
We just comment here on the method adopted to invert the system 
(\ref{eq:cons1})--(\ref{eq:cons3}), which is needed to recover the primitive variables 
$(p,\rho,v_i,B_i)$ from the conserved variables $(D,S_i,U,B_i)$. 
As well known, in the relativistic framework such a conversion is not analytic, and a numerical root-finding 
approach is therefore needed.  In our numerical code we have at disposal two alternative strategies.
The first one is due to \citet{DelZanna2007}, and it can be used both in the case of purely RHD systems and 
of RMHD systems.
It is based on  the property that, after introducing the variables $x=v^2$, $y=\rho h W^2$, the whole problem can be reduced to the solution of the system $F_1=0$ and $F_2=0$, where
\bea
\label{F1}
F_1(x,y)=x (y+B^2)^2 - y^{-2}(S_i B^i)^2(2y + B^2)-  S^2\,,\\
\label{F2}
F_2(x,y)=y - p + \frac{1}{2}(1+x)B^2-\frac{1}{2}y^{-2} (S_i B^i)^2- U\,,
\eea
where $S^2=S^i S_i$, $B^2=B^i B_i$. Moreover, since the pressure of the ideal gas EOS (\ref{eq:EOS})
can be written as
\be
p=\frac{\gamma-1}{\gamma}\left[y(1-x)-D\sqrt{1-x}\right]\,,
\ee
the second equation of the system above, namely Eq.~(\ref{F2}), further reduces to an equation in the single unknown $y$, which can be solved
with any root-finding algorithm. We emphasize that this strategy is our only choice when magnetic fields are present.
The second strategy that we have implemented can instead be used only for RHD systems
and it is  very similar to that of \citet{Baiotti2004thesis} [see also Appendix D of 
\citet{Rezzolla_book:2013}]. It requires the numerical solution of the following equation
\begin{equation}
\label{cons2prim-second-choice}
p-\bar{p}[\rho({\bf u},p),\epsilon({\bf u},p)]=0\,,
\end{equation}
where $p$ is the unknown pressure and $\bar{p}$ is the pressure as obtained through the equation of state (\ref{eq:EOS}) 
written in terms of the conserved variables  ${\bf u}$ and of $p$ itself. In both the strategies just described, the standard Newton--Raphson method has been adopted, combined to a bisection step for those cases when the Newton--Raphson fails.
We emphasize that, for flows with large Lorentz factors, i.e. $W\geq 10$, the second strategy behaves better that the first one, which instead manifests a higher failure rate.
In Sect.~\ref{sec:fva} we 
discuss the strategy undertaken when such failures occur.

%------------------------------------
\section{Numerical method}
\label{Numerical_method}
%------------------------------------

As written like in Eq.~(\ref{eq:UFS}), the equations of special
relativistic hydrodynamics can be solved through a
modern version of the ADER approach, namely by combining a
high order WENO reconstruction in space with a
local space-time Galerkin predictor strategy for the time
update. In the following we provide a concise
explanation of the overall procedure, addressing to the
relevant literature for an extended discussion, and
especially to \citet{DET2008}, \citet{Dumbser2009} and 
\citet{Dumbser2012b}.

%----------------------------------------
\subsection{The Finite volume approach}
\label{sec:fva}
According to the standard finite volume approach,
a one-step discretization of the system of hyperbolic 
conservation laws (\ref{eq:UFS}) 
is obtained by integration of the PDE over a space-time control volume 
$\mathcal{C}_{ijkn}=[x_{i-\halb};x_{i+\halb}] \times [y_{j-\halb};y_{j+\halb}] \times [z_{k-\halb};z_{k+\halb}] \times [t^n;t^{n+1}]$ as 
\footnote{To keep the
notation simpler, in the following we avoid the Einstein summation
convention for the flux components, by writing
explicitly ${\bf f}={\bf f}^x$,${\bf g}={\bf f}^y$,${\bf
  h}={\bf f}^z$.} 
\begin{eqnarray}
\label{eq:finite_vol}
{\bf \bar u}_{ijk}^{n+1}&=&{\bf \bar u}_{ijk}^{n}-\frac{\Delta t}{\Delta x_i}\left({\bf f}_{i+\halb,jk}-{\bf f}_{i-\halb,jk} \right)-\frac{\Delta t}{\Delta y_j}\left({\bf g}_{i,j+\halb,k}-{\bf g}_{i,j-\halb,k} \right)
-\frac{\Delta t}{\Delta z_k}\left({\bf h}_{ij,k+\halb}-{\bf h}_{ij,k-\halb} \right)
\end{eqnarray}
where
\begin{equation}
{\bf \bar u}_{ijk}^{n}=\frac{1}{\Delta x_i}\frac{1}{\Delta y_j}\frac{1}{\Delta z_k}\int \limits_{x_{i-\halb}}^{x_{i+\halb}}\int \limits_{y_{j-\halb}}^{y_{j+\halb}}\int \limits_{z_{k-\halb}}^{z_{k+\halb}}{ \bf u}(x,y,z,t^n)dz\,\,dy\,\,dx
\end{equation}
is the spatial average of the solution in the element 
$I_{ijk}=[x_{i-\halb};x_{i+\halb}]\times[y_{j-\halb};y_{j+\halb}]\times[z_{k-\halb};z_{k+\halb}]$
at time $t^n$, while 
\begin{equation}
\label{flux:F}
{\bf f}_{i+\halb,jk}= \frac{1}{\Delta t}\frac{1}{\Delta y_j}\frac{1}{\Delta z_k} \hspace{-1mm}  \int \limits_{t^n}^{t^{n+1}} \! \int \limits_{y_{j-\halb}}^{y_{j+\halb}}\int \limits_{z_{k-\halb}}^{z_{k+\halb}} \hspace{-1mm} 
{\bf \tilde f}_{{\rm RP}}(x_{i+\halb},y,z,t) \, dz \, dy \, dt\,,
%\! \left({\bf q}_h^-(x_{i+\halb},y,z,t),{\bf q}_h^+(x_{i+\halb},y,z,t)\right) dz \, dy \, dt, 
\end{equation}
\begin{equation}
\label{flux:G}
{\bf g}_{i,j+\halb,k}=\frac{1}{\Delta t}\frac{1}{\Delta x_i}\frac{1}{\Delta z_k} \hspace{-1mm}  \int \limits_{t^n}^{t^{n+1}} \! \int \limits_{x_{i-\halb}}^{x_{i+\halb}}\int \limits_{z_{k-\halb}}^{z_{k+\halb}} \hspace{-1mm} 
{\bf \tilde g}_{{\rm RP}}(x,y_{j+\halb},z,t)\!dz\,dx\,dt,
% \! \left({\bf q}_h^-(x,y_{j+\halb},z,t),{\bf q}_h^+(x,y_{j+\halb},z,t)\right) dz\,dx\,dt, \\
\end{equation}
\begin{equation}
\label{flux:H}
{\bf h}_{ij,k+\halb}=\frac{1}{\Delta t}\frac{1}{\Delta x_i}\frac{1}{\Delta y_j} \hspace{-1mm}  \int \limits_{t^n}^{t^{n+1}} \! \int \limits_{x_{i-\halb}}^{x_{i+\halb}}\int \limits_{y_{j-\halb}}^{y_{j+\halb}} \hspace{-1mm}  
{\bf \tilde h}_{{\rm RP}}(x,y,z_{k+\halb},t)\!dy\,dx\,dt
%\! \left({\bf q}_h^-(x,y,z_{k+\halb},t),{\bf q}_h^+(x,y,z_{k+\halb},t)\right) dy\,dx\,dt, 
\end{equation}
are the space--time averaged numerical fluxes, with 
${\bf \tilde f}_{{\rm RP}}(x_{i+\halb},y,z,t)$ being a 
numerical flux function, which is obtained by solving the 
Riemann problem at an element interface, either exactly or approximately, 
at a generic time $t$. 
Equation (\ref{eq:finite_vol}) has been derived directly from 
the conservation equations and is an exact integral relation. 
In order to be used as a high order numerical one-step scheme in space and 
time, a number of actions must be taken,  which are discussed in more detail in 
Sect.~\ref{sec:A dimension-by-dimension WENO reconstruction} and Sect.~\ref{sec:The local space-time DG approach}.
The first one consists of 
a high order reconstruction of the solution within each 
control volume $I_{ijk}$, starting from the known cell averages 
${\bf \bar  u}_{ijk}^{n}$, which are the only quantities that are 
directly evolved in time via Eq.~(\ref{eq:finite_vol}). 
Such a reconstruction of the conserved variables
is performed through a high order 
weighted essentially non-oscillatory (WENO) interpolation, 
and provides a polynomial representation of the solution, 
as we will discuss below. 
In some cases, in order to reduce possible oscillations and to increase the accuracy, the 
reconstruction has been performed on the characteristic variables \citep{eno, Jiang1996} which are then transformed back to the conserved ones~\citep{Balsara2001,Toro09}.
However, for challenging problems such as the one described in Sect.~\ref{sec:RP2D}, the reconstruction (either in conserved or in characteristic variables) can 
cause failures when the conversion to the primitive variables is performed. Under these circumstances, 
it is advantageous to reduce the order of accuracy in the troubled cells, following 
the "a-posteriori'' MOOD strategy proposed by \citet{MOOD}
and applied here to the relativistic framework for the first time.

The second action required by our one-step time-update numerical scheme 
consists of computing, locally for each
space-time cell $\mathcal{C}_{ijkn}$, the time evolution of the 
reconstructed polynomial, which in the following we denote as 
$\mathbf{q}_h(x,y,z,t)$. This, in turns, allows to compute the 
solution of the Riemann problem at the cell interfaces as 
\begin{equation}
{\bf \tilde f}_{{\rm RP}}={\bf \tilde f} \! \left({\bf q}_h^-(x_{i+\halb},y,z,t),{\bf q}_h^+(x_{i+\halb},y,z,t)\right)\,,
\end{equation}
where ${\bf q}_h^-$ and ${\bf q}_h^+$ are the left and 
right boundary extrapolated states at the interface, 
respectively. The
form of the function ${\bf \tilde f}$ depends, of course,
on the specific choice of the Riemann solver
adopted. 
Although exact Riemann solvers exist both for
special relativistic hydrodynamics and magnetohydrodynamics
\citep{Marti94,Rezzolla01,Giacomazzo:2005jy}, 
faster approximate Riemann solvers are typically
preferred. In our calculations we have adopted two of them. The first one 
is the popular HLL Riemann solver by \cite{Harten83}, which does not rely on the
characteristic structure of the equations and only needs the knowledge of the fastest and of the slowest of the eigenvalues.
The second one is the Osher-type Riemann solver in the version proposed by 
\cite{OsherUniversal} and it is based on the knowledge of the full spectral decomposition of the equations. 
Unlike the Roe Riemann solver, however, it is entropy-satisfying and thus does not produce unphysical rarefaction 
shocks. 
Because of the higher computational cost and complexity, 
we have used the Osher-type Riemann solver only for RHD systems, while for RMHD systems we have used the simpler 
HLL Riemann solver.

%-----------------------------------------------
\subsection{A dimension-by-dimension WENO reconstruction}
\label{sec:A dimension-by-dimension WENO reconstruction}
The WENO reconstruction that we have implemented is 
genuinely multi-dimensional, but, from an operational point of view, it works in a
dimension-by-dimension fashion. In practice,  we first 
introduce spacetime reference coordinates $\xi,\eta,\zeta,\tau\in[0,1]$, which are defined by
\begin{equation}
\label{eq:xi}
x = x_{i-\halb} + \xi   \Delta x_i, \quad 
y = y_{j-\halb} + \eta  \Delta y_j, \quad 
z = z_{k-\halb} + \zeta \Delta z_k, \quad
t = t^n + \tau \Delta t\,.
\end{equation} 
After that, focusing on
the $x$ direction for convenience,
we choose the degree $M$ of the polynomial approximating the
solution, and an orthogonal basis of polynomials~\citep{Solin2006},
all of degree $M$, rescaled on the same unit interval
$[0,1]$. The basis is formed by the $M+1$
Lagrange interpolation polynomials, 
$\{\psi_l(\lambda)\}_{l=1}^{M+1}$, passing through the $M+1$
Gauss-Legendre quadrature nodes  
$\{\lambda_k\}_{k=1}^{M+1}$ and with the standard property that
\begin{equation}
\psi_l(\lambda_k)=\delta_{lk}\hspace{1cm}l,k=1,2,\ldots, M+1\,,
\end{equation}
where $\delta_{lk}$ is the "Kronecker delta'', i.e. $\delta_{lk}=1$ if $l=k$, $\delta_{lk}=0$ otherwise.
Having done that, a (small) number of one-dimensional
reconstruction stencils is adopted, each of them formed by
the union of $M+1$ adjacent cells, i.e.
\begin{equation}
\label{eqn.stencildef}  
\mathcal{S}_{ijk}^{s,x} = \bigcup \limits_{e=i-L}^{i+R}
        {I_{ejk}}, \quad 
\end{equation}
where $L=L(M,s)$ and $R=R(M,s)$ are the 
spatial extension of the stencil to the left and to the
right. In practice, we have adopted the pragmatic
approach for which 
odd order schemes (even polynomials
of degree $M$) always use three stencils ($N_s=3$),
while 
even order schemes (odd polynomials of degree $M$) always
adopt four stencils ($N_s=4$), 
with the exception of the $M=1$ case, for which there are only two stencils.
For the sake of clarity, in Appendix A we have reported the coordinates of the Gaussian 
points, the nodal basis polynomials and the corresponding stencils for a few values of 
$M$ up to $M=4$.

With all this machinery at hands, we use the polynomial 
basis functions to reconstruct the solution at time $t^n$
as\footnote{Note that here, and in the following, the 
Einstein summation convention is used over the repeated
index $p$, even if such an index does not denote the
covariant and contravariant
components of a tensor.}
\begin{equation}
\label{eqn.recpolydef.x} 
 \w^{s,x}_h(x,t^n) = \sum \limits_{p=0}^M \psi_p(\xi) \hat \w^{n,s}_{ijk,p} := \psi_p(\xi) \hat \w^{n,s}_{ijk,p}\,, 
\end{equation}
As usual for finite volume methods, the reconstructed
polynomial must preserve the cell-average of the solution
over each element $I_{ijk}$, namely
\begin{equation}
 \frac{1}{\Delta x_e} \int _{x_{e-\halb}}^{x_{e+\halb}} \psi_p(\xi(x)) \hat \w^{n,s}_{ijk,p} \, dx = {\bf \bar u}^n_{ejk}, \qquad \forall {I}_{ejk} \in \mathcal{S}_{ijk}^{s,x}\,,      
 \label{eqn.rec.x} 
\end{equation}
which provide a system of linear equations for the
unknown coefficients $\hat \w^{n,s}_{ijk,p}$. 
This operation is repeated for each 
stencil relative to the element $I_{ijk}$.
After that, we can construct a data-dependent nonlinear combination of the 
polynomials computed from each stencil, i.e. 
\begin{equation}
\label{eqn.weno} 
 \w_h^x(x,t^n) = \psi_p(\xi) \hat \w^{n}_{ijk,p}, \quad \textnormal{ with } \quad  
 \hat \w^{n}_{ijk,p} = \sum_{s=1}^{N_s} \omega_s \hat \w^{n,s}_{ijk,p}\,.
\end{equation}   
The nonlinear weights $\omega_s$ are 
computed following the same logic as for the optimal WENO of
\cite{Jiang1996}, i.e. 
\begin{equation}
\omega_s = \frac{\tilde{\omega}_s}{\sum_k \tilde{\omega}_k}\,,  \qquad
\tilde{\omega}_s = \frac{\lambda_s}{\left(\sigma_s + \epsilon \right)^r}\,. 
\label{eqn.omegas}  
\end{equation} 
However, the actual values of the linear weights ${\lambda}_s$ are not the same as
those of the optimal WENO and they are chosen according to
a more pragmatic approach. In fact, the weight 
of the central stencils is  given a very large value, 
${\lambda}_s=10^5$, while the weight of the one-sided stencils is set to
${\lambda}_s=1$. 
Moreover, in our implementation we have used\footnote{It has been shown that
the numerical results are substantially independent of $\epsilon$ and $r$~\citep{Liu1994}.}
$\epsilon=10^{-14}$ and $r=8$.
The oscillation indicator $\sigma_s$ of Eq.~(\ref{eqn.omegas}) is
\begin{equation}
\sigma_s = \Sigma_{pm} \hat \w^{n,s}_{ijk,p} \hat \w^{n,s}_{ijk,m}\,,
\label{eqn.sigmas} 
\end{equation}
and it requires the computation of the oscillation indicator matrix \citep{DET2008}
\begin{equation}
\Sigma_{pm} = \sum \limits_{\alpha=1}^M \int \limits_0^1 \frac{\partial^\alpha \psi_p(\xi)}{\partial \xi^\alpha} \cdot \frac{\partial^\alpha \psi_m(\xi)}{\partial \xi^\alpha} d\xi\,,
\end{equation}
which, compared to alternative expressions proposed in the literature, has the advantage that it does not depend on the grid spacing, and is therefore "universal''.
We emphasize 
that the reconstruction polynomial $\w_h^x(x,t^n)$ resulting from Eq.~(\ref{eqn.weno}) 
 is still an average in the $y$ and $z$
directions. Hence, the procedure explained so far is
repeated along the two missing directions $y$ 
and $z$.
The net effect of this approach is to provide a genuine 
multidimensional reconstruction, although in such a way that each direction is treated separately.

%------------------------------------------------------
\subsection{The local space-time DG approach}
\label{sec:The local space-time DG approach}
The high order computation of the numerical fluxes
(\ref{flux:F})--(\ref{flux:H}), which contain a time
integration from $t^n$ to $t^{n+1}$,
requires the numerical flux (Riemann solver) ${\bf \tilde f}_{{\rm
RP}}$ to be computed with high accuracy at any time
in the interval $t\in[t^n;t^{n+1}]$. To this extent,
we need an operation, to be performed locally for each
cell, which uses as input the high order polynomial
$\w_h$ obtained from the WENO reconstruction, and gives
as output its evolution in time, namely
\begin{equation}
\label{LSDG}
\w_h(x,y,z,t^n)\longrightarrow \q_h(x,y,z,t)\,.
\end{equation} 
Unlike the original ADER approach, where this operation
was obtained through the so called Cauchy-Kovalewski
procedure~\citep{Toro2005,Dumbser2007}, in the modern
version of ADER the transformation represented by
(\ref{LSDG}) is obtained through an 
element--local space--time Discontinuous Galerkin predictor that 
is based on the \textit{weak} integral form of
Eq.~(\ref{eq:UFS})~\citep{DET2008}.
The basic idea can be summarized as follows.
The sought polynomial $\q_h(x,y,z,t)$ is supposed to be expanded in space and time as
\begin{equation}
 \mathbf{q}_h = \mathbf{q}_h(\boldsymbol{\xi},\tau) = \theta_\mathfrak{p}\left(\boldsymbol{\xi},\tau \right) \hat \q_\mathfrak{p}\,,   
 \label{eqn.st.q} 
\end{equation}
where the polynomial basis functions
$\theta_\mathfrak{p}$ are given by a tensor--product of the basis functions 
$\psi_l$ already used for the WENO reconstruction, namely
\begin{equation}
  \theta_\mathfrak{p}(\boldsymbol{\xi},\tau) = \psi_p(\xi) \psi_q(\eta) \psi_r(\zeta) \psi_s(\tau)\,. 
\end{equation} 
The terms $\hat \q_\mathfrak{p}$ are the so-called
degrees of freedom and they are the unknowns of the problem.
After multiplying the governing 
PDE of Eq.~\eqref{eq:UFS}, written in the reference coordinates $(\xi,\eta,\zeta,\tau)$, 
with the space--time test functions $\theta_\mathfrak{q}$ and integrating over
the space--time reference control volume, we obtain\footnote{Here we have defined
\begin{equation}
{\bf f}^\ast= \frac{\Delta t}{\Delta x_i} \, {\bf f}, \quad 
{\bf g}^\ast= \frac{\Delta t}{\Delta y_j} \, {\bf g}, \quad 
{\bf h}^\ast= \frac{\Delta t}{\Delta z_k} \, {\bf h}\,. \quad 
\end{equation}
}
\begin{equation}
 \int \limits_{0}^{1} \int \limits_{0}^{1}  \int \limits_{0}^{1}   \int \limits_{0}^{1}   
\theta_\mathfrak{q} \left( 
  \frac{\partial{\bf u}}{\partial \tau} + \frac{\partial \mathbf{f}^\ast}{\partial \xi} + \frac{\partial \mathbf{g}^\ast}{\partial \eta} + \frac{\partial \mathbf{h}^\ast}{\partial \zeta} \right) d\xi d\eta d\zeta d\tau = 0.  
\label{eqn.pde.weak1} 
\end{equation}
The key aspect of the whole strategy is to perform an integration by parts in time, while keeping the 
treatment local in space. After doing so, we get
\begin{eqnarray}
 && \int \limits_{0}^{1} \int \limits_{0}^{1}  \int \limits_{0}^{1} \theta_\mathfrak{q}(\boldsymbol{\xi},1) \u(\boldsymbol{\xi},1) d\xi d\eta d\zeta - 
  \int \limits_{0}^{1} \int \limits_{0}^{1}  \int \limits_{0}^{1}   \int \limits_{0}^{1} \left( \frac{\partial}{\partial \tau} \theta_\mathfrak{q} \right) \u d\xi d\eta d\zeta d\tau   \nonumber \\ 
 && + \int \limits_{0}^{1} \int \limits_{0}^{1}  \int \limits_{0}^{1}   \int \limits_{0}^{1} \left[   
    \theta_\mathfrak{q} \left( 
   \frac{\partial \mathbf{f}^\ast}{\partial \xi} + \frac{\partial \mathbf{g}^\ast}{\partial \eta} + \frac{\partial \mathbf{h}^\ast}{\partial \zeta}  \right) \right] d\xi d\eta d\zeta d\tau 
   = \int \limits_{0}^{1} \int \limits_{0}^{1}  \int \limits_{0}^{1} \theta_\mathfrak{q}(\boldsymbol{\xi},0) \u(\boldsymbol{\xi},t^n) d\xi d\eta d\zeta.  
\label{eqn.pde.weak2} 
\end{eqnarray}
In the first two integrands of Eq.~(\ref{eqn.pde.weak2}) we can perform the replacement $\u \rightarrow \mathbf{q}_h$, 
since $\mathbf{q}_h$ is the discrete space-time solution we are looking  for.
In the integrand on the right hand side of Eq.~(\ref{eqn.pde.weak2}), on the other hand, we 
can perform the replacement $\u(\boldsymbol{\xi},t^n) \rightarrow \w_h(\boldsymbol{\xi},t^n)$,
since at time $t^n$ the solution is known, and it is represented by the reconstructed polynomial computed according to the procedure described in Sect.~\ref{sec:A dimension-by-dimension WENO reconstruction}. In addition to this,
we assume that the fluxes too can be expanded over the basis as we did in Eq.~(\ref{eqn.st.q}), namely 
\begin{equation}
 \mathbf{f}^{\ast}_h = \theta_\mathfrak{p} \hat{\mathbf{f}}^{\ast}_\mathfrak{p},   \qquad 
 \mathbf{g}^{\ast}_h = \theta_\mathfrak{p} \hat{\mathbf{g}}^{\ast}_\mathfrak{p},   \qquad 
 \mathbf{h}^{\ast}_h = \theta_\mathfrak{p} \hat{\mathbf{h}}^{\ast}_\mathfrak{p}\,.   \qquad 
  \label{eqn.st.fs} 
\end{equation}
From the computational point of view, the advantage of the nodal basis becomes apparent at this stage. In fact,
the above degrees of freedom for the fluxes are simply the point--wise evaluation of the physical fluxes, hence 
\begin{equation}
 \hat{\mathbf{f}}^{\ast}_\mathfrak{p} = {\mathbf{f}}^{\ast}\left( \hat \q_\mathfrak{p} \right), \qquad 
 \hat{\mathbf{g}}^{\ast}_\mathfrak{p} = {\mathbf{g}}^{\ast}\left( \hat \q_\mathfrak{p} \right), \qquad 
 \hat{\mathbf{h}}^{\ast}_\mathfrak{p} = {\mathbf{h}}^{\ast}\left( \hat \q_\mathfrak{p} \right). \qquad 
  \label{eqn.nodal.eval} 
\end{equation}
Had a {\emph modal basis} been adopted instead, $\hat{\mathbf{f}}^{\ast}_\mathfrak{p}$, $\hat{\mathbf{g}}^{\ast}_\mathfrak{p}$
 and  $\hat{\mathbf{h}}^{\ast}_\mathfrak{p}$
 could have been obtained only after performing a
  time-consuming ``L2-projection''.
Inserting Eqns. \eqref{eqn.st.q} and \eqref{eqn.st.fs} into \eqref{eqn.pde.weak2} yields 
\begin{eqnarray}
 && \int \limits_{0}^{1} \int \limits_{0}^{1}  \int \limits_{0}^{1} \theta_\mathfrak{q}(\boldsymbol{\xi},1) \theta_\mathfrak{p}(\boldsymbol{\xi},1) \hat \q_\mathfrak{p}
  \, d\xi d\eta d\zeta 
 - \int \limits_{0}^{1} \int \limits_{0}^{1}  \int \limits_{0}^{1}   \int \limits_{0}^{1} \left(\frac{\partial}{\partial \tau} \theta_\mathfrak{q} \right) \theta_\mathfrak{p} \hat \q_\mathfrak{p}  
 \, d\xi d\eta d\zeta d\tau 
   \nonumber \\ 
 && + \int \limits_{0}^{1} \int \limits_{0}^{1}  \int \limits_{0}^{1}   \int \limits_{0}^{1} \left[   
     \theta_\mathfrak{q} \left( 
     \frac{\partial}{\partial \xi}   \theta_\mathfrak{p} \hat{\mathbf{f}}^{\ast}_\mathfrak{p} 
   + \frac{\partial}{\partial \eta}  \theta_\mathfrak{p} \hat{\mathbf{g}}^{\ast}_\mathfrak{p} 
   + \frac{\partial}{\partial \zeta} \theta_\mathfrak{p} \hat{\mathbf{h}}^{\ast}_\mathfrak{p} 
     \right) \right] \, d\xi d\eta d\zeta d\tau 
   \nonumber \\ 
   && = \int \limits_{0}^{1} \int \limits_{0}^{1}  \int \limits_{0}^{1} \theta_\mathfrak{q}(\boldsymbol{\xi},0) \w_h(\boldsymbol{\xi},t^n) \, d\xi d\eta d\zeta.  
\label{eqn.pde.weak3} 
\end{eqnarray}
It may be noticed that Eq.~(\ref{eqn.pde.weak3}) contains several integrals which only involve the basis functions and their derivatives, and which can be pre-computed in the code. 
Hence, after defining the integrals 
\begin{eqnarray}
 \mathbf{K}^1_{\mathfrak{q} \mathfrak{p}} &=& \int \limits_{0}^{1} \int \limits_{0}^{1}  \int \limits_{0}^{1} \theta_\mathfrak{q}(\boldsymbol{\xi},1) \theta_\mathfrak{p}(\boldsymbol{\xi},1) d \boldsymbol{\xi} - 
                \int \limits_{0}^{1} \int \limits_{0}^{1}  \int \limits_{0}^{1}   \int \limits_{0}^{1} \left(\frac{\partial}{\partial \tau} \theta_\mathfrak{q} \right) \theta_\mathfrak{p} d \boldsymbol{\xi} d\tau\,,\\
 \mathbf{K}^{\boldsymbol{\xi}}_{\mathfrak{q} \mathfrak{p}} &=& \left( \mathbf{K}^{\xi}_{\mathfrak{q} \mathfrak{p}} , \mathbf{K}^{\eta}_{\mathfrak{q} \mathfrak{p}}, \mathbf{K}^{\zeta}_{\mathfrak{q} \mathfrak{p}} \right) =  \int \limits_{0}^{1} \int \limits_{0}^{1}  \int \limits_{0}^{1} \int \limits_{0}^{1} \theta_\mathfrak{q}  
     \frac{\partial}{\partial \boldsymbol{\xi}}  \theta_\mathfrak{p} d \boldsymbol{\xi} d\tau\,,\\
  \mathbf{F}^0_{\mathfrak{q} \mathfrak{p}} &=& \int \limits_{0}^{1} \int \limits_{0}^{1}  \int \limits_{0}^{1} \theta_\mathfrak{q}(\boldsymbol{\xi},0) \psi_\mathfrak{m}(\boldsymbol{\xi}) d \boldsymbol{\xi}\,, 
\end{eqnarray}
where $d \boldsymbol{\xi} = d\xi d\eta d\zeta$, we can rewrite the system (\ref{eqn.pde.weak3})
in a compact form as an algebraic equation system for the unknown coefficients $\hat \q_\mathfrak{p}$, i.e.
\begin{equation}
 \mathbf{K}^1_{\mathfrak{q} \mathfrak{p}}   \hat \q_\mathfrak{p} + 
 \mathbf{K}^\xi  _{\mathfrak{q} \mathfrak{p}} \cdot \hat{\mathbf{f}}^\ast_\mathfrak{p} + 
 \mathbf{K}^\eta _{\mathfrak{q} \mathfrak{p}} \hat{\mathbf{g}}^\ast_\mathfrak{p} + 
 \mathbf{K}^\zeta_{\mathfrak{q} \mathfrak{p}} \hat{\mathbf{h}}^\ast_\mathfrak{p} = 
 \mathbf{F}^0_{\mathfrak{q} \mathfrak{m}} \hat{\mathbf{w}}_{\mathfrak{m}}^n\,.
\label{eqn.pde.weak4} 
\end{equation} 
This system of equations must be solved approximately through standard iterative procedures up to a desired tolerance, and it represents the single most
expensive routine in a typical simulation, amounting to $\sim [24-28]\%$ of the total CPU time.
It is also interesting to note that the ADER approach becomes particularly convenient in case of relativistic hydrodynamics. In fact, by avoiding 
the sub-steps typical of Runge--Kutta schemes, it  avoids the conversion from conserved to primitive variables that would be required at each sub-step, and that is always a delicate and time-consuming operation. Again in a typical simulation, the CPU time spent in the conversion from the conserved to the primitive variables is $~\sim[12-16]\%$ of the total CPU time.

The proposed approach has been shown to cope
successfully even with stiff source
terms~\citep{Dumbser2009, Zanotti2011b}. However, there is no such 
a necessity in the case of special relativistic hydrodynamics, whose
source terms vanish in Cartesian coordinates.
%------------------------------------------------------
\subsection{Adaptive mesh refinement}
AMR algorithms can be roughly divided in two main categories. 
In the first category, nested arrays of logically rectangular grid patches are used, according to the 
original Berger-Colella-Oliger approach~\citep{Berger84,Berger89}. 
In the second category, a 'cell-by-cell'  
refinement is instead adopted, and the resulting cells are managed as elements of a tree-data structure.
It is in this version that AMR techniques were first applied to astrophysics, for performing N-body cosmological simulations~\citep{Khokhlov1997}, and this is also the
strategy that we have implemented. We refer to \citet{Dumbser2012b} for a detailed description of our AMR algorithm, which was extensively validated for the classical Euler and magnetohydrodynamics equations.
We just recall here that, as in any other AMR algorithm, there are a number of 
free parameters, namely
\begin{itemize}
\item
The maximum level of refinement $\ell_{\rm max}$, typically $2$ or $3$ in our tests.
\item
The refinement factor $\mathfrak{r}$, governing the number of sub-cells that are generated according to 
\begin{equation}
\label{refine-factor}
\Delta x_{\ell} = \mathfrak{r} \Delta x_{\ell+1}\, \quad \Delta y_{\ell} = \mathfrak{r} \Delta y_{\ell+1} \, \quad 
\Delta z_{\ell} = \mathfrak{r} \Delta z_{\ell+1}, 
\end{equation}
where $\Delta x_{\ell}$ is size of the cell, at $\ell$ refinement level, along the $x$-direction, and similarly for the other directions. 
\item
The choice of the refinement criterion, which in our approach is simply feature-based, uses the calculation of a second derivative, see \citet{Lohner1987}.  
In practice, a cell ${\mathcal C}_m$ is marked for refinement  if $\chi_m>\chi_{\rm ref}$, while it is marked for 
re-coarsening if $\chi_m<\chi_{\rm rec}$, where
\begin{equation}
\chi_m=\sqrt{\frac{\sum_{k,l} (\partial^2 \Phi/\partial x_k \partial x_l)^2 }{\sum_{k,l}[(|\partial \Phi/\partial x_k|_{i+1}+|\partial \Phi/\partial x_k|_i)/\Delta x_l+\varepsilon|(\partial^2 /\partial x_k \partial x_l )||\Phi|]^2} }\,.
\label{eqn.indicator}
\end{equation}
The summation $\sum_{k,l}$ is taken over the number of space dimension of the problem in order to include the cross term derivatives. The function $\Phi=\Phi(\u)$ can be any suitable indicator 
function of the conserved variables $\u$ and in all our tests we have used $\Phi(\u)=D=\rho W$.
We have observed that the threshold values   $\chi_{\rm ref}$ and $\chi_{\rm rec}$ can be slightly model dependent. In most of our tests 
we have chosen $\chi_{\rm ref}$ in the range $\sim[0.2,0.25]$ and $\chi_{\rm rec}$ in the 
range $\sim[0.05,0.15]$. 
Moreover, the first and second derivatives involved in the definition of $\chi_m$ are computed through standard finite-differencing based on the
cell averages of $\u$.
Finally, the parameter $\varepsilon$ acts as a filter preventing  refinement in regions of small  ripples and is given the value $\varepsilon = 0.01$.
In alternative, we refer the reader to the use of error based adaptation as investigated by \citet{Mulet2} in the context of the high order WENO AMR schemes 
first proposed by \citet{Mulet1}. 
\end{itemize}

%-----------------------------------------------------------
\section{Numerical tests}
\label{Results}
%-----------------------------------------------------------
In this Section we consider a wide set of numerical test cases, both for  RHD systems, discussed in Sect.~\ref{sec:isen1D}--\ref{EP3D}, and for
RMHD systems, discussed in Sect.~\ref{RMHD-Alfven}--\ref{RMHD-OT}. The analysis of the order of convergence is deferred to Sect.~\ref{RMHD-Alfven}.
%-------------------------------------------------------------------------
\subsection{RHD one-dimensional isentropic flow}
\label{sec:isen1D}
%
%

%
%----------------------------------------------------------
\begin{figure}
\begin{center}
{\includegraphics[angle=0,width=6.5cm,height=6.5cm]{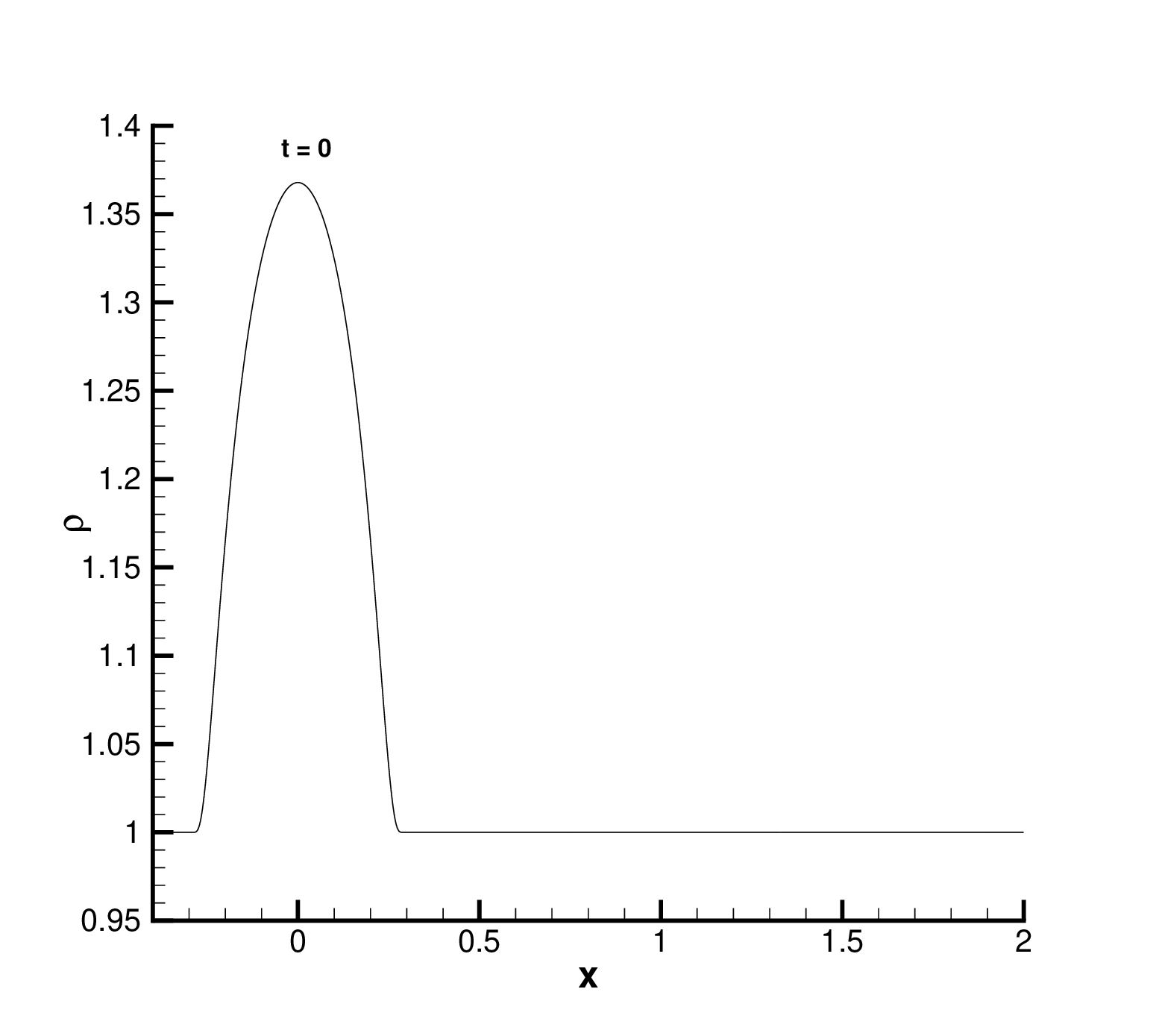}}
{\includegraphics[angle=0,width=6.5cm,height=6.5cm]{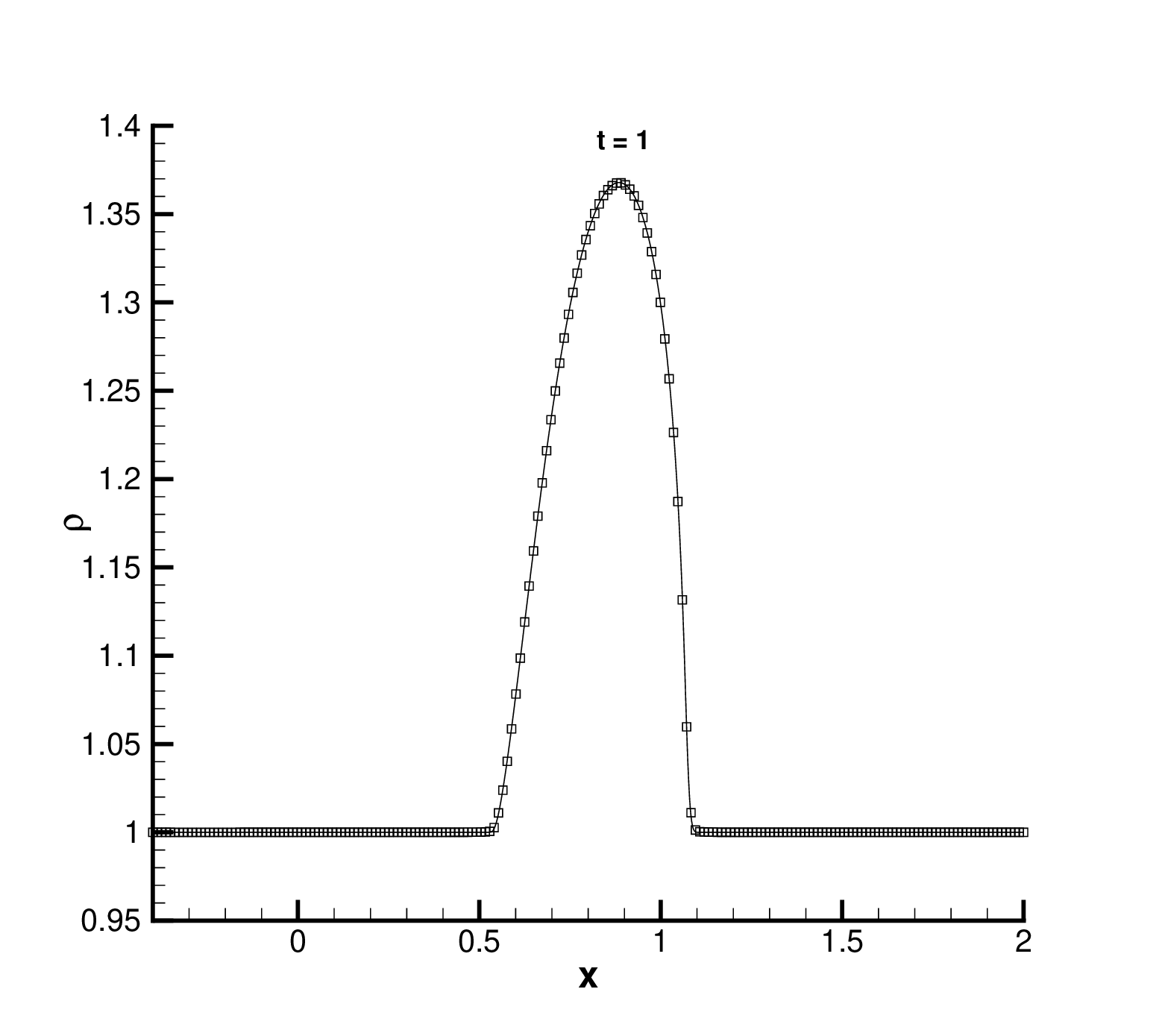}}
\end{center}
\caption{Isentropic one-dimensional flow. Left panel: initial profile of the rest-mass density. Right panel: 
solution at time $t=1.0$ obtained 
with the fourth order ADER-WENO scheme.}
\label{fig:isentropic}
\end{figure}
%
%----------------------------------------------------------
The first test that we have considered, already studied by \citet{Zhang2006} and \citet{Radice2012a},
describes a one-dimensional simple wave \citep{Anile_book}, namely a smooth solution 
for which two, out of the three, Riemann invariants associated to the acoustic eigenvalues are constant. 
Such Riemann invariants are the specific entropy $s$ and one of the two quantities
\begin{equation}
\label{RI}
J_{\pm}=\frac{1}{2}\ln{\left(\frac{1+v}{1-v}\right)}\pm \frac{1}{\sqrt{\gamma-1}}\ln{\left(\frac{\sqrt{\gamma-1}+c_s}{\sqrt{\gamma-1}-c_s}\right)}\,,
\end{equation}
where the sound speed $c_s$ is given by
\begin{equation}
c_s^2=\frac{\gamma p}{h\rho}=
\frac{\gamma(\gamma-1)p}{\gamma p + (\gamma-1)\rho}\,.
\end{equation} 
In our calculations, we have chosen $J_-$ as constant, hence describing an isentropic right-propagating simple wave~\citep{Anile1983}.
Consistently with these assumptions,
the fluid obeys a polytropic equation of state  of the type
\begin{equation}
p=K\rho^\gamma\,.
\end{equation}
The initial profile of the rest-mass density is chosen  as in \citet{Radice2012a}, i.e.
\begin{equation}\label{Isen-rho0}
  \rho(x, 0) = \left\{\begin{array}{ll}
  1+\exp[-1/(1-x^2/L^2)]& \;\textrm{for}\quad |x| < L\,, \\
 \noalign{\medskip}
 1 & \;\textrm{for}\quad |x|>L    \,, \\
 \noalign{\medskip}
 \end{array}\right.
\end{equation}
where $L$ determines the width of the profile. 
Since $\rho(x,0)=1$ and $v(x,0)=0$ for $|x|>L$, Eq.~(\ref{RI}) can be used to compute the value of $J_-$ and, consequently, the profile of the initial velocity in the range $-L<x<L$.
As the wave propagates towards the right, steepening occurs until a shock wave forms. 
In our calculations we have solved the problem over the computational domain $-0.4<x<2.0$ up to time $t=1.0$
(before the shock formation) and we have
used $K=100$, $\gamma=5/3$, $L=0.3$.  

Fig.\ref{fig:isentropic} shows the results of our calculation for the representative model with 40 cells on the level zero grid 
(see Tab.~\ref{tab.conv1}), for which a fourth order ADER-WENO scheme with $\ell_{\rm max}=2$ and $\mathfrak{r}=3$ have been adopted.
The left panel reports the initial condition, while the right panel depicts the solution at the final time, compared to the  reference solution.
\begin{table}[!t]
\caption{Initial left (L) and right (R) states of the relativistic shock tube problems. 
  The last two columns report the adiabatic index and the final time $t_f$ of the simulation.} 
\vspace{0.5cm}
\renewcommand{\arraystretch}{1.0}
\begin{center}
\begin{tabular}{ccccccccc}
\hline
\hline
 Problem  & $\rho_L$  & $v_L$ & $p_L$  & $\rho_R$  & $v_R$ & $p_R$      & $\gamma$ & $t_f$ \\
\hline
1         & 1         & -0.6  & 10     & 10        & 0.5   &  20        & 5/3      & 0.4   \\
\hline
2         & $10^{-3}$ & 0     & 1      & $10^{-3}$ & 0     &  $10^{-5}$ & 5/3      & 0.4 \\
\hline
3         & 1         & 0.9   & 1      & 1         & 0     &  10        & 4/3      & 0.4 \\
\end{tabular}
\end{center}
\label{tab.RP.ic}
\end{table}

%-------------------------------------------------------------------------
\subsection{RHD one-dimensional Riemann problems}
To further validate the new scheme, we have considered a set of relativistic shock tubes, with initial conditions 
reported in Table~\ref{tab.RP.ic}. In all cases the numerical solution has been computed 
both with the HLL and with the Osher-type Riemann solver, and has been
compared to the exact solution computed according to the procedure outlined in \citet{Rezzolla01}. 
In the three tests analyzed below the numerical domain $[0,1]$ is covered by an initial uniform mesh composed of $100$ cells.

\begin{figure}
{\includegraphics[angle=0,width=4.4cm,height=5.0cm]{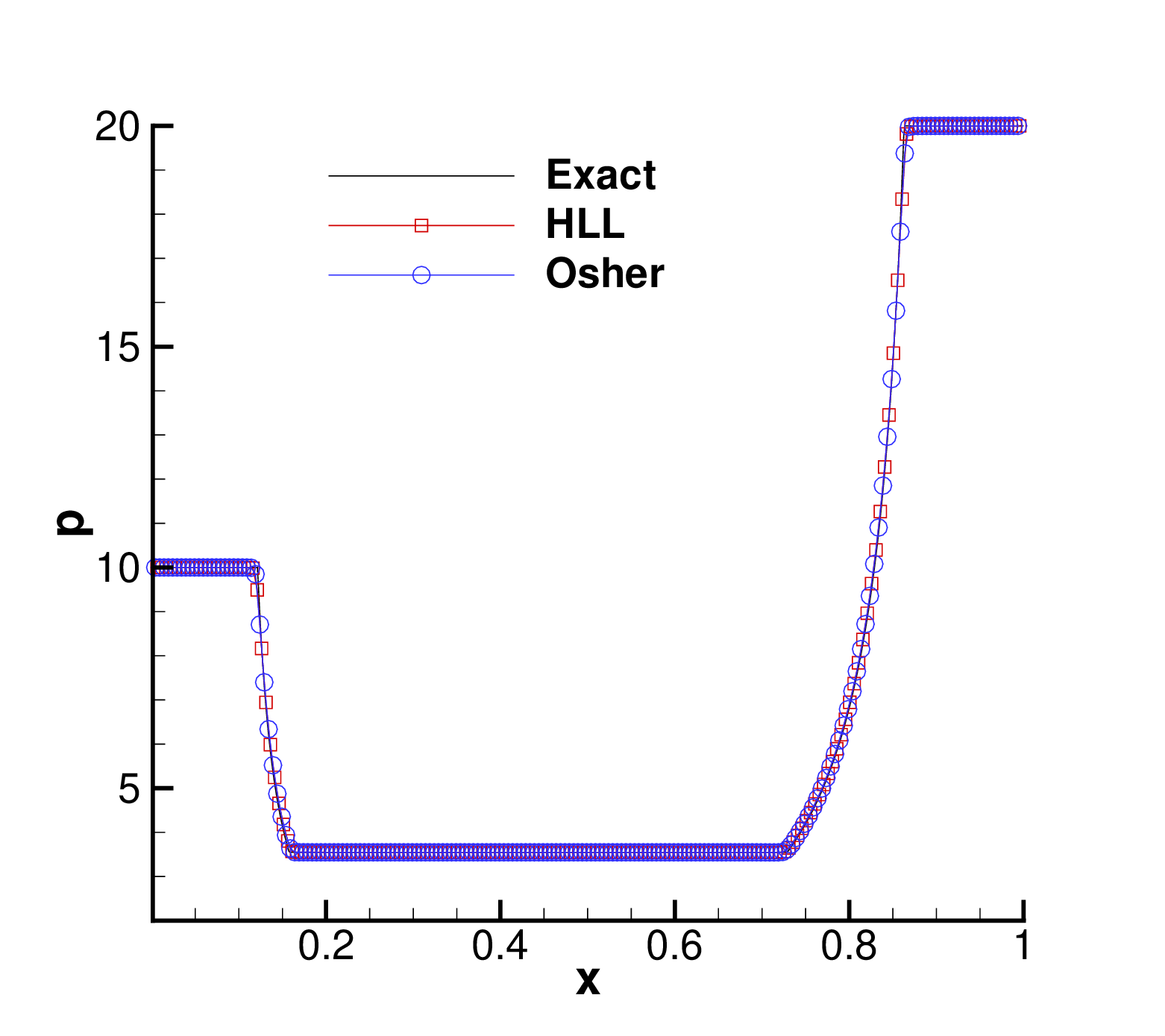}}
{\includegraphics[angle=0,width=4.4cm,height=5.0cm]{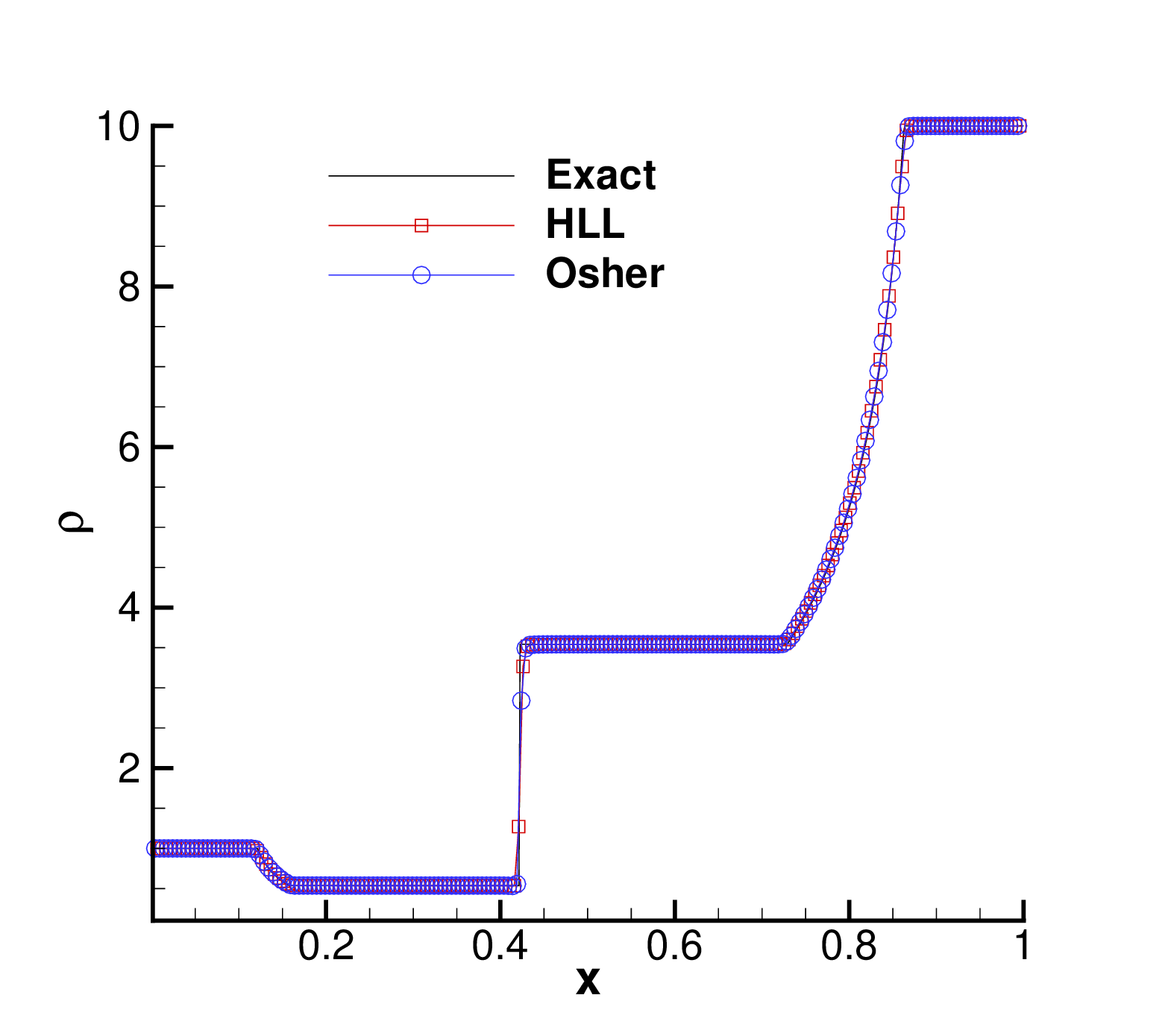}}
{\includegraphics[angle=0,width=4.4cm,height=5.0cm]{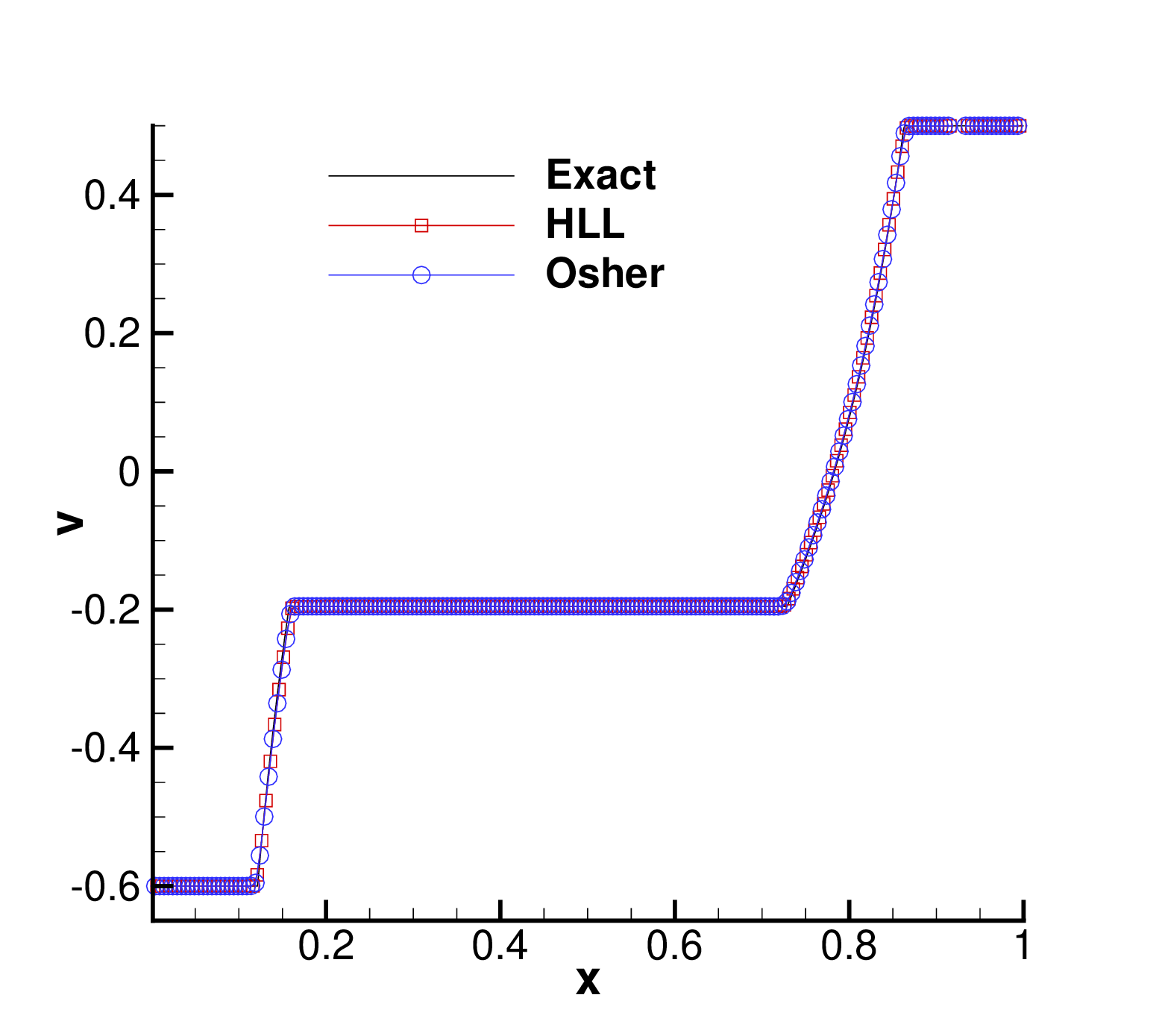}}
\caption{Solution of Problem 1 (see Table~\ref{tab.RP.ic}) with the
  fourth order ADER-WENO scheme at time $t=0.4$.}
\label{fig:shock-tube-2R}
\end{figure}

Problem 1, considered also by \citet{Mignone2005}, has initial conditions producing a wave-pattern that consists of two rarefaction waves (henceforth, a $2{\cal R}$ wave-pattern), propagating to the left and to the right, and the usual contact discontinuity between them. A 
fourth order WENO scheme with 
$\ell_{\rm max}=2$, $\mathfrak{r}=3$ and $\rm{CFL}=0.6$ has been used. 
The results of this test are shown in Fig.~\ref{fig:shock-tube-2R}, with HLL and Osher fluxes reproducing the exact solution with
essentially the same accuracy.

%----------------------------------------------------------
%
%----------------------------------------------------------
%----------------------------------------------------------
\begin{figure}
{\includegraphics[angle=0,width=4.4cm,height=5.0cm]{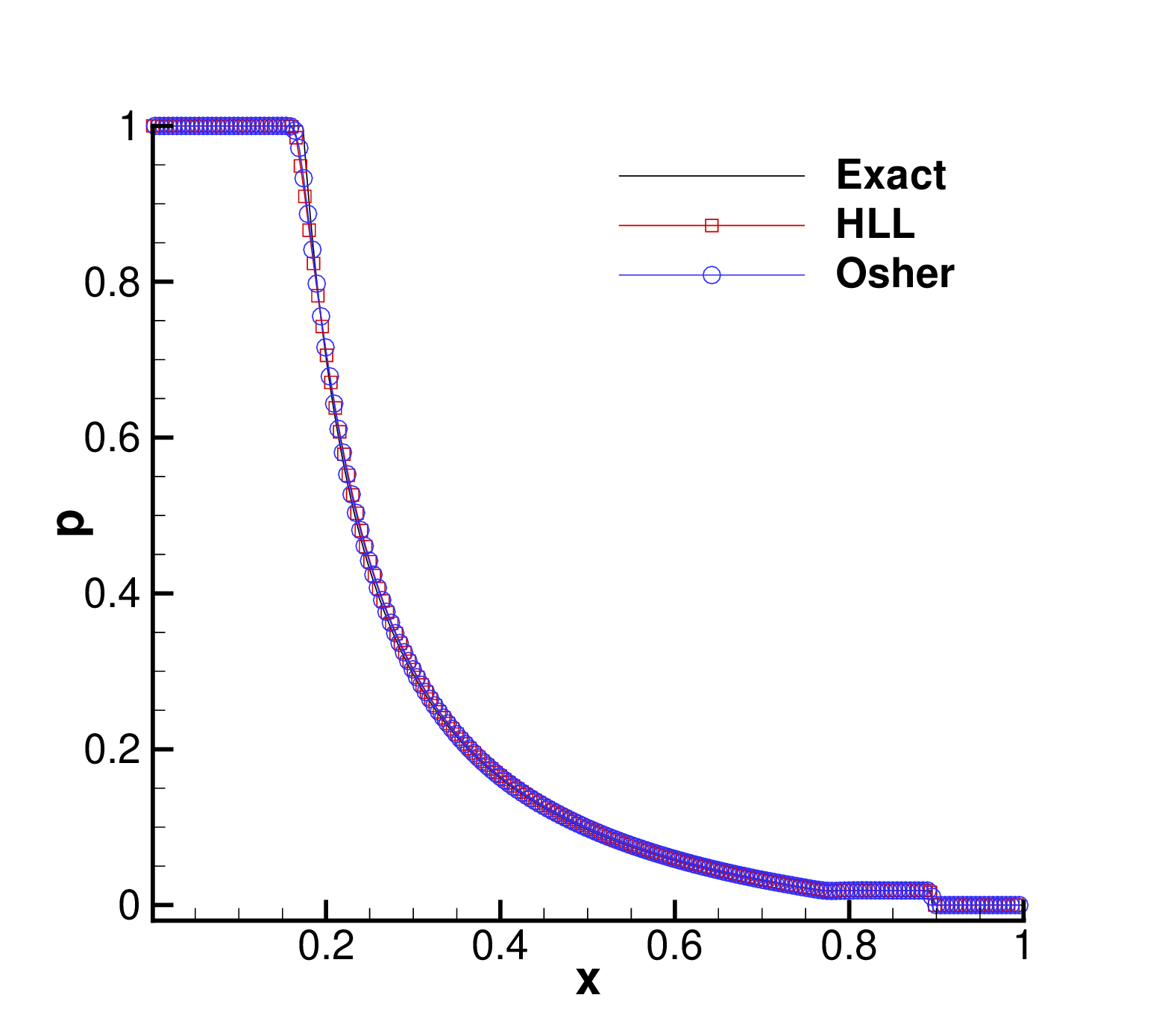}}
{\includegraphics[angle=0,width=4.4cm,height=5.0cm]{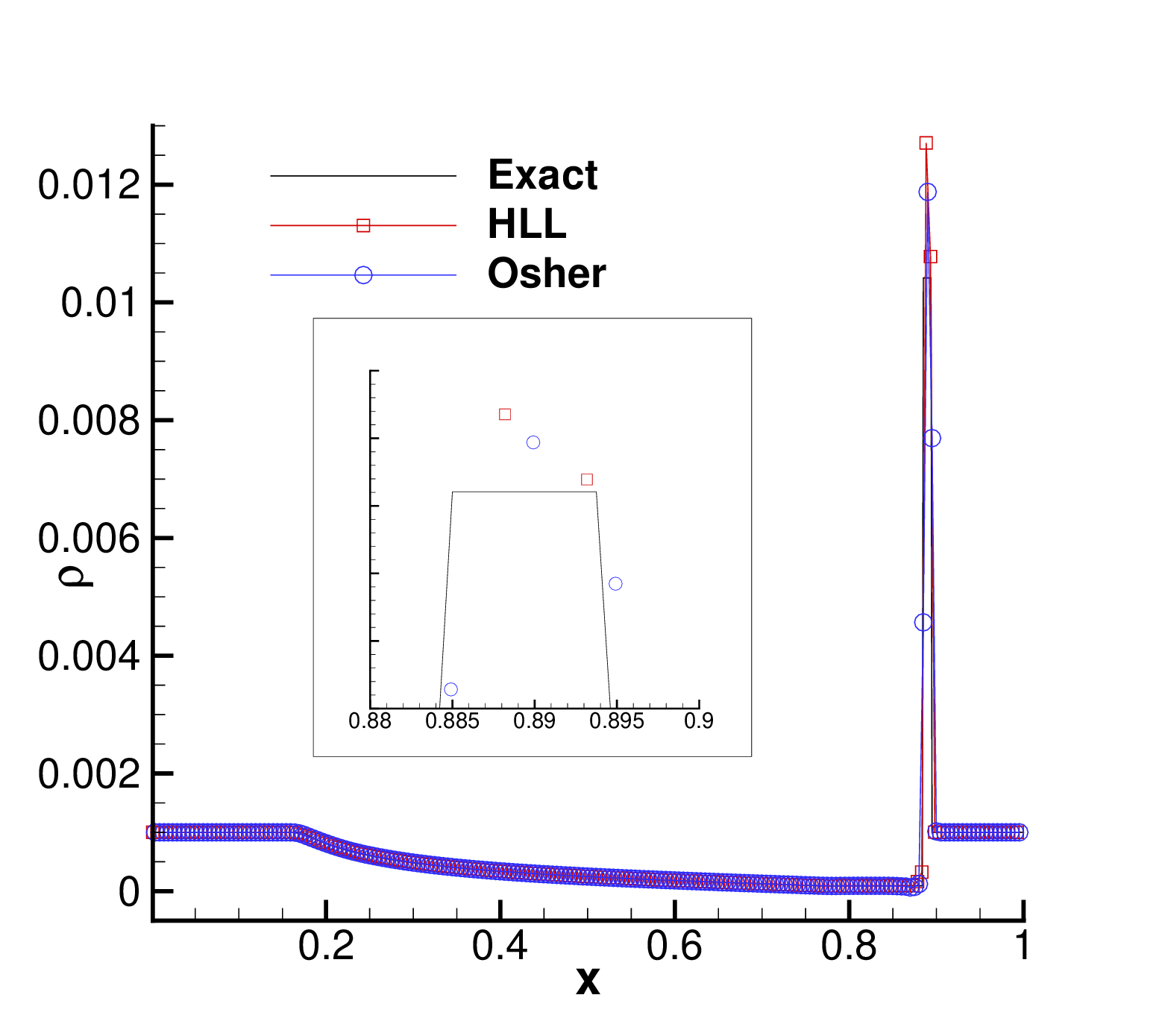}}
{\includegraphics[angle=0,width=4.4cm,height=5.0cm]{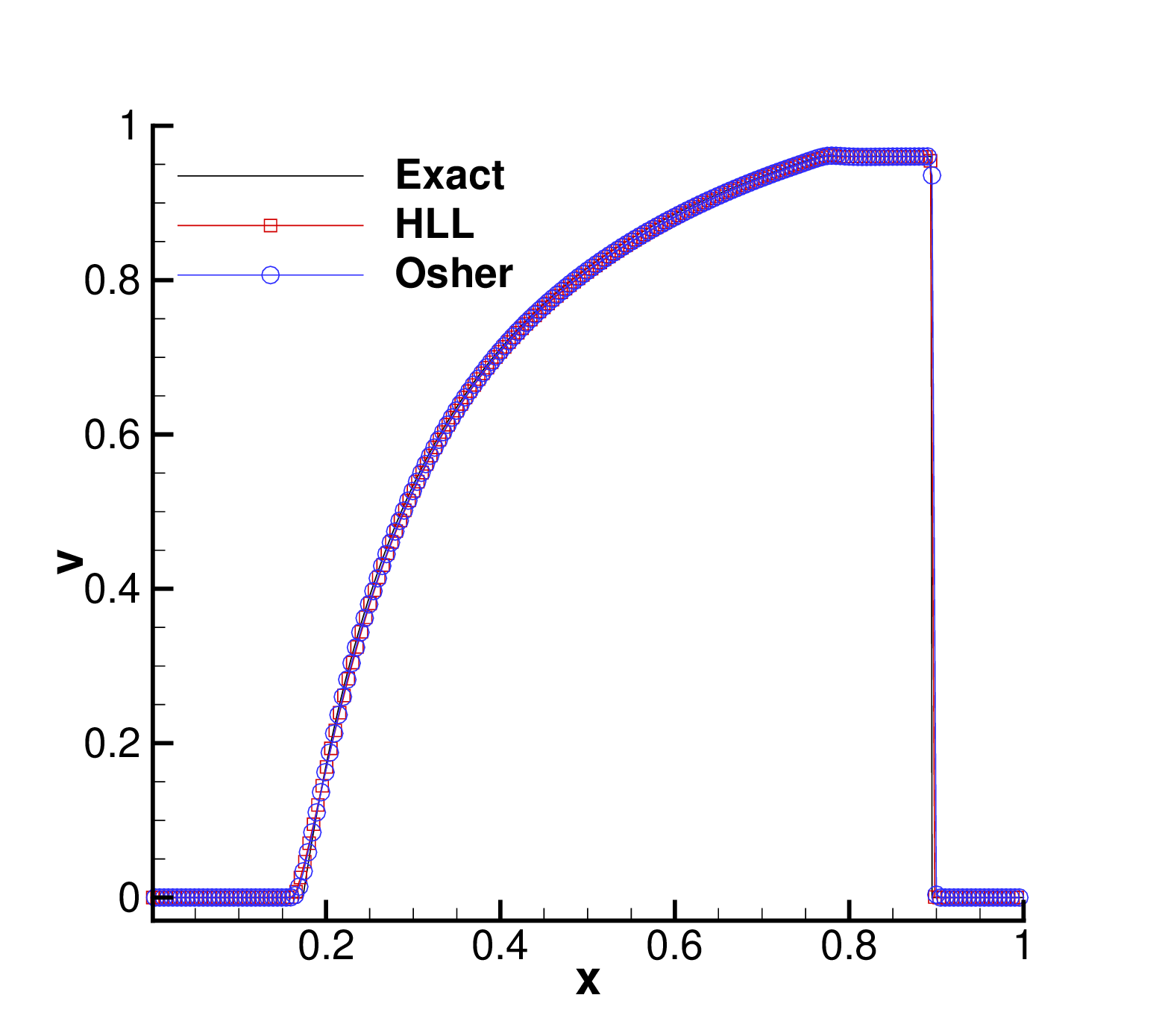}}
\caption{Solution of Problem 2 (see Table~\ref{tab.RP.ic}) with the
  fourth order ADER-WENO scheme at time $t=0.4$.}
\label{fig:shock-tube-RS}
\end{figure}
%
%----------------------------------------------------------
\begin{figure}
{\includegraphics[angle=0,width=4.4cm,height=5.0cm]{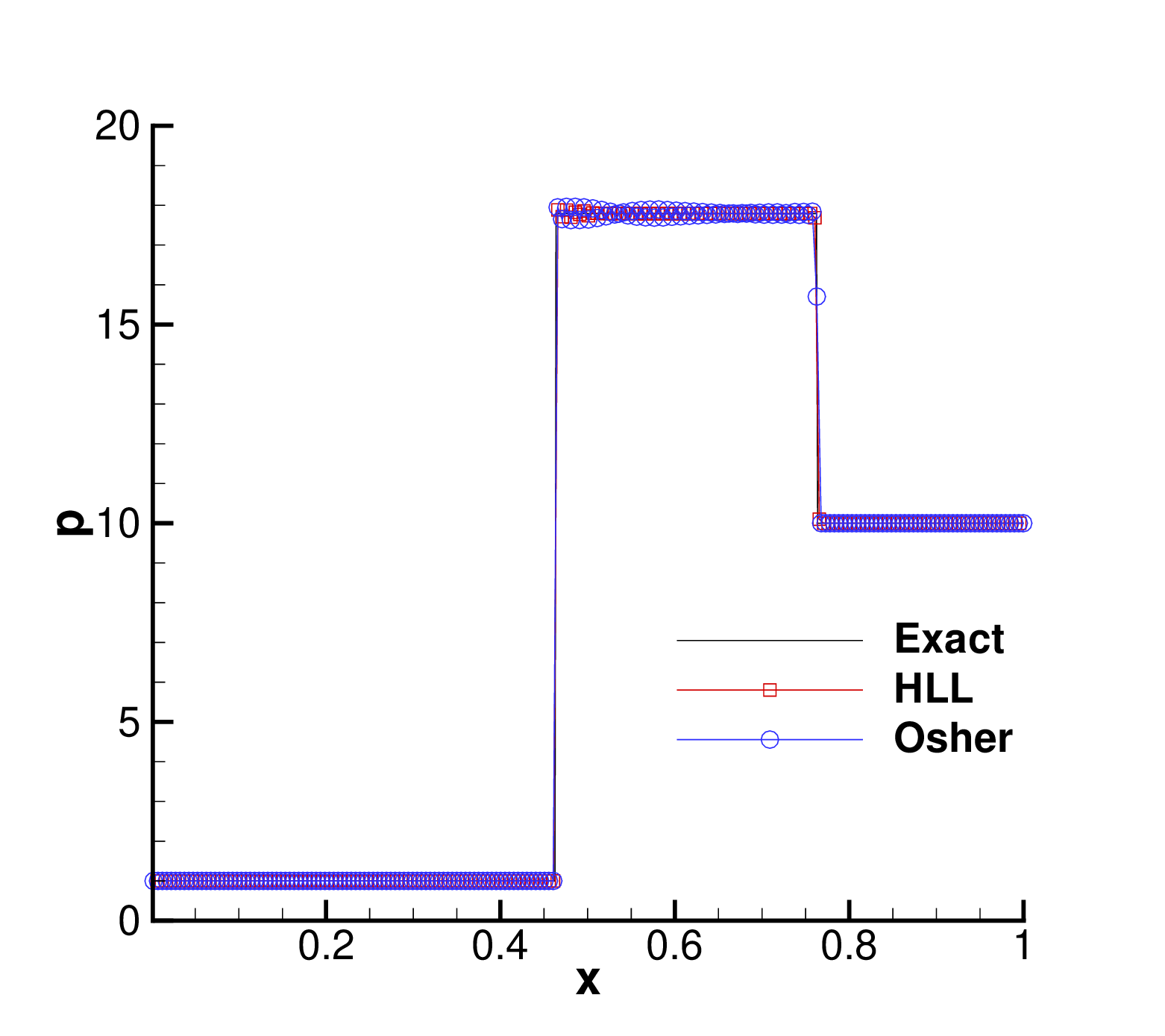}}
{\includegraphics[angle=0,width=4.4cm,height=5.0cm]{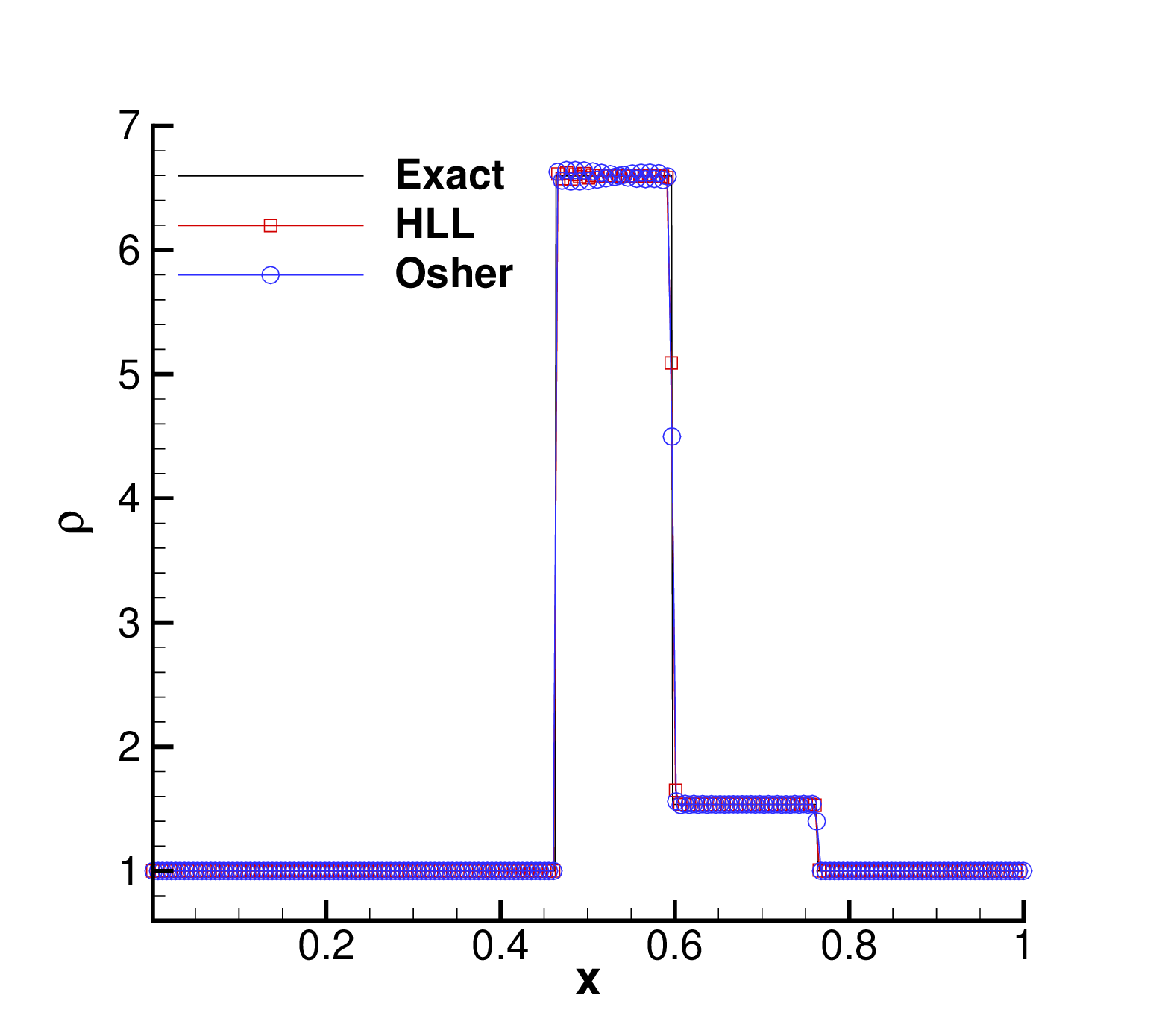}}
{\includegraphics[angle=0,width=4.4cm,height=5.0cm]{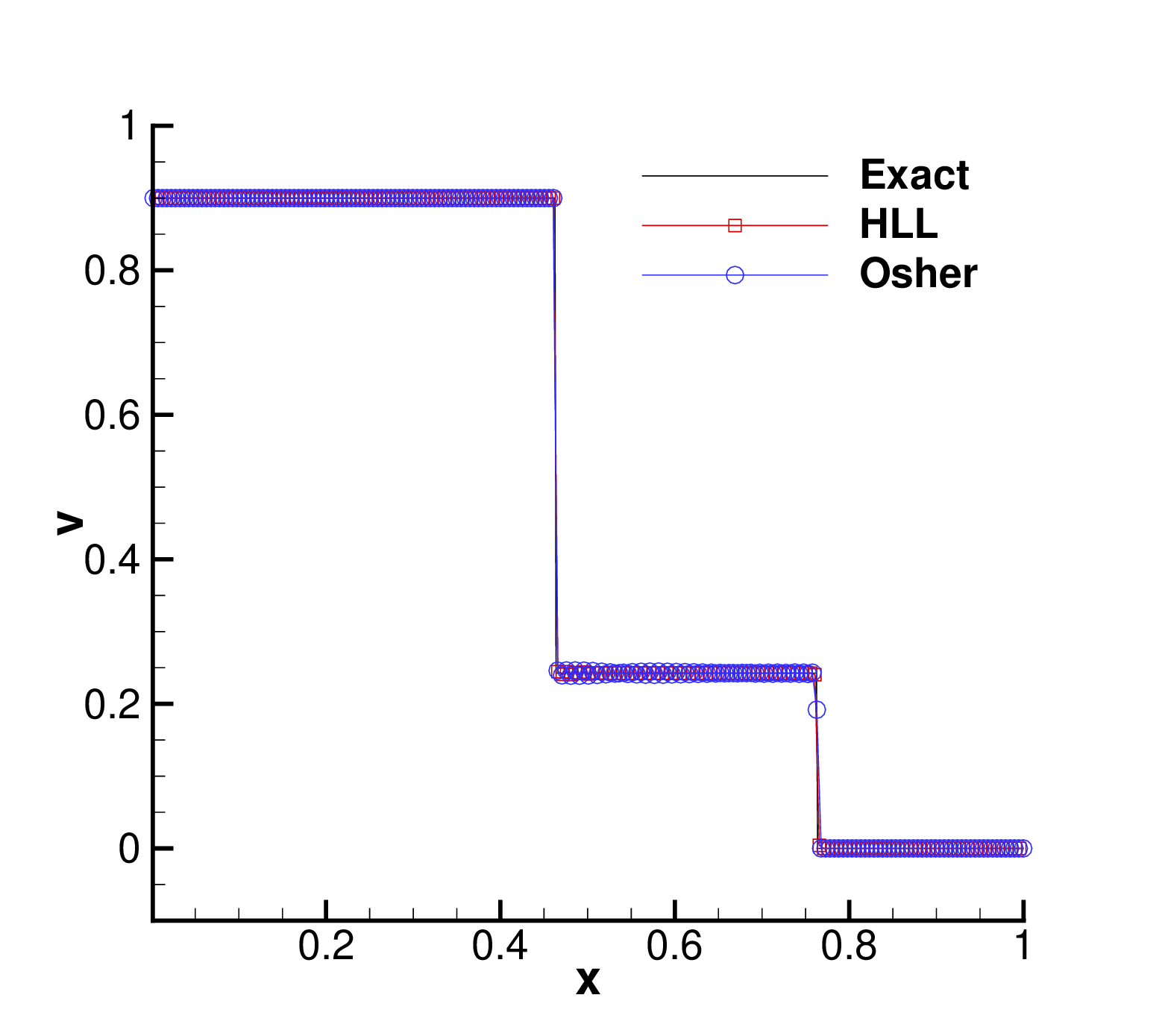}}
\caption{Solution of Problem 3 (see Table~\ref{tab.RP.ic}) with the
  fourth order ADER-WENO scheme at time $t=0.4$.}
\label{fig:shock-tube-2S}
\end{figure}

Problem 2, considered also by \citet{Radice2012a}, has initial conditions producing a wave-pattern that consists of a rarefaction wave, propagating to the left, a contact discontinuity and a shock wave propagating to the right (henceforth, a ${\cal RS}$ wave-pattern).
The shock wave is characterized by a strong density contrast.
A fourth order WENO scheme with 
$\ell_{\rm max}=2$, $\mathfrak{r}=3$ and $\rm{CFL}=0.6$ has been used. 
The results of this test are shown in Fig.~\ref{fig:shock-tube-RS}. 
We note that in this case the Osher flux gives a slightly better 
representation of the blast wave, as can be seen from the inset in the central panel.

Problem 3, considered also by \citet{Zhang2006}, has initial conditions producing a wave-pattern that consists of two shock waves, propagating to the left and to the right, plus again the contact discontinuity between them (henceforth, a $2{\cal S}$ wave-pattern).
A fourth order WENO scheme with 
$\ell_{\rm max}=2$, $\mathfrak{r}=3$ and $\rm{CFL}=0.6$ has been used. 
The results of this test are shown in Fig.~\ref{fig:shock-tube-2S}. We note that some oscillations appear in the post-shock region of the right-propagating shock, which are particularly visible in the rest mass density and in the pressure. Such oscillations are stronger for the Osher flux 
than for the more dissipative HLL flux.

\begin{table}[!t]   
\caption{Numerical convergence results for the shock tube problems using the second, the third and the fourth order version of the scheme. The error $L_2$-norms refer to the variable $\rho$ (density) at the final time $t_f=0.4$. 
}
\begin{center} 
\renewcommand{\arraystretch}{0.8}
\begin{tabular}{lcccccc} 
\hline
%\hline
%  $\ell_{\rm max}=1$ & $N_G\times N_G$  & $\epsilon_{L_2}$ & $\mathcal{O}(L_2)$ & & $N_G\times N_G$ & $\epsilon_{L_2}$ & $\mathcal{O}(L_2)$  \\ 
  \hline
  $\ell_{\rm max}=0$, &  Problem 1 &  &  & & & \\ 
  \hline
  $N_G$  & $\epsilon_{L_2}$ & $\mathcal{O}(L_2)$   & $\epsilon_{L_2}$ & $\mathcal{O}(L_2)$  & $\epsilon_{L_2}$ & $\mathcal{O}(L_2)$  \\ 
\hline
            &{$\mathcal{O}2$} &{$\mathcal{O}2$} &  {$\mathcal{O}3$} &  {$\mathcal{O}3$} & {$\mathcal{O}4$} & {$\mathcal{O}4$}  \\
\hline
  48         & 0.1040E+0  &          & 8.1633E-02 &          & 8.0836E-02 &       \\ 
  96         & 7.2390E-02 & 0.52     & 5.0825E-02 & 0.68     & 4.8533E-02 & 0.73   \\ 
  192        & 4.9323E-02 & 0.55     & 3.6793E-02 & 0.46     & 3.2851E-02 & 0.56  \\ 
  288        & 3.9808E-02 & 0.53     & 2.8863E-02 & 0.60     & 2.6741E-02 & 0.51   \\  
  384        & 3.4400E-02 & 0.51     & 2.6361E-02 & 0.31     & 2.3341E-02 & 0.47   \\
\hline
  $\ell_{\rm max}=0$, &  Problem 2 &  &  & & & \\ 
  \hline
  $N_G$  & $\epsilon_{L_2}$ & $\mathcal{O}(L_2)$   & $\epsilon_{L_2}$ & $\mathcal{O}(L_2)$  & $\epsilon_{L_2}$ & $\mathcal{O}(L_2)$  \\ 
\hline
            & {$\mathcal{O}2$} &{$\mathcal{O}2$} &  {$\mathcal{O}3$}  &  {$\mathcal{O}3$} & {$\mathcal{O}4$} & {$\mathcal{O}4$}  \\
\hline
  48         & 2.3516E-04 &          & 2.7214E-04 &          & 2.1798E-04 &       \\ 
  96         & 2.2984E-04 & 0.03     & 2.3015E-04 & 0.24     & 1.9878E-04 & 0.13   \\ 
  192        & 2.2027E-04 & 0.06     & 1.8076E-04 & 0.35     & 1.6360E-04 & 0.28  \\ 
  288        & 2.0329E-04 & 0.19     & 1.5462E-04 & 0.38     & 1.4305E-04 & 0.33   \\  
  384        & 1.8807E-04 & 0.27     & 1.3605E-04 & 0.44     & 1.2920E-04 & 0.35   \\
\hline
  $\ell_{\rm max}=0$, &  Problem 3 &  &  & & & \\ 
  \hline
  $N_G$  & $\epsilon_{L_2}$ & $\mathcal{O}(L_2)$   & $\epsilon_{L_2}$ & $\mathcal{O}(L_2)$  & $\epsilon_{L_2}$ & $\mathcal{O}(L_2)$  \\ 
\hline
            &{$\mathcal{O}2$} &{$\mathcal{O}2$} &   {$\mathcal{O}3$}  &  {$\mathcal{O}3$} & {$\mathcal{O}4$} & {$\mathcal{O}4$}  \\
\hline
  48         & 0.1375E-00 &          & 0.1084E-00 &          & 9.8280E-02 &       \\ 
  96         & 9.0721E-02 & 0.60     & 7.6012E-02 & 0.51     & 7.9195E-02 & 0.31   \\ 
  192        & 6.0782E-02 & 0.58     & 5.2709E-02 & 0.53     & 5.1727E-02 & 0.61  \\ 
  288        & 6.0567E-02 & 0.01     & 4.9566E-02 & 0.15     & 5.1590E-02 & 0.01   \\  
  384        & 4.9106E-02 & 0.73     & 4.3106E-02 & 0.48     & 3.8403E-02 & 1.03   \\
\hline
\hline
  $\ell_{\rm max}=2$, & Problem 1  & &   & & & \\ 
  \hline
  $N_G$  & $\epsilon_{L_2}$ & $\mathcal{O}(L_2)$   & $\epsilon_{L_2}$ & $\mathcal{O}(L_2)$  & $\epsilon_{L_2}$ & $\mathcal{O}(L_2)$  \\ 
\hline
            & &{$\mathcal{O}2$} &         & {$\mathcal{O}3$} &  & {$\mathcal{O}4$}  \\
\hline

  50         & 3.2766E-02 &        & 2.5525E-02 &      & 2.2031E-02 &       \\ 
  100        & 2.4211E-02 & 0.43   & 1.7979E-02 & 0.50 & 1.5174E-02 & 0.54   \\ 
  150        & 2.0189E-02 & 0.45   & 1.4871E-02 & 0.47 & 1.3248E-02 & 0.33  \\ 
  200        & 1.7895E-02 & 0.42   & 1.2979E-02 & 0.47 & 1.1311E-02 & 0.54   \\  
  250        & 1.6391E-02 & 0.39   & 1.2376E-02 & 0.21 & 1.0004E-02 & 0.55   \\
\hline
  $\ell_{\rm max}=2$, & Problem 2  & &   & & & \\ 
  \hline
  $N_G$  & $\epsilon_{L_2}$ & $\mathcal{O}(L_2)$   & $\epsilon_{L_2}$ & $\mathcal{O}(L_2)$  & $\epsilon_{L_2}$ & $\mathcal{O}(L_2)$  \\ 
\hline
            & &{$\mathcal{O}2$} &         & {$\mathcal{O}3$} &  & {$\mathcal{O}4$}  \\
\hline

  50$^\ast$  & 1.8277E-04 &        & 1.3901E-04 &      & 1.3543E-04 &       \\ 
  100        & 1.3735E-04 & 0.41   & 9.1050E-05 & 0.61 & 8.4308E-05 & 0.68   \\ 
  150        & 1.1605E-04 & 0.41   & 7.6561E-05 & 0.43 & 6.9318E-05 & 0.43  \\ 
  200        & 1.0367E-04 & 0.39   & 6.9863E-05 & 0.32 & 5.8261E-05 & 0.60   \\  
  250        & 9.5207E-05 & 0.38   & 6.0027E-05 & 0.68 & 5.1095E-05 & 0.59   \\
\hline
 $\ell_{\rm max}=2$, & Problem 3  & &   & & & \\ 
  \hline
  $N_G$  & $\epsilon_{L_2}$ & $\mathcal{O}(L_2)$   & $\epsilon_{L_2}$ & $\mathcal{O}(L_2)$  & $\epsilon_{L_2}$ & $\mathcal{O}(L_2)$  \\ 
\hline
            & &{$\mathcal{O}2$} &         & {$\mathcal{O}3$} &  & {$\mathcal{O}4$}  \\
\hline

  50$^\ast$  & 5.0893E-02 &        & 4.1120E-02 &      & 4.1825E-02 &       \\ 
  100        & 3.6623E-02 & 0.47   & 2.9813E-02 & 0.46 & 2.7244E-02 & 0.62   \\ 
  150        & 2.9965E-02 & 0.49   & 2.7059E-02 & 0.24 & 2.3018E-02 & 0.41  \\ 
  200        & 2.9446E-02 & 0.06   & 2.4190E-02 & 0.39 & 2.0958E-02 & 0.32   \\  
  250        & 2.5318E-02 & 0.67   & 1.9509E-02 & 0.96 & 1.6690E-02 & 1.02   \\
\hline
\end{tabular} 
\end{center}
\label{tab.conv2}
\end{table} 
Table~\ref{tab.conv2} reports an analysis of 
the $L_2$ norms of the error (computed with respect to the available exact solution)
and the corresponding orders of convergence. This analysis has been performed 
with the Osher flux and with reconstruction in characteristic variables, for both the cases $\ell_{\rm max}=0$ (no AMR)
and $\ell_{\rm max}=2$.
As expected for solutions with discontinuities, the order of convergence is usually smaller or equal to unity. 
In a few cases, e.g. for Problem 2 with $l_{max}=0$ and $N_G=96$, the convergence is anomalously low and very closed to zero.
This typically happens when there are a few cells manifesting oscillations at discontinuities.
Moreover,
the absolute value of the error decreases as the nominal order of the scheme is increased.
\subsection{RHD one-dimensional Riemann problems with tangential velocities}
In spite of the obvious additional complications introduced by special relativity, the one-dimensional Riemann problem without tangential velocities  does not introduce qualitative differences with respect to its Newtonian analog. The situation changes drastically when tangential  velocities are present.\footnote{
A velocity component is tangential if it belongs to the plane normal to the discontinuity front.
} 
As shown by~\citet{Rezzolla02},
the coupling of the velocity components through the Lorentz factor is responsible of a new effect which is not present in Newtonian hydrodynamics. Namely, given a Riemann problem with initial conditions producing a ${\cal RS}$
wave pattern and a non zero relative velocity along the normal direction,
it is always possible to transform it into a $2{\cal S}$ wave-pattern by increasing the value of the initial tangential velocity in the state of initial {\em lower} pressure, while keeping the rest of the variables unmodified. Similarly, 
given a Riemann problem with initial conditions producing a ${\cal RS}$
wave pattern and a non zero relative velocity along the normal direction, 
it is always possible to transform it into a $2{\cal R}$ wave-pattern by increasing the value of the initial tangential velocity in the state of initial {\em higher} pressure. Besides being at the heart of a possible acceleration mechanism in relativistic jets~\citep{Aloy:2006rd}, this physical effect is also interesting from a numerical point of view, since it can be used to test the
ability of a relativistic code in treating dynamical systems with fully coupled velocities.

\begin{figure}
{\includegraphics[angle=0,width=4.4cm,height=5.0cm]{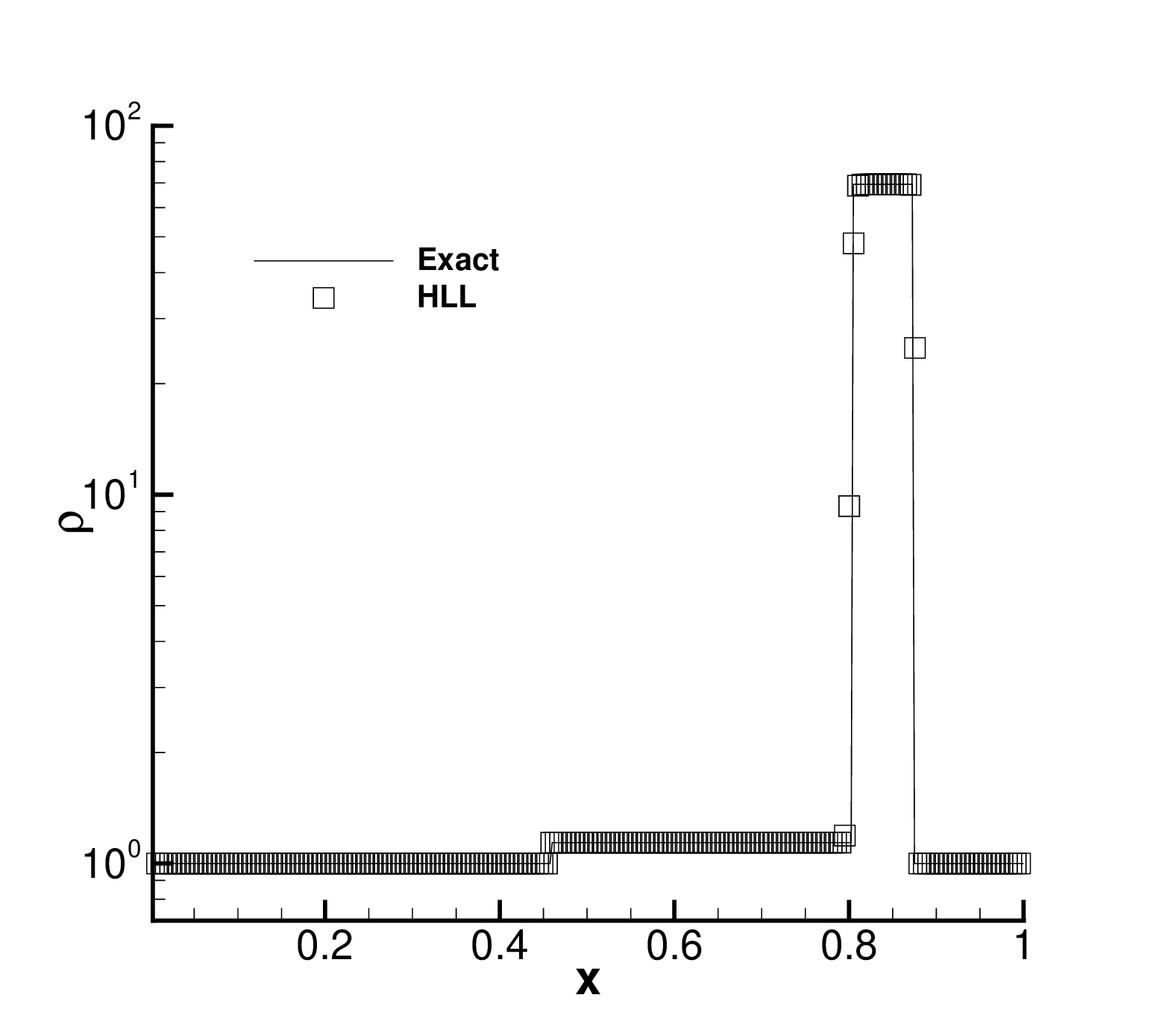}}
{\includegraphics[angle=0,width=4.4cm,height=5.0cm]{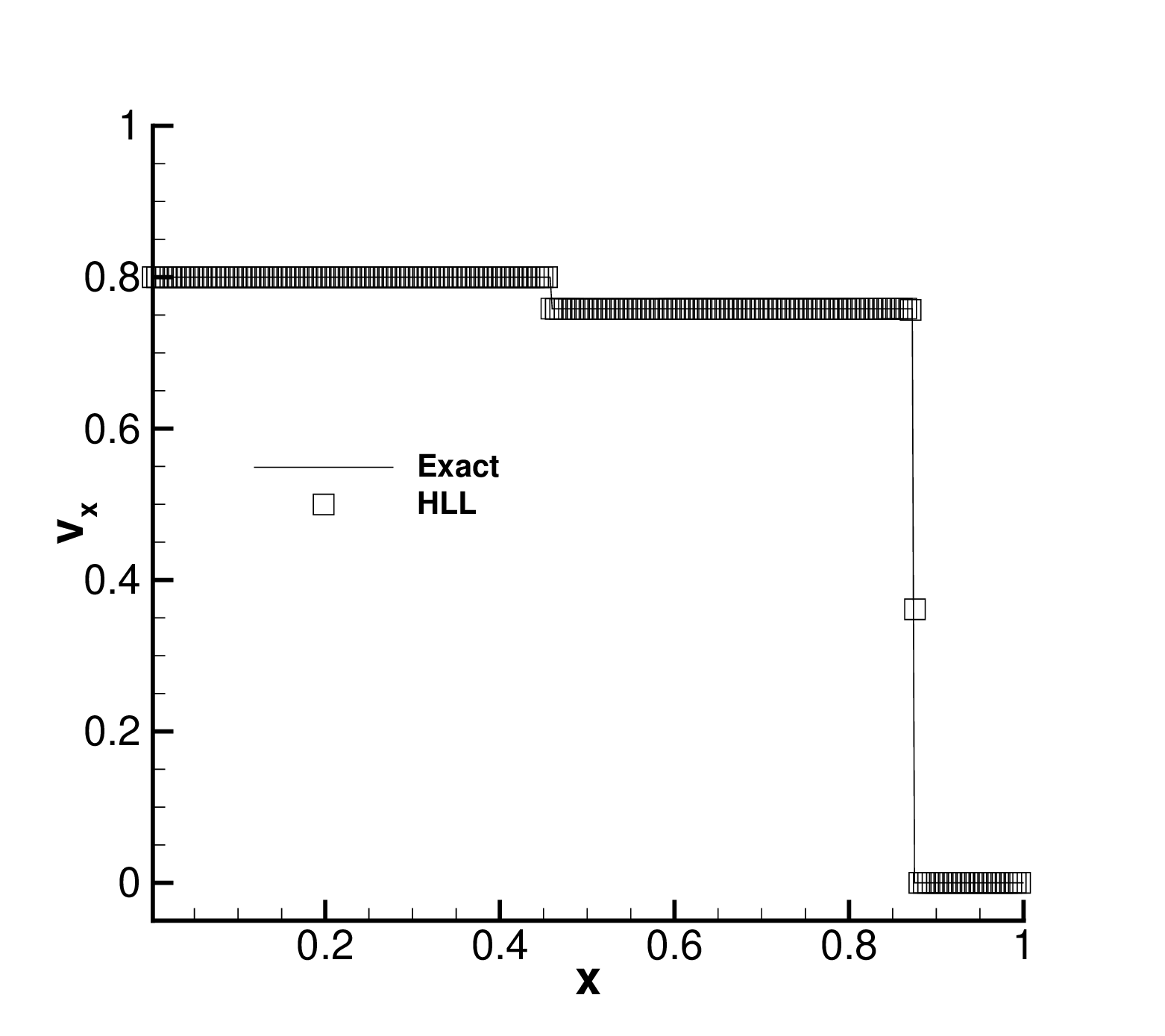}}
{\includegraphics[angle=0,width=4.4cm,height=5.0cm]{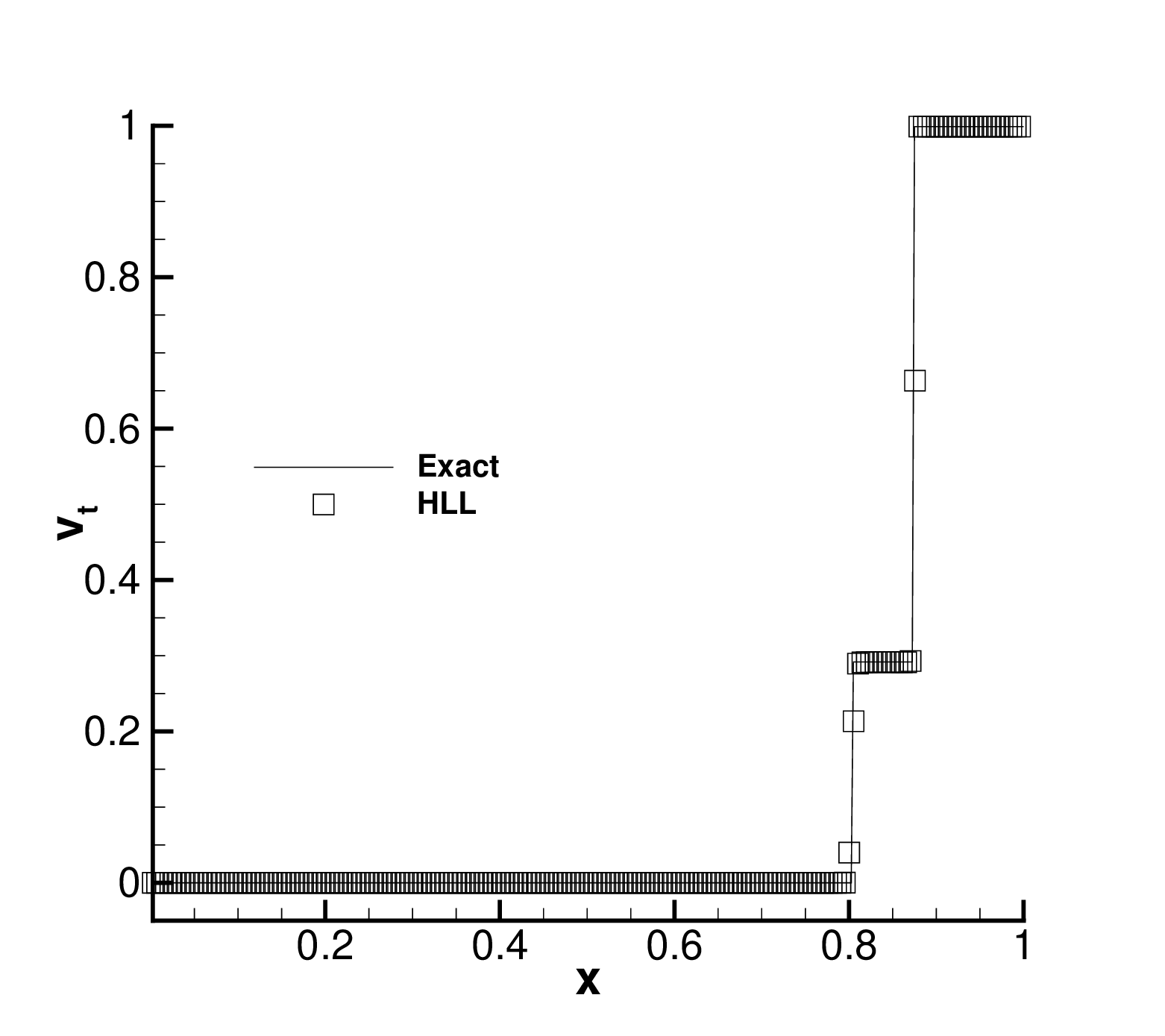}}
\caption{Solution of the Riemann problem with a non-zero tangential velocity at time $t=0.4$. A
  third order ADER-WENO scheme has been used. }
\label{fig:RS-2S}
\end{figure}

We have simulated one such relevant example that has the following initial conditions
\begin{equation}\label{RP-transverse}
  \big(\rho, v_x, v_t, p\big) = \left\{\begin{array}{ll}
\big(1.0,    0.8, 0.0,   1000.0\big) & \;\textrm{for}\quad x < 0.5\,, \\
 \noalign{\medskip}
 \big(1.0,   0.0, 0.999, 0.01\big) & \;\textrm{for}\quad x>0.5    \,, \\
 \noalign{\medskip}
 \end{array}\right.
 \end{equation}
and adiabatic index $\gamma=5/3$.
In the absence of tangential velocities, namely when $v_t=0$ on both initial states, the solution to this Riemann problem would 
consist of a ${\cal RS}$ wave-pattern. However, due to the high value of $v_t$ on the right state, the wave-pattern changes to a $2{\cal S}$ one. The solution is reported in Fig.~\ref{fig:RS-2S}, which also shows the 
reference solution computed according to \citet{Rezzolla03}.
A third order WENO scheme with  $\ell_{\rm max}=2$, $\mathfrak{r}=3$ and $\rm{CFL}=0.6$ has been used, proving the ability of the code in treating configurations with highly coupled velocity components. We note that for this test the HLL and the Osher-type Riemann solver provides essentially the same accuracy and only the
data obtained with HLL have been plotted in the figure.

%----------------------------------------------------------
%
%
%
%----------------------------------------------------------
\begin{figure}
{\includegraphics[angle=0,width=7.0cm,height=7.0cm]{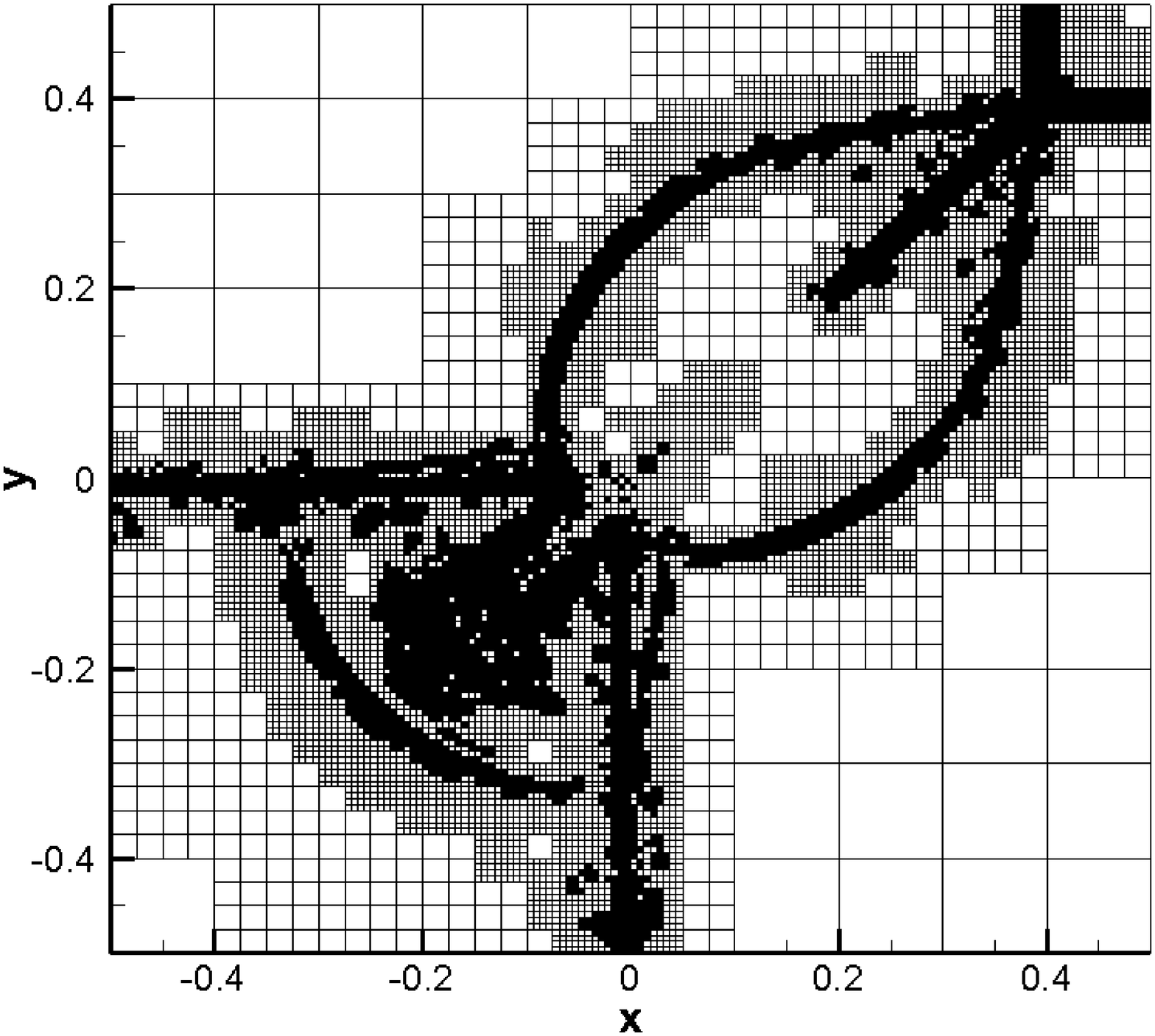}}
{\includegraphics[angle=0,width=7.0cm,height=7.0cm]{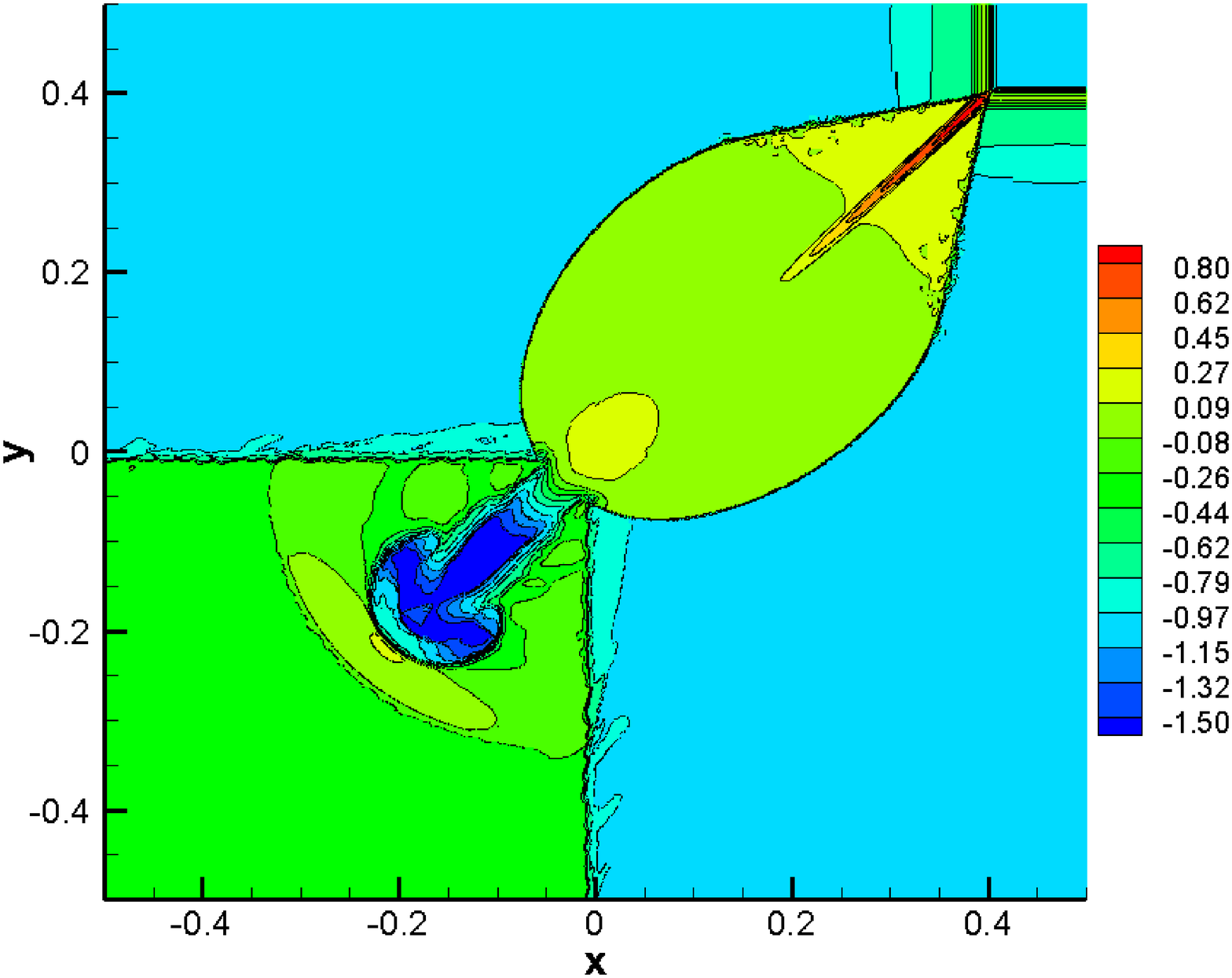}}
\caption{Solution of the 2D Riemann Problem obtained
with a  third order ADER-WENO scheme at time $t=0.4$. 
Left panel: snapshot of the AMR grid.
Right panel: logarithm of the rest mass density.
}
\label{fig:RP2D}
\end{figure}
%
%

%-----------------------------------------------
\subsection{RHD two-dimensional Riemann problem}
\label{sec:RP2D}
In the first of our multi-dimensional tests
we solve the two-dimensional Riemann problem originally proposed by \citet{DelZanna2002}, and later reproduced by 
\citet{Lucas-Serrano:2004aq},
\citet{Mignone2005},
\citet{Zhang2006}.
We emphasize that in this problem the order of accuracy is the third, corresponding to $M=2$,
essentially motivated by the necessity to reduce the computational time due to the constraint $\mathfrak{r}\geq M$ required by our scheme.
The initial conditions are prescribed 
over the square $[-1,1]\times[-1,1]$ and are given by
\begin{equation}\label{RP2D}
  \big(\rho, v_x, v_y, p\big) = \left\{\begin{array}{ll}
\big(0.1,0.99, 0   ,    1\big) & \;\textrm{for}\quad x < 0, y > 0\,, \\
 \noalign{\medskip}
 \big(0.1,   0, 0   , 0.01\big) & \;\textrm{for}\quad x>0,y >0    \,, \\
 \noalign{\medskip}
  \big(0.5,   0, 0   ,    1\big) & \;\textrm{for}\quad x<0,y < 0   \,, \\
 \noalign{\medskip}
  \big(0.1,   0, 0.99,    1\big) & \;\textrm{for}\quad x>0, y < 0\,,\\
 \noalign{\medskip}
 \end{array}\right.
 \end{equation}
with adiabatic index $\gamma=5/3$. As the system evolves in time, two curved shock fronts propagate in the upper-right quadrant, while 
a jet-like structure emerges along the main diagonal.
We have solved this problem with a third order WENO scheme applied to the characteristic variables
and using the HLL Riemann solver. The grid is initially uniform with  
$20\times20$ cells. Three levels of refinement have been adopted, with $\mathfrak{r}=4$ and $\rm{CFL}=0.4$.  
The results of our computations are reported in Fig.\ref{fig:RP2D}, which shows 
the contour-plot of the rest mass density (right panel) and the corresponding AMR grid (left panel) at the final time $t=0.4$. 
We stress that, in a limited number of troubled cells close to the corner $(0,0)$, the accuracy of the solution has been intentionally 
reduced to first-order following a MOOD-type approach [see \citet{MOOD}], to avoid the appearance of strong oscillations 
which make the recovering of the primitive variables fail.
In spite of 
this, the solution is very well reproduced by our numerical scheme and it can be compared with that obtained by \citet{Lucas-Serrano:2004aq} 
using the same order of accuracy on a uniform mesh composed of $400\times400$ elements.
%
%
%-----------------------------------------------
\subsection{RHD Kelvin--Helmholtz instability}
%----------------------------------------------------------
%
\begin{figure}
\begin{center}
{\includegraphics[angle=0,width=4.0cm,height=8.0cm]{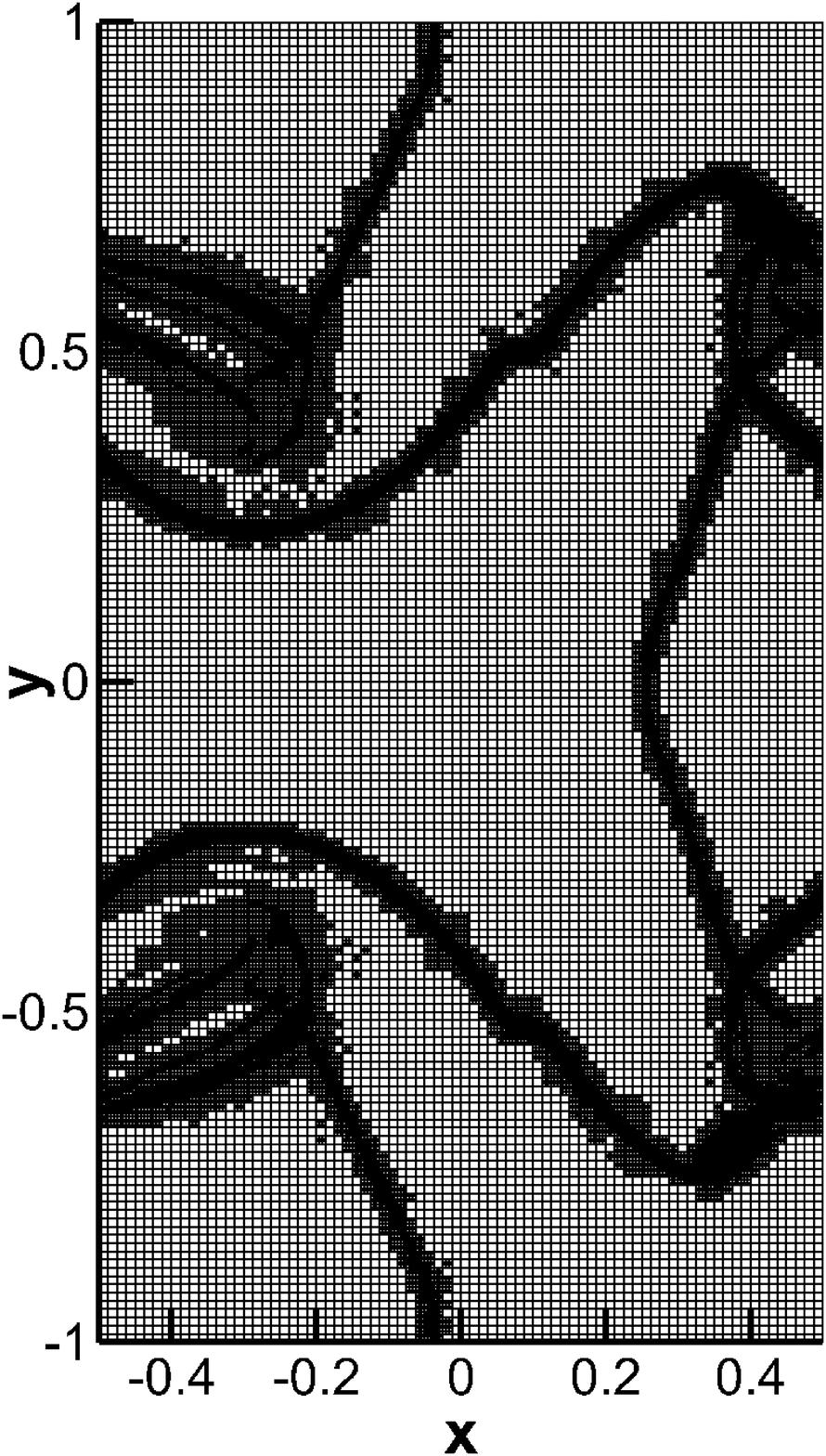}}
{\includegraphics[angle=0,width=4.0cm,height=8.0cm]{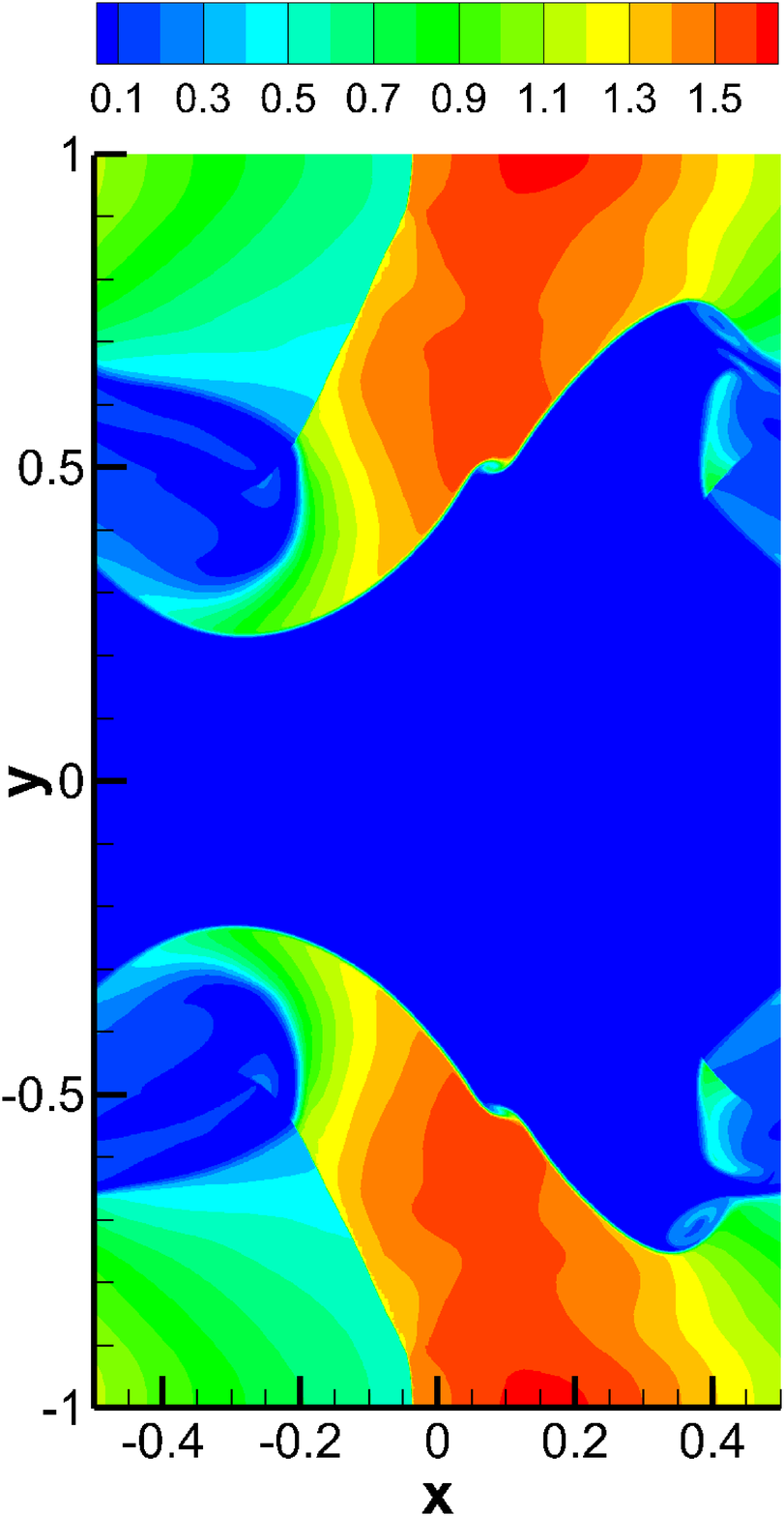}}
\end{center}
\hspace{1cm}
\begin{center}
{\includegraphics[angle=0,width=4.0cm,height=8.0cm]{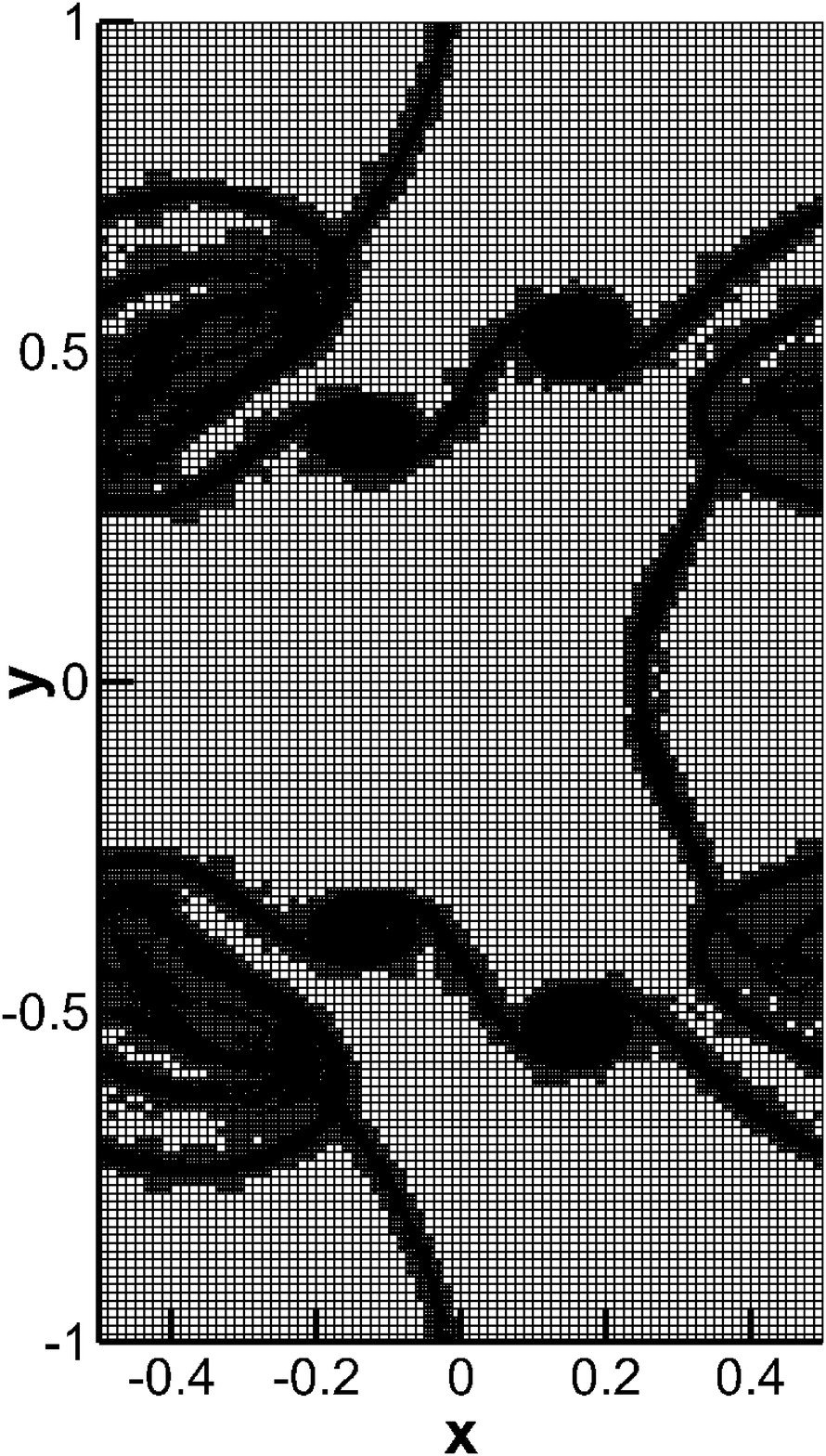}}
{\includegraphics[angle=0,width=4.0cm,height=8.0cm]{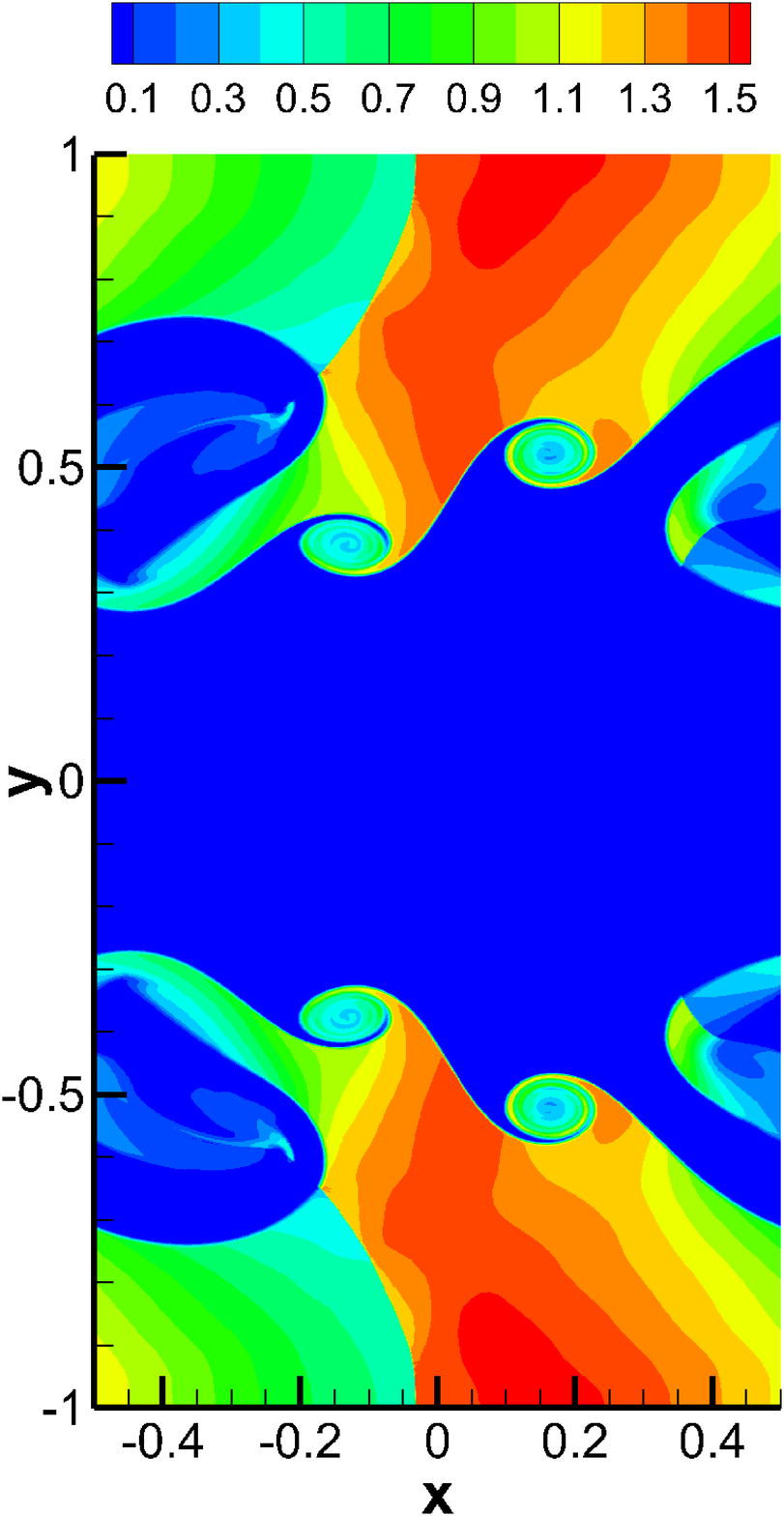}}
\end{center}
\caption{
Two-dimensional Kelvin-Helmholtz instability at $t=3.0$ with {\bf refinement factor} $\boldsymbol{\mathfrak{r}=3}$.
Left panels: the AMR grid.
Right panels: the distribution of the rest-mass density.
Top panels: results obtained with the HLL Riemann solver.
Bottom panels: results obtained with the Osher Riemann solver.
In both cases a third order ADER-WENO with CFL=0.4 has been adopted.
}
\label{fig:KH-RHD-reffactor3}
\end{figure}
\begin{figure}
\begin{center}
{\includegraphics[angle=0,width=4.0cm,height=8.0cm]{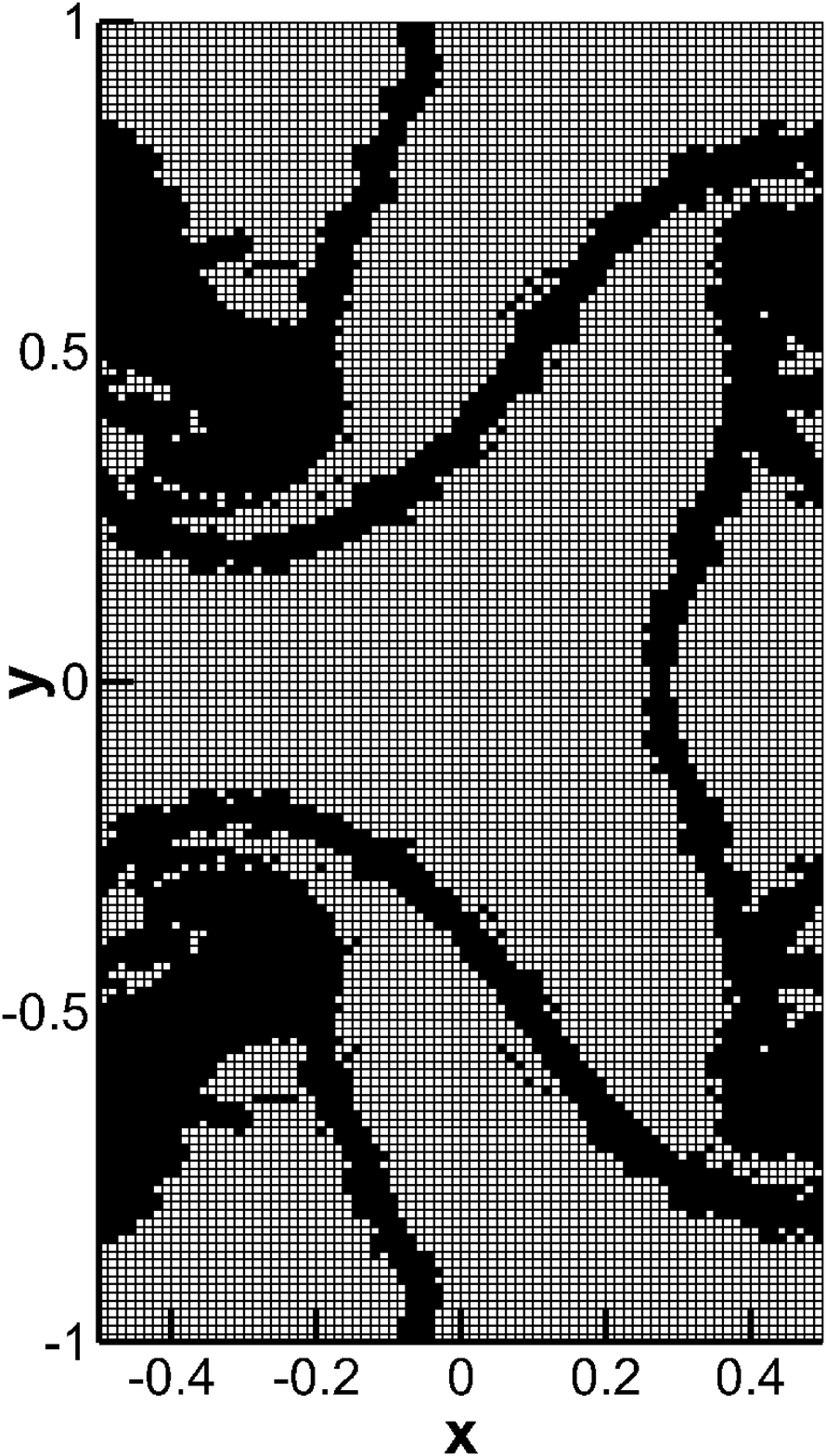}}
{\includegraphics[angle=0,width=4.0cm,height=8.0cm]{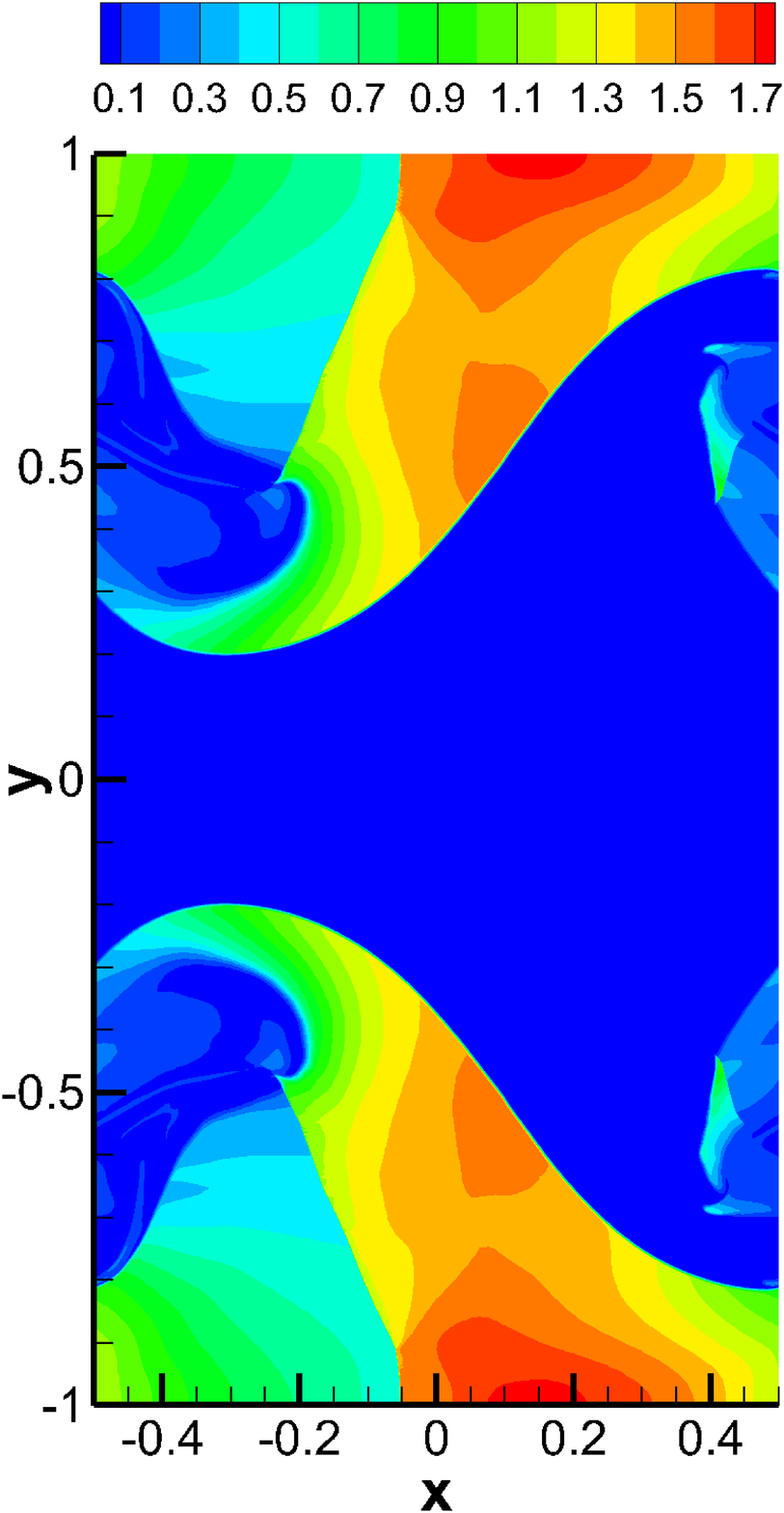}}
\end{center}
\hspace{1cm}
\begin{center}
{\includegraphics[angle=0,width=4.0cm,height=8.0cm]{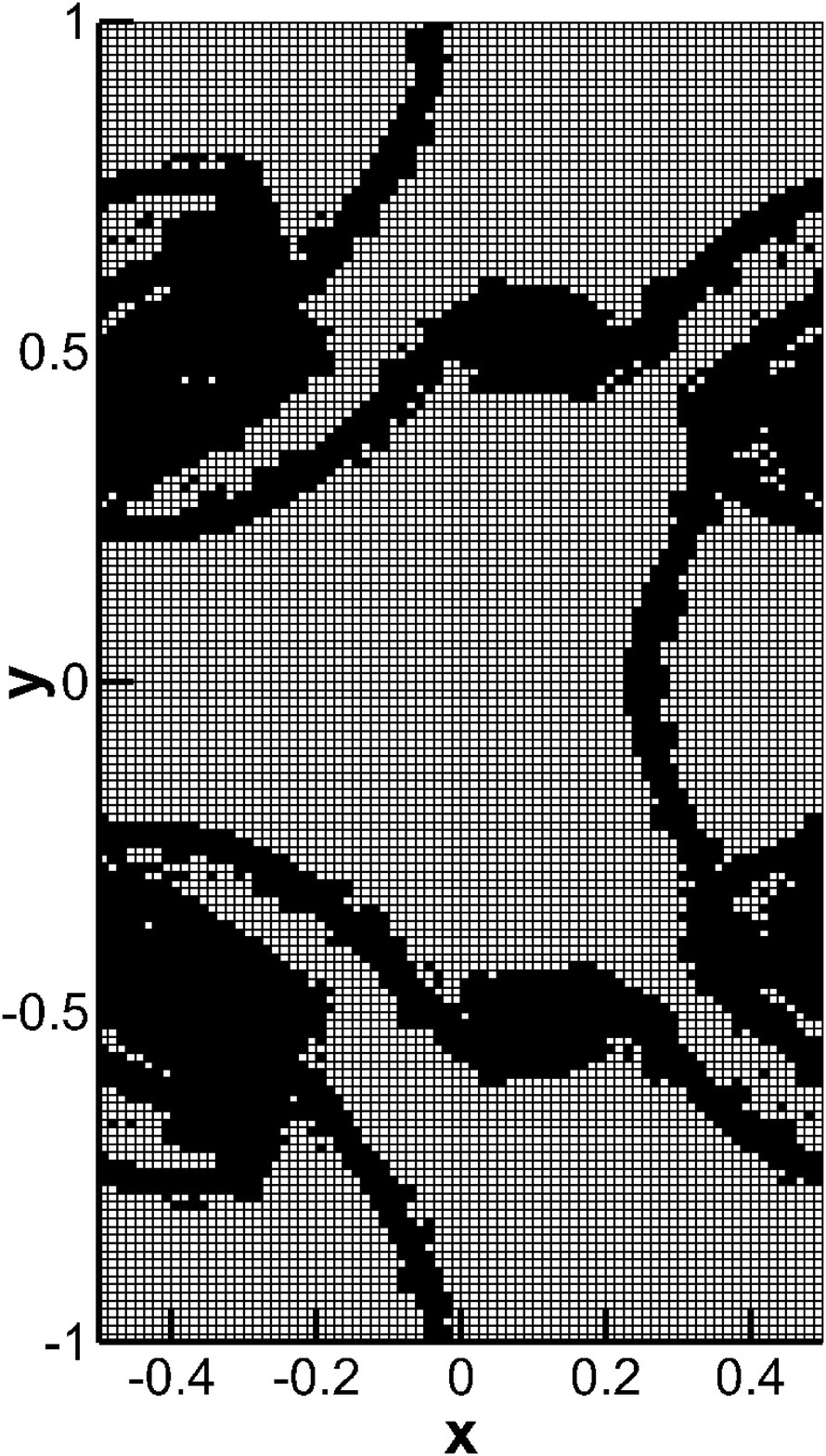}}
{\includegraphics[angle=0,width=4.0cm,height=8.0cm]{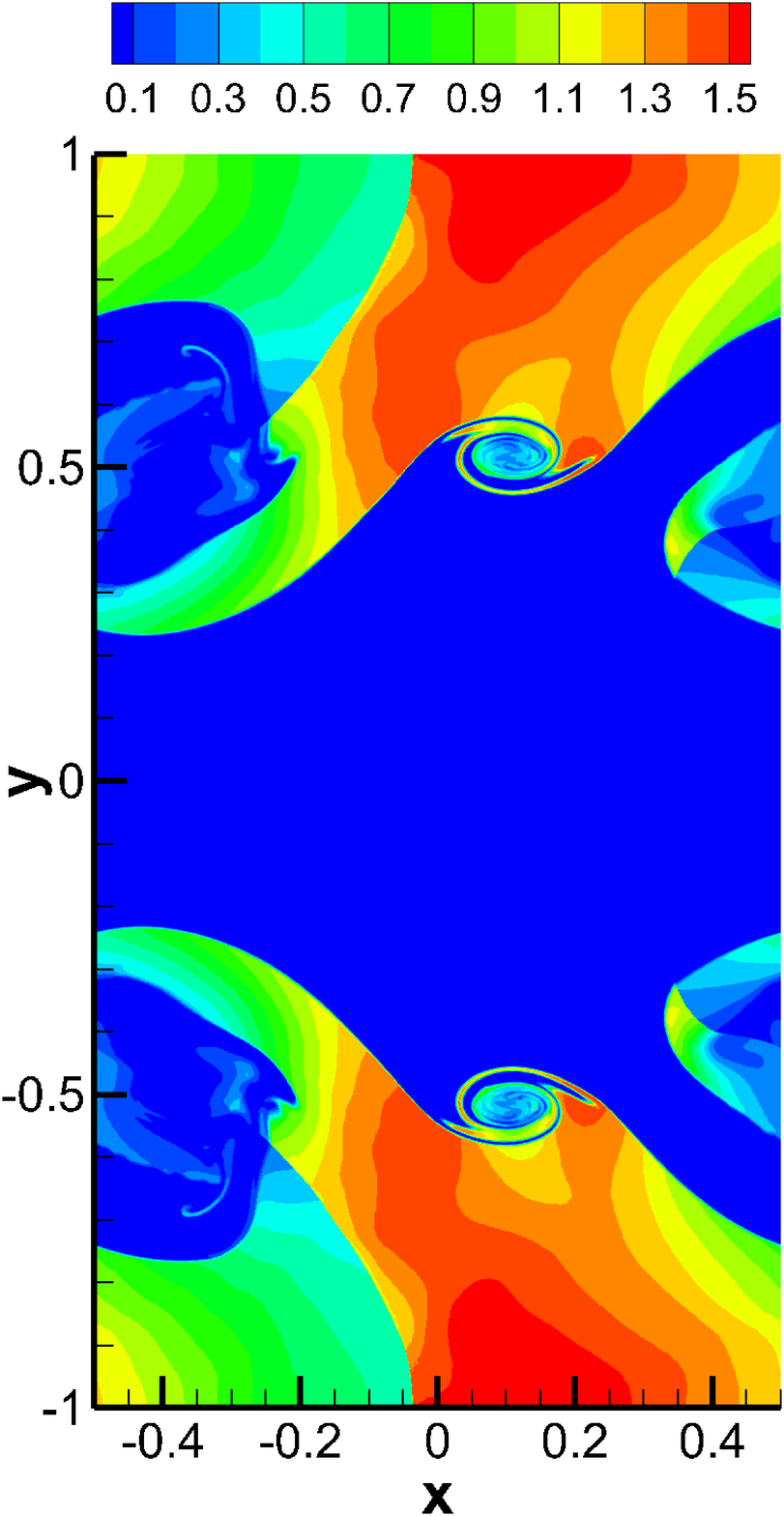}}
\end{center}
\caption{
Two-dimensional Kelvin-Helmholtz instability at $t=3.0$ with {\bf refinement factor} $\boldsymbol{\mathfrak{r}=4}$.
Left panels: the AMR grid.
Right panels: the distribution of the rest-mass density.
Top panels: results obtained with the HLL Riemann solver.
Bottom panels: results obtained with the Osher Riemann solver.
In both cases a third order ADER-WENO with CFL=0.4 has been adopted.
}
\label{fig:KH-RHD-reffactor4}
\end{figure}
The Kelvin--Helmholtz (KH) instability is a classical instability of fluid dynamics and 
in the astrophysical context is currently invoked  to explain the observed phenomenology of extended radio-jets [see \citet{Marti03} and references therein]. 
In this section we present a simple test performed in two dimensions, where the linear growth phase of the KH instability is reproduced.
Following the works of \citet{Mignone2009}, \citet{Beckwith2011} and \citet{Radice2012a}, we choose the initial conditions as
\begin{equation}\label{KHI-vx}
  v_x = \left\{\begin{array}{ll}
  v_s \tanh{[(y-0.5)/a]} & \quad y > 0\,, \\
 \noalign{\medskip}
 -v_s \tanh{[(y+0.5)/a]}  & \quad y \leq 0    \,, \\
 \noalign{\medskip}
 \end{array}\right.
\end{equation}
where $v_s=0.5$ is the velocity of the shear layer and $a=0.01$ is its characteristic size. In order to trigger the instability, a perturbation in the transverse velocity is introduced as
\begin{equation}\label{KHI-vy}
  v_y = \left\{\begin{array}{ll}
  \eta_0 v_s \sin{(2\pi x)} \exp{[-(y-0.5)^2/\sigma]} & \quad y > 0\,, \\
 \noalign{\medskip}
 -\eta_0 v_s \sin{(2\pi x)} \exp{[-(y+0.5)^2/\sigma]}  & \quad y \leq 0    \,, \\
 \noalign{\medskip}
 \end{array}\right.
\end{equation}
where $\eta_0=0.1$ is the amplitude of the perturbation, while $\sigma=0.1$ is its length scale. Finally, the rest-mass density is chosen as
\begin{equation}\label{KHI-rho}
  \rho = \left\{\begin{array}{ll}
  \rho_0 + \rho_1 \tanh{[(y-0.5)/a]} & \quad y > 0\,, \\
 \noalign{\medskip}
 \rho_0 - \rho_1 \tanh{[(y+0.5)/a]}  & \quad y \leq 0    \,, \\
 \noalign{\medskip}
 \end{array}\right.
\end{equation}
with $\rho_0=0.505$ and $\rho_1=0.495$.  The adiabatic index is $\gamma=4/3$ and the pressure is $p=1$ everywhere.
\begin{figure}
\begin{center}
{\includegraphics[angle=0,width=7.0cm,height=7.0cm]{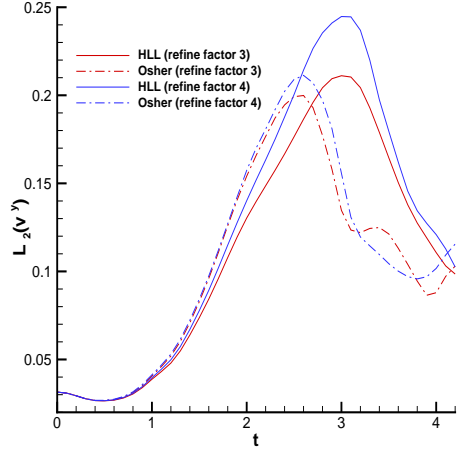}}
\caption{
Time evolution of the L2 norm of $v_y$ showing the transition from the linear to the non-linear phase of the KH instability, occurring for $t\in [2,3]$. 
The red and blue curves refer to the refinement factor $\mathfrak{r}=3$ and $\mathfrak{r}=4$, respectively.
The comparison between HLL and Osher is also shown.  
}
\label{fig:KH-RHD-linear-phase}
\end{center}
\end{figure}
We have solved this problem on a rectangular domain $[-0.5,0.5]\times[-1,1]$, starting from an initial uniform grid with resolution $80\times160$
and adopting periodic boundary conditions both in $x$ and in $y$ directions. 
A third order ADER-WENO has been chosen 
and AMR is activated with relevant parameters $\ell_{\rm max}=2$ and CFL=0.4. 
The results of our computations are reported in Fig.~\ref{fig:KH-RHD-reffactor3} and in Fig.~\ref{fig:KH-RHD-reffactor4}, 
which refer to a refinement factor $\mathfrak{r}=3$ and $\mathfrak{r}=4$, respectively.
In each figure the left panels display 
the AMR grid at time $t=3$ while the right panels display the corresponding
distribution of the rest--mass density. Finally,
the top and the bottom panels refer to the HLL Riemann solver and to the Osher-type Riemann solver, respectively. In all cases we have performed  reconstruction
on the characteristic variables. 
At least two comments can be made about these results. First,
the more diffusive HLL Riemann solver does not allow for the development of secondary small-scale instabilities along the shear layer (compare the top and the bottom panels in each figure).
Second, although the Osher-type Riemann solver is clearly able to capture the formation of tiny structures, by increasing the resolution the number of 
produced vortices decreases. This is evident after comparing the bottom panels of Fig.~\ref{fig:KH-RHD-reffactor3} and Fig.~\ref{fig:KH-RHD-reffactor4}
which, at the reported time $t=3$, have 269,858 and 627,024 cell elements, respectively. This result suggests that these secondary instabilities do not have a clear physical origin, and confirms what already found by \citet{Radice2012a}.

The growth scale of the instability is instead shown in Fig.~\ref{fig:KH-RHD-linear-phase}, where we have plotted, as a function of time,
the $L_2$ norm of the transverse velocity $v_y$. This is in fact a  global quantity suitable to monitor the development of the instability.
The red and the blue curves refer to the refinement factor $\mathfrak{r}=3$ and $\mathfrak{r}=4$, respectively, while the continuous and the dashed lines are used to distinguish among the two Riemann solvers adopted.
The transition from the linear to the non-linear regime  occurs for $t\in [2,3]$ and it is slightly delayed when the diffusive HLL solver is used.
Moreover, in the higher resolution simulations obtained with $\mathfrak{r}=4$, the steepness of the linear phase is larger, producing  higher maxima of  $L_2(v_y)$.
Since the HLL solver fails to capture contact discontinuities, which are certainly produced in a complex flow, this test shows the importance of using an advanced Riemann solver like the Osher-type that we have adopted.

%----------------------------------------------------------
\subsection{RHD Richtmyer--Meshkov instability}
\label{RHD-RMI}
%----------------------------------------------------------
%
\begin{figure}
\begin{center}
{\includegraphics[angle=0,width=12.0cm,height=1.9cm]{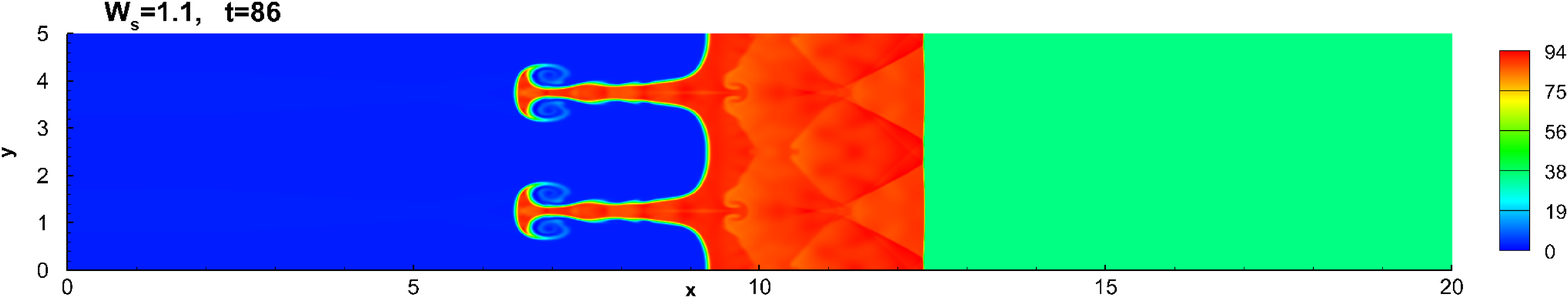}}
{\includegraphics[angle=0,width=12.0cm,height=1.9cm]{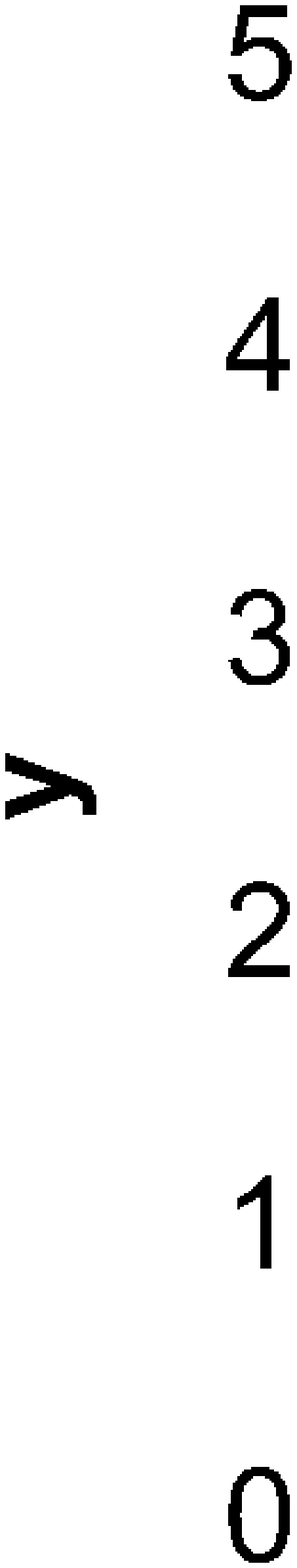}}
\end{center}
\caption{
Third order simulation of the RM instability in a configuration with Lorentz factor $W_s=1.1$. The instability is fully developed.
Only a portion of the numerical grid is shown.
}
\label{fig:RMW1.1}
\end{figure}
\begin{figure}
\begin{center}
{\includegraphics[angle=0,width=12.0cm,height=1.9cm]{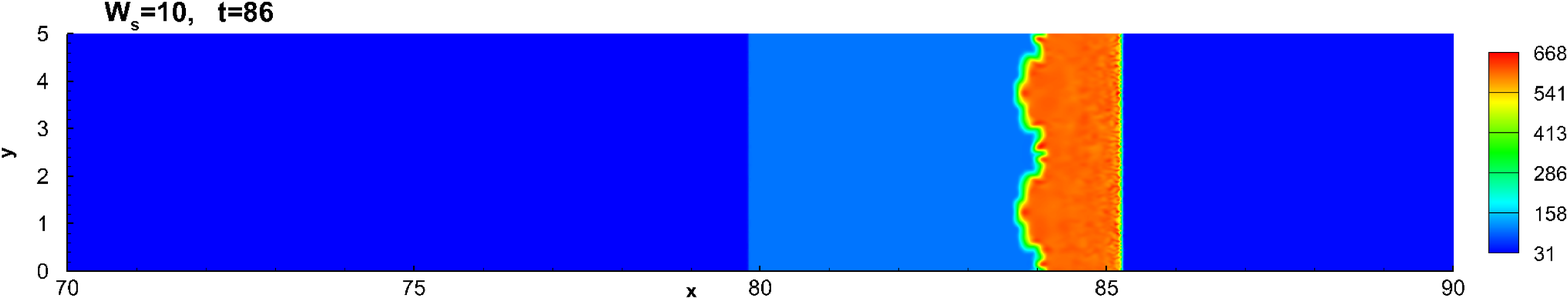}}
{\includegraphics[angle=0,width=12.0cm,height=1.9cm]{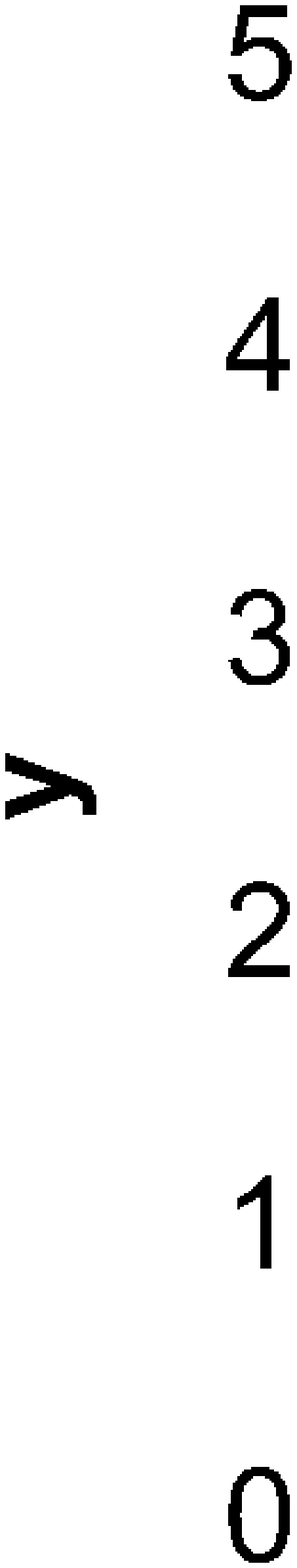}}
\end{center}
\caption{
Third order simulation RM instability in a configuration with Lorentz factor $W_s=10$. The instability is essentially suppressed.
Only a portion of the numerical grid is shown.
}
\label{fig:RMW10}
\end{figure}
As a last test in two space dimensions, we have considered a simple case of the Richtmyer--Meshkov instability, which arises when a shock wave 
crosses a contact discontinuity within a fluid~\citep{Richtmyer1960,Meshkov1969}. 
The numerical domain is $\Omega=[0,100]\times[0,5]$, and it is initially covered by 
a level zero grid with $600 \times 30$ cells.
The fluid is initially at rest ($v_x=0,v_y=0$), with a jump in the density between two states 
separated by a sinusoidal perturbation at $x_0+a\sin(\pi/2+2\pi y/\lambda)$, where $x_0=3$, $a=0.25$ and $\lambda=2.5$. 
At time $t=0$ a single shock wave is placed at $x=1$ and propagates towards the right with a prescribed Lorentz factor $W_s$. The shock wave is built after solving the relativistic Rankine--Hugoniot conditions~\citep{Taub1948}. The purpose of this test is not to perform a detailed physical analysis of the 
Richtmyer--Meshkov instability in the relativistic regime, but rather to show the ability of the numerical scheme in handling large Lorentz factor flows. After the initial shock wave hits the sinusoidal interface, the standard development of the instability, well documented in Newtonian experiments and simulations [see \cite{Brouillette2002} for a review], generates a series of "mushroom'' structures, which have the net effect of mixing the fluid at either sides of the initial discontinuity.

This is also the case in the relativistic regime, but only for moderate Lorentz factors. Indeed, as 
reported in Fig.~\ref{fig:RMW1.1}, the typical phenomenology of the Richtmyer--Meshkov instability is produced 
in a model with $W_s=1.1$. 
On the contrary,  Fig.~\ref{fig:RMW10} corresponds to a model with 
$W_s=10$ and relativistic Mach number ${\cal M}_{\rm s}=32.2$, for which  
the Richtmyer--Meshkov instability is essentially suppressed (see also \cite{Mohseni2013}). 
We emphasize that, in this test with high Lorentz factor, the conversion from the conserved to the primitive variables 
is better obtained when the second strategy described in Section~\ref{srh} is adopted, namely when Eq.~(\ref{cons2prim-second-choice}) is solved.
%----------------------------------------------------------
\subsection{RHD three-dimensional spherical explosion}
\label{EP3D}
%----------------------------------------------------------
%
%
\begin{figure}
\begin{center}
{\includegraphics[angle=0,width=7.0cm,height=7.0cm]{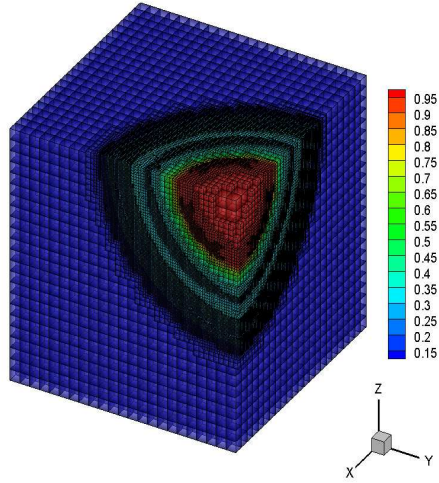}}
\caption{
AMR grid structure at time $t=0.25$ in the spherical explosion problem. The color contour of the rest--mass density is also shown.
}
\label{fig:Explosion3D-RHD-grid}
\end{center}
\end{figure}
\begin{figure}
\begin{center}
\begin{tabular}{lr} 
\includegraphics[angle=0,width=0.5\textwidth]{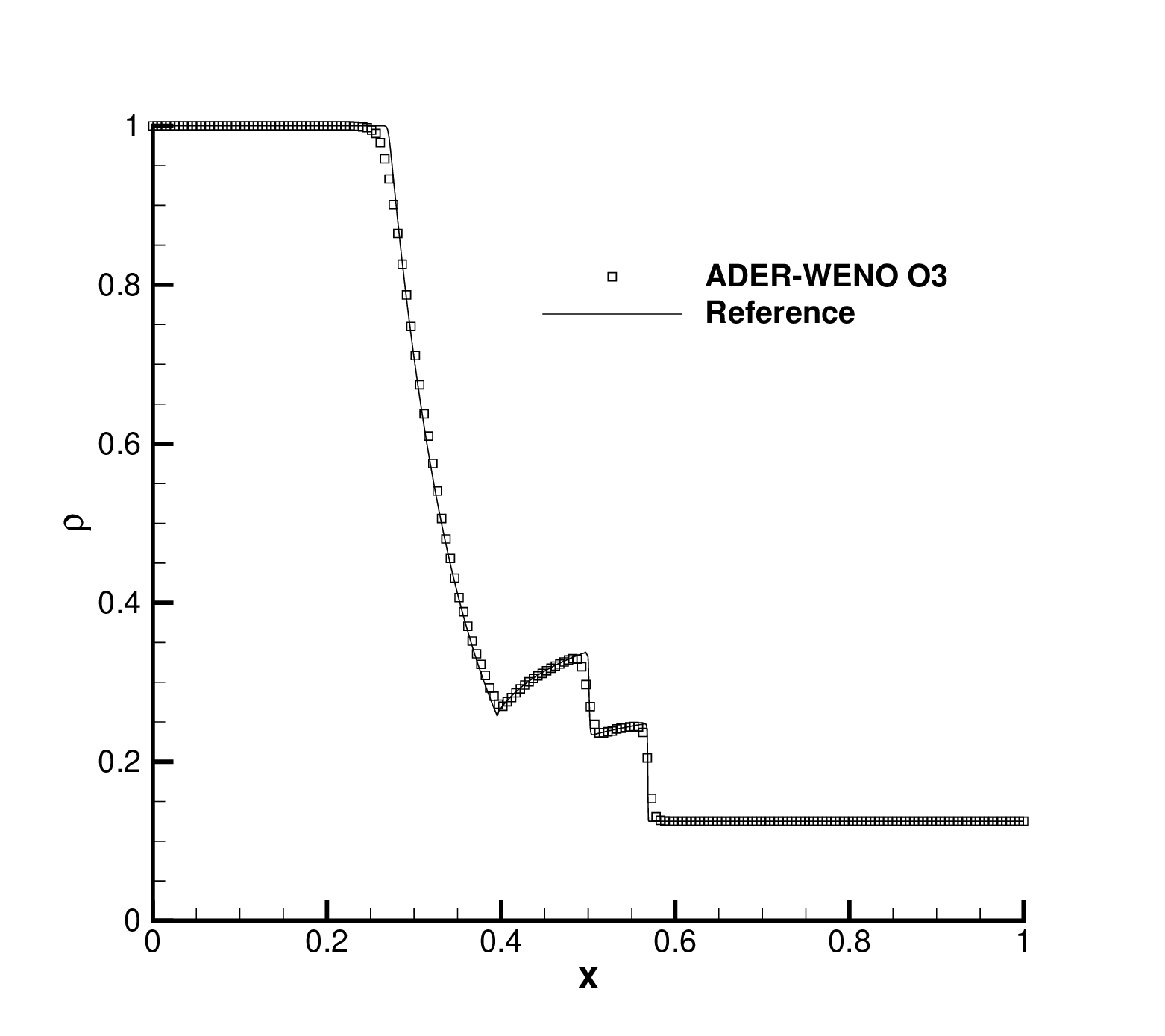} &
\includegraphics[angle=0,width=0.5\textwidth]{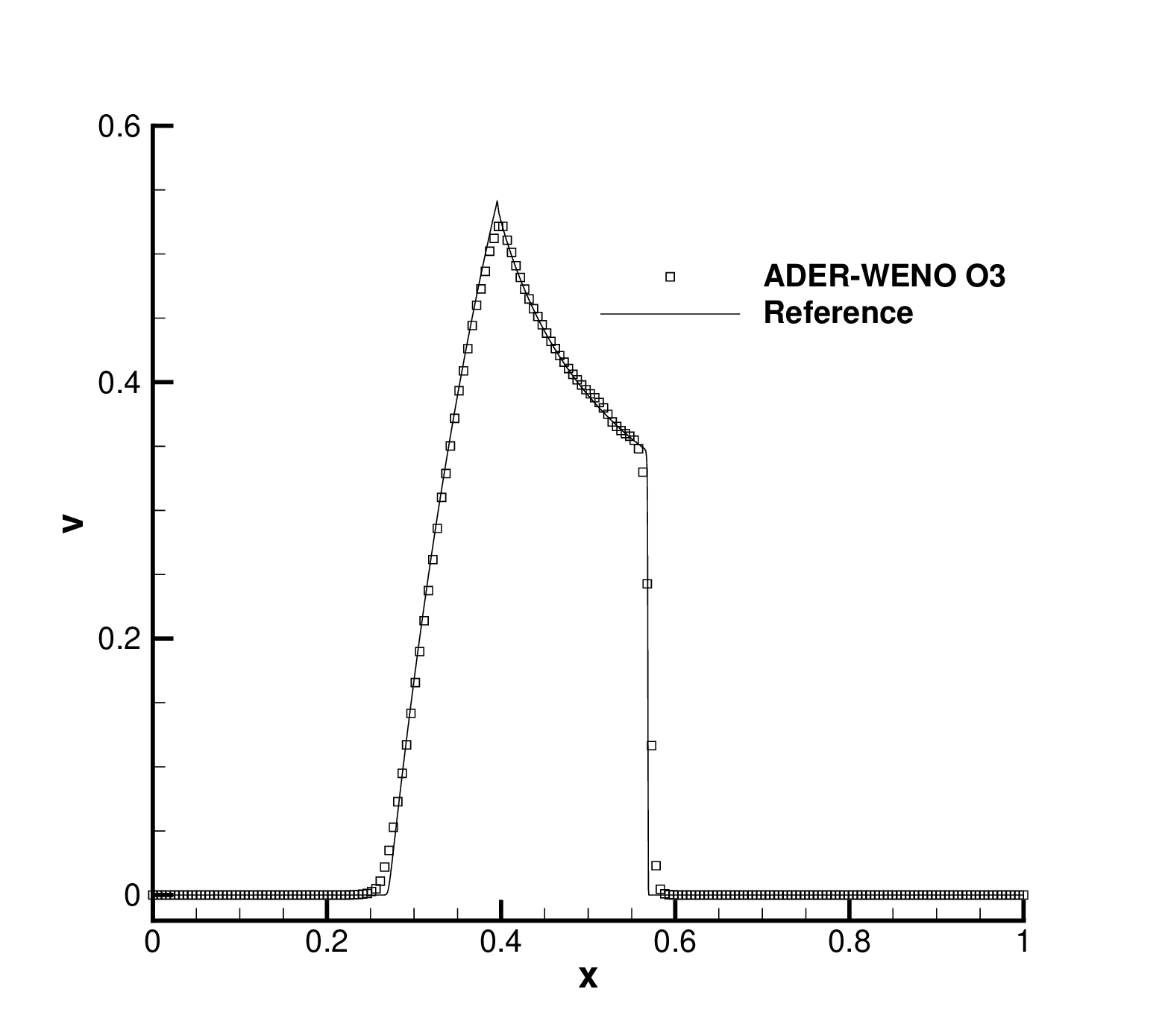} \\
\includegraphics[angle=0,width=0.5\textwidth]{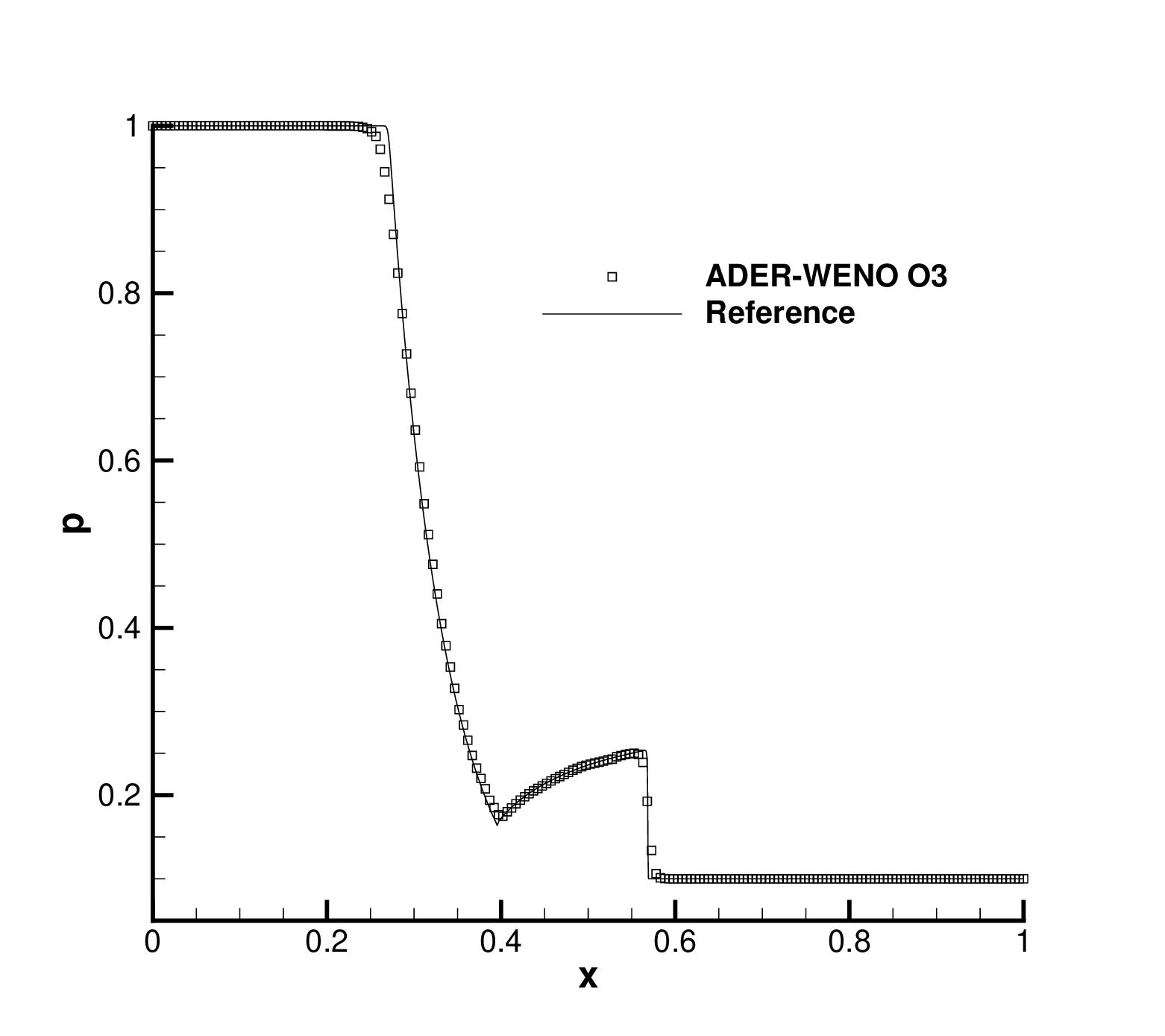} &
\includegraphics[angle=0,width=0.5\textwidth]{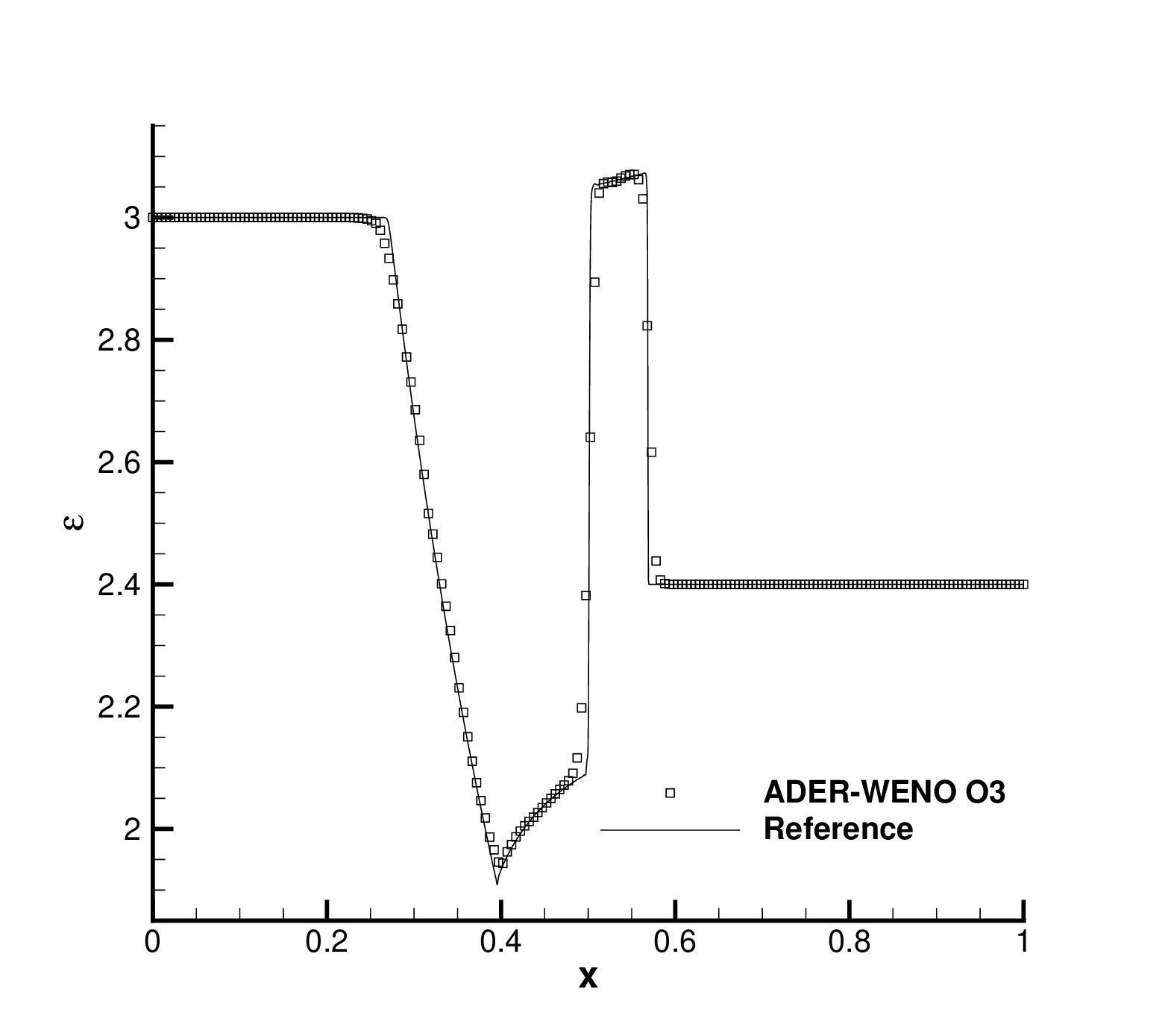} 
\end{tabular}  
\caption{{Explosion test in three space dimensions.} From top left to bottom right: comparison of the 1D reference solution with the numerical solution obtained 
with a third order ADER-WENO scheme on an AMR mesh at time $t=0.25$. One-dimensional cuts along the $x$-axis are shown for density, velocity, pressure and specific internal energy. Note that the plotted numerical data are the result of an interpolation, and are uniformly distributed. 
}
\label{Explosion3D-RHD-hydro}
\end{center}
\end{figure}
In order to validate the numerical scheme also in three spatial dimensions, 
we solve the special relativistic explosion problem on the computational domain $\Omega=[-1; 1]^3$, with initial conditions given by 
\begin{equation}\label{RP2D}
  \big(\rho, v,  p\big) = \left\{\begin{array}{ll}
\big(1.0, 0   ,    1\big) & \;\textrm{for}\quad r \leq R\,, \\
 \noalign{\medskip}
 \big(0.125,   0,    , 0.1\big) & \;\textrm{for}\quad r>R    \,, \\
 \noalign{\medskip}
 \end{array}\right.
\end{equation} 
where the radial coordinate $r = \sqrt{x^2+y^2+z^2}$, while $R=0.4$ denotes the radius of the initial discontinuity. In practice, the
initial flow variables take the same constant values of the classical Sod's problem \citep{Sod1978}.
The adiabatic index of the ideal-gas equation of state has been set to $\gamma=4/3$. Due to the symmetry of the problem,  the solution can be compared with an equivalent one dimensional problem 
in the radial direction $r$, which we have solved using the ECHO code by \cite{DelZanna2007}
with a high resolution uniform grid composed of $3000$ cells. 
The initial grid at level zero of this three-dimensional problem uses $48\times48\times48$ cells, while during the evolution we have chosen 
$\mathfrak{r}=2$ and $\ell_{\rm max}=3$. We have performed a calculation at third order accuracy 
with the Osher-type Riemann solver and reconstruction in characteristic variables.
The AMR grid at the final time $t=0.25$ is shown in Fig.~\ref{fig:Explosion3D-RHD-grid} and is formed by 7,044,376
elements. 
We stress that the a totally refined uniform grid corresponding to this configuration would be composed by 56,623,104 elements. 
One dimensional profiles of the most relevant quantities along the $x$ axis 
are reported in the four panels of Fig.~\ref{Explosion3D-RHD-hydro}. 
A very good agreement with the one-dimensional reference solution is obtained.

%-----------------------------------------------------------
\subsection{RMHD circularly-polarized Alfven Wave: convergence test}
\label{RMHD-Alfven}
%-----------------------------------------------------------

\begin{figure}
{\includegraphics[angle=0,width=6.5cm,height=6.5cm]{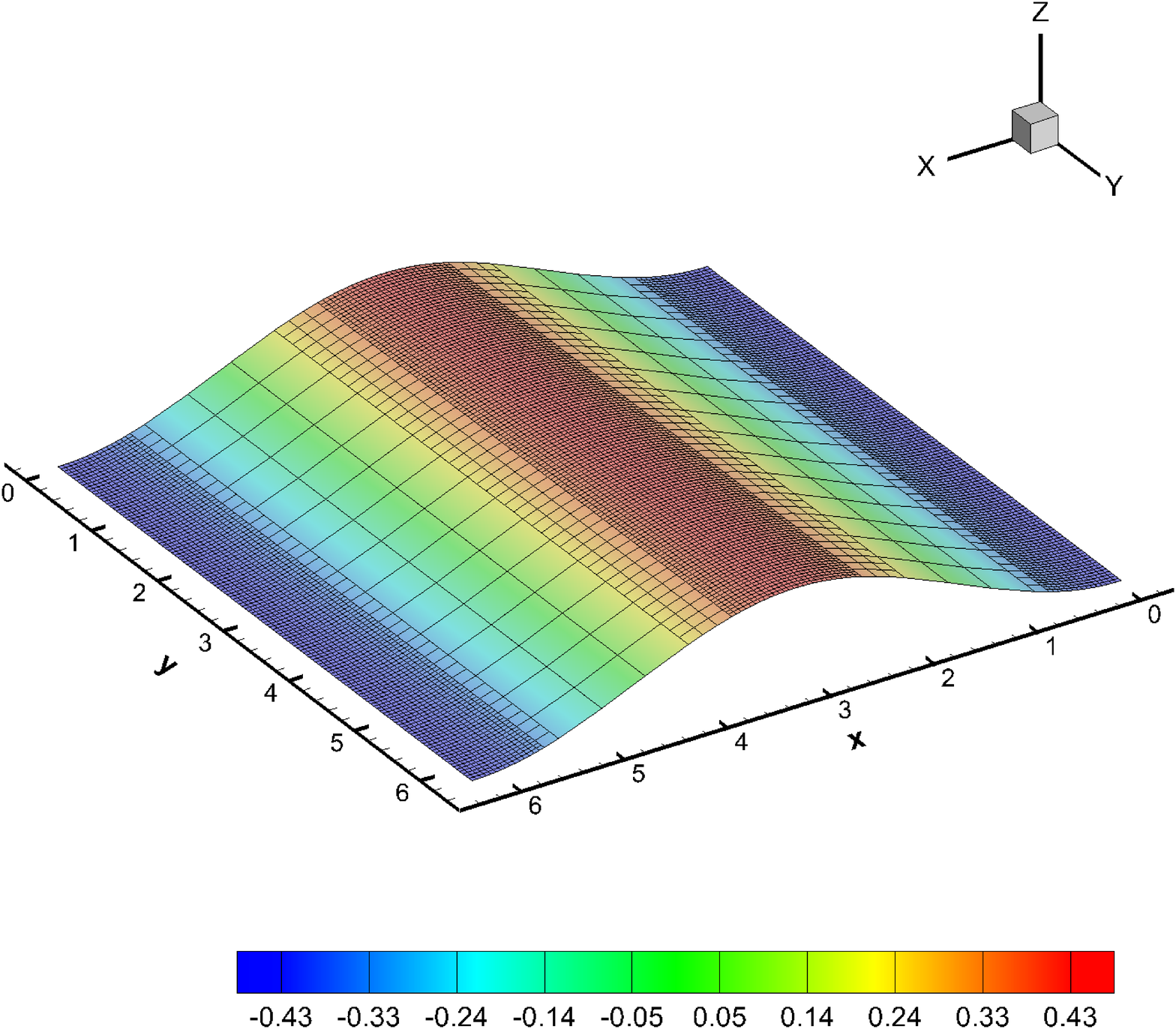}}
{\includegraphics[angle=0,width=6.5cm,height=6.5cm]{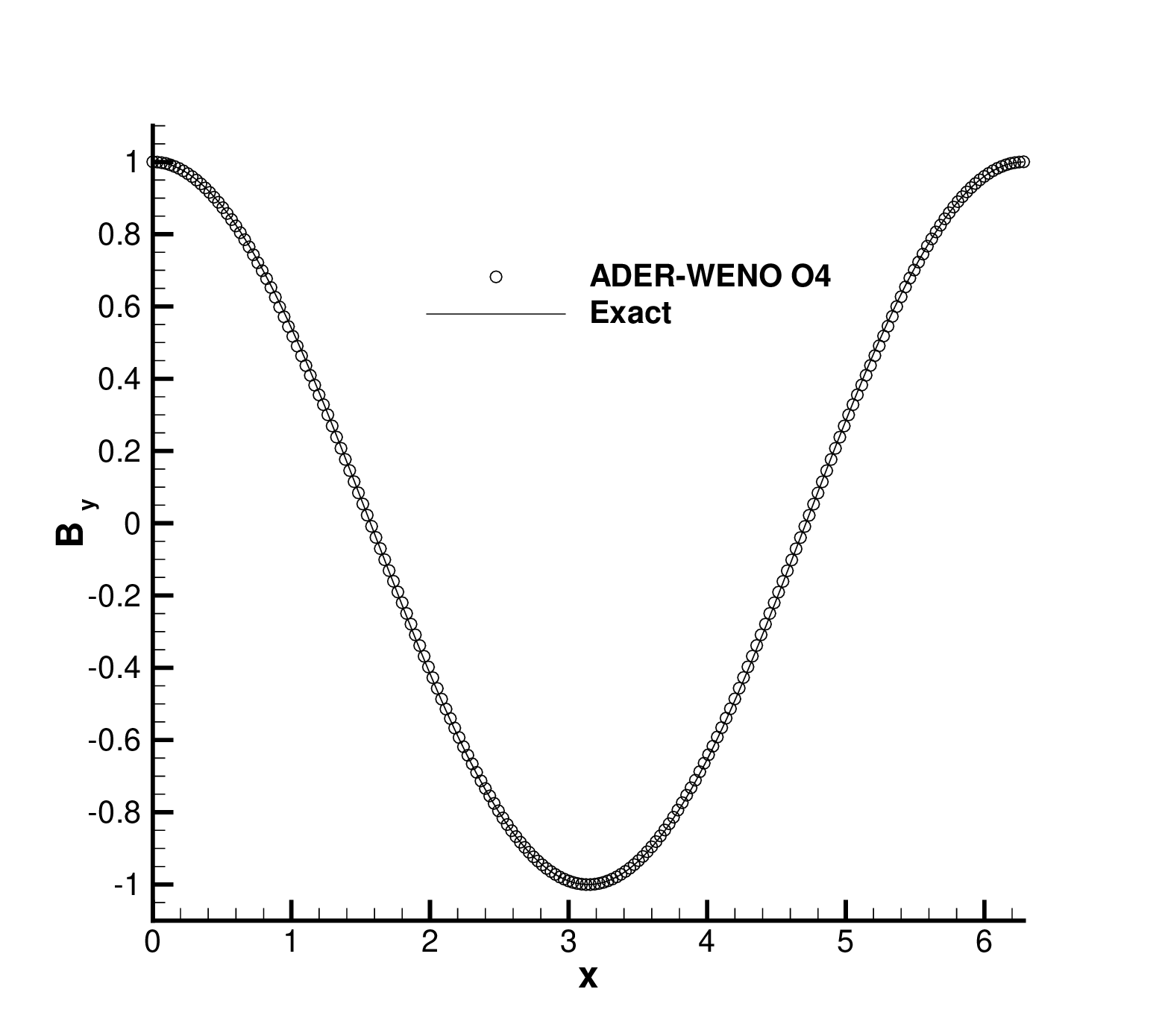}}
\caption{Solution of the circularly-polarized Alfven Wave problem with the
  fourth order ADER-WENO scheme at time $t=T$. Left panel: distribution of $v_y$ over the AMR grid.
	Right panel: comparison of the numerical and of the exact solution for $B_y$.
	}
\label{fig:Alfven3D}
\end{figure}

Alfven waves in ideal relativistic MHD have been studied in detail by \cite{Komissarov1997}, while the solution for a large amplitude circularly-polarized
Alfven wave was described by \cite{DelZanna2007}. Being analytic, the latter solution is particularly convenient to perform an analysis of the convergence properties of our numerical scheme. In practice, the magnetic field is given by 
\begin{eqnarray}
B_x&=&B_0 \\
B_y&=&\eta B_0\cos[k(x-v_A t)]\\
B_z&=&\eta B_0\sin[k(x-v_A t)]\,,
\end{eqnarray}
where $B_0$ is the uniform magnetic field in the direction of propagation of the wave (the $x$ direction), $\eta$ is the amplitude of the wave, which can be arbitrarily large, $k$ is the wave number, while $v_A$ is the Alfven speed, which can be shown to be \citep{DelZanna2007}
\begin{equation}
v_A^2=\frac{B_0^2}{\rho h\!+\!B_0^2(1\!+\!\eta^2)}\left[\frac{1}{2}\left(1\!+\sqrt{1\!-\left(\frac{2\eta B_0^2}{\rho h\!+\!B_0^2(1\!+\!\eta^2)}\right)^2}\,\right)\right]^{-1}\!\!.
\end{equation}
The transverse velocity field, with vector tips describing circles in the plane normal to $\vec{B}_0$ (the $yz$ plane), is 
\begin{equation}
v_y=-v_A B_y/B_0,~~~v_z=-v_A B_z/B_0\,.
\end{equation}
Since the wave is incompressible, the background vales of $\rho$ and $p$ are not affected, and the wave propagates with speed given by $v_A$. 
In our calculation we have used $\rho=p=B_0=\eta=1$. We have performed this test in two spatial dimensions, using periodic boundary conditions, over the computational domain $\Omega=[0; 2\pi]\times[0; 2\pi]$. After one period $T=L/v_A=2\pi/v_A$, we compare the numerical solution with the analytic one at time $t=0$.
Fig.~\ref{fig:Alfven3D} shows the result of a representative calculation at the fourth order of accuracy, and the comparison with the exact solution.
In order to verify the effective convergence rates of the numerical scheme, 
we have reported in Table~\ref{tab.conv1} an analysis of 
the $L_2$ norms of the error of $v_y$ (computed with respect to the reference solution as explained above)
and the corresponding orders of convergence. 
This analysis has been performed 
with the HLL flux, $\ell_{\rm max}=2$, $\rm{CFL}=0.8$ and using the minimum refinement factor allowed by the numerical scheme, which, we recall, is 
$\mathfrak{r}=2$ for $M=2$ and $\mathfrak{r}=3$ for $M=3$.
Tab.~\ref{tab.conv1} shows that  the nominal order of convergence is essentially confirmed. The fourth order version of the scheme performs
comparatively better than the third order one.  \\
In addition, we have also compared a simple AMR simulation, that uses $\ell_{\rm max}=1$, $\mathfrak{r}=2$ starting from an initial grid with $20\times20$ cells, with two simulations performed over uniform grids, the first one corresponding to the coarse grid of the AMR case, and the second one corresponding to the 
fine grid of the AMR case.  The results of this comparison are reported in Tab.~\ref{tab.ot.compare}. As expected, the error of the AMR simulation,
denoted as (b) in the Table, falls between those performed over the two uniform grids, denoted as (a) and (c), respectively.
The last column of the table shows the CPU time, normalized to the simulation on the coarse uniform mesh, and it gives evidence that 
the effort inherent to the AMR implementation is largely justified.
%
%
%----------------------------------------------------------
\begin{table}[!t]   
\caption{Numerical convergence results for the circularly-polarized Alfven wave test using the third and the fourth 
order version of the one--step ADER-WENO finite volume scheme.
The error norms refer to the $v_y$ component of the velocity at the final time $t=T$. 
}
\begin{center} 
\renewcommand{\arraystretch}{0.8}
\begin{tabular}{lcclcc} 
\hline
 $\ell_{\rm max}=2$ &   &  & & & \\ 
  \hline
  $N_G$    & $\epsilon_{L_2}$ & $\mathcal{O}(L_2)$ & $N_G$ & $\epsilon_{L_2}$ & $\mathcal{O}(L_2)$  \\ 
\hline
           &         & {$\mathcal{O}3$} & & & {$\mathcal{O}4$}  \\
\hline

  25       & 4.1159E-02 &      & 15        & 6.7463E-03 &       \\ 
  30       & 2.6681E-02 & 2.37 & 20        & 2.1505E-03 & 3.97  \\ 
  35       & 1.8184E-02 & 2.48 & 25        & 7.6842E-04 & 4.61  \\ 
  40       & 1.3008E-02 & 2.51 & 30        & 3.1766E-04 & 4.84  \\  
  45       & 9.7234E-03 & 2.47 & 35        & 1.5723E-04 & 4.56  \\
	50       & 7.5300E-03 & 2.42 & 40        & 8.5133E-05 & 4.59  \\
%	55       & 2.9564E-04 & 2.65 &           &            & \\
%	60       & 2.0361E-04 & 2.79 &           &            & \\
\hline
\end{tabular} 
\end{center}
\label{tab.conv1}
\end{table} 
\begin{table}[!t]   
\caption{Comparison of the error and of the CPU time using (a) a uniform coarse grid with $20\times20$ elements, (b) an AMR grid with 
$\ell_{\rm max}=1$, $\mathfrak{r}=2$ and (c) a uniform fine grid with $40\times40$ elements.
The test adopts the circularly-polarized Alfven wave using a third order ADER-WENO scheme.
The CPU time is normalized with respect to the simulation on the coarse uniform mesh.} 
\begin{center} 
\renewcommand{\arraystretch}{1.0}
\begin{tabular}{lccc} 
\hline
                         & Cells at $t=T$ & $\epsilon_{L_2}(v_y)$ & CPU time \\ 
\hline
(a) Uniform Coarse       & 400            & 8.7703E-02            & 1.0   \\ 
(b) AMR                  & 1000           & 3.8007E-02            & 4.98  \\ 
(c) Uniform Fine         & 1600           & 1.3273E-02            & 7.91 \\ 
\hline
\end{tabular} 
\end{center}
\label{tab.ot.compare}
\end{table} 
%
%
%-----------------------------------------------------------
\subsection{RMHD relativistic shock tube problem}
\label{sec:RMHDtube}
\begin{figure}
\begin{center}
\vspace{-1.2cm}
{\includegraphics[angle=0,width=12.0cm,height=11.0cm]{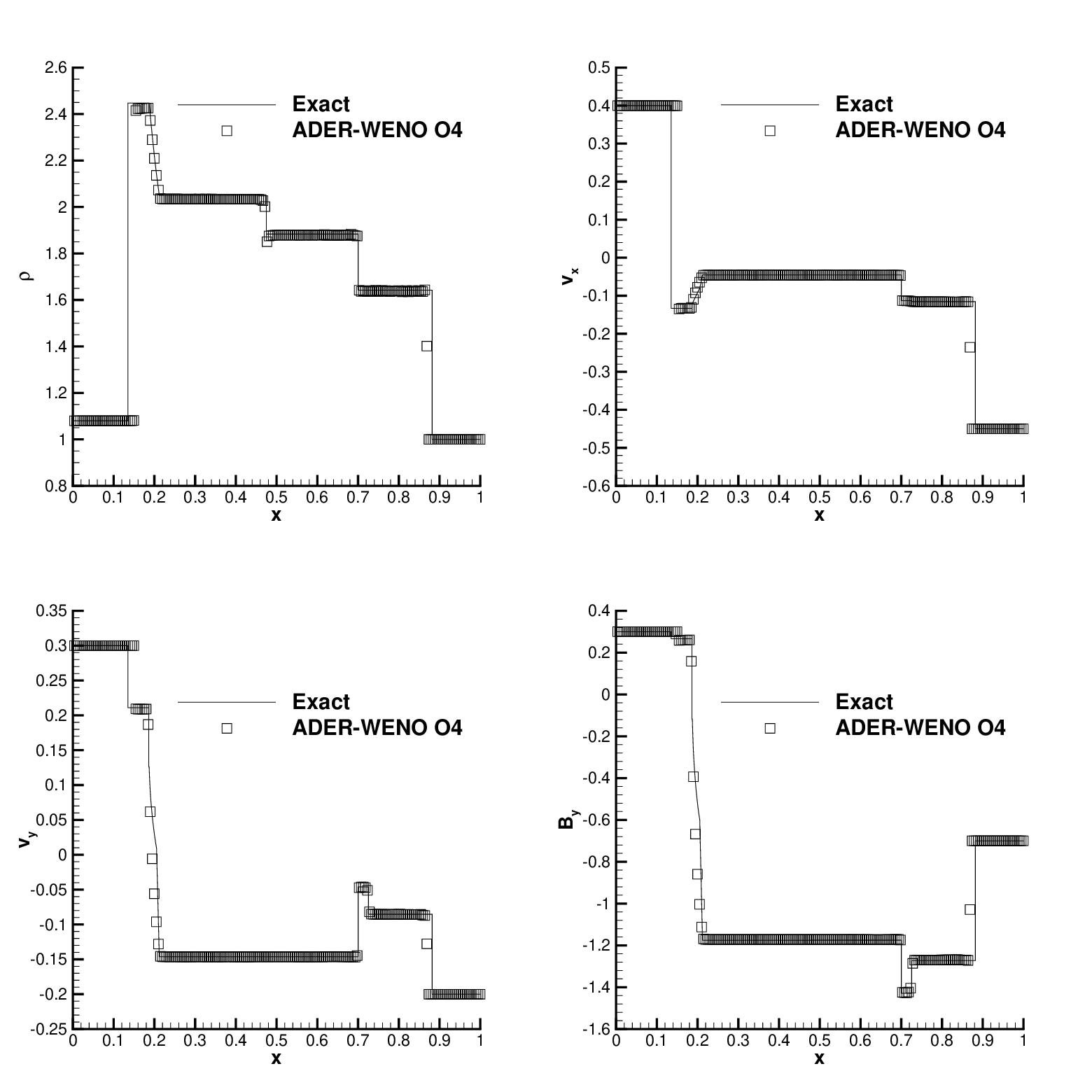}}
\caption{Solution of the RMHD shock tube with the
  fourth order ADER-WENO scheme at time $t=0.55$.}
\label{fig:shock-tube-B5}
\vspace{-1.1cm}
\end{center}
\end{figure}

%-----------------------------------------------------------
%
The new scheme has been validated also through one-dimensional shock tube problems. For brevity, we report here only the 
fifth test among those presented by \cite{Balsara2001}, with initial conditions given by 
\begin{equation}
\label{blast-wave}
(\rho,v_x,v_y,v_z,p,B_y,B_z)= \left\{
\begin{array}{lllllllll}
(1.08,&  0.4,  &  0.3, & 0.2, & 0.95, & 0.3, & 0.3)    &   {\rm if} & x \in[0;0.5]  \\ 
(1.0, & -0.45, & -0.2, & 0.2, & 1.0,  & -0.7, & 0.5)   &   {\rm if} & x \in[0.5;1] \,. \\ 
\end{array} \right.
\end{equation}
The magnetic field along the $x$ direction is continuous across the discontinuity, and it is given by $B_x=2.0$. 
The adiabatic index of the ideal gas is $\gamma=1.6666$, and the problem is evolved to the final time $t_f=0.55$.
Seven waves are produced as the Riemann fan opens: a fast shock, an Alfven wave and a slow rarefaction, which propagate to the left; the usual contact discontinuity; and a slow shock, an Alfven wave and a fast shock, which propagate to the right. We have performed this test at the fourth order of accuracy with the HLL Riemann solver, using $\ell_{\rm max}=2$, $\mathfrak{r}=3$, starting from  an initial uniform grid with 180 cells. The results of the calculations 
are reported in Fig.~\ref{fig:shock-tube-B5}, where they are compared with the exact solution obtained with the Riemann solver of \citet{Giacomazzo:2005jy}.
%
%----------------------------------------------------------
%
%
%-----------------------------------------------------------
\subsection{RMHD Rotor Problem}
\label{RMHD-ROTOR}
\begin{figure}
\begin{center}
{\includegraphics[angle=0,width=6.7cm,height=6.5cm]{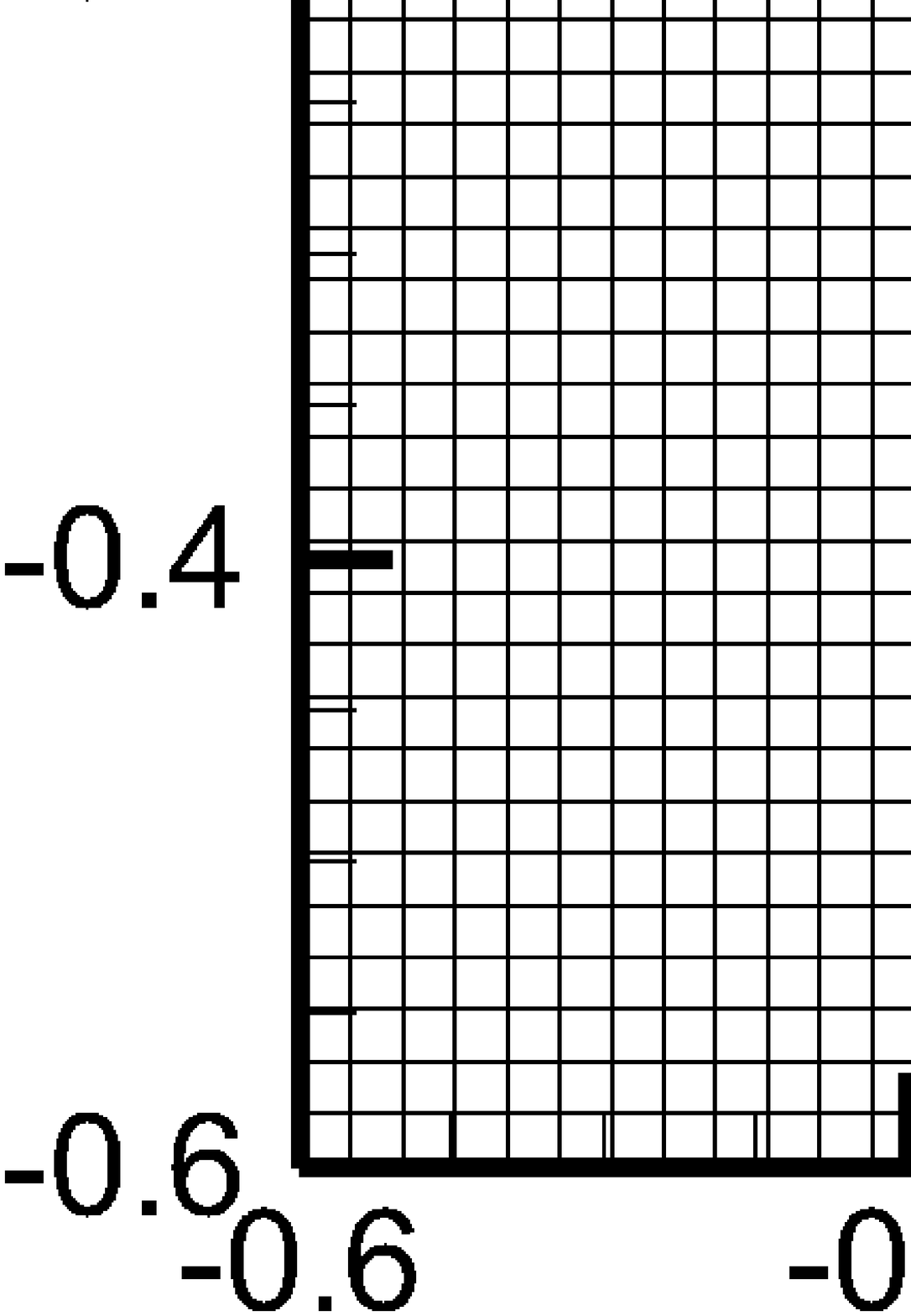}}
{\includegraphics[angle=0,width=6.7cm,height=6.5cm]{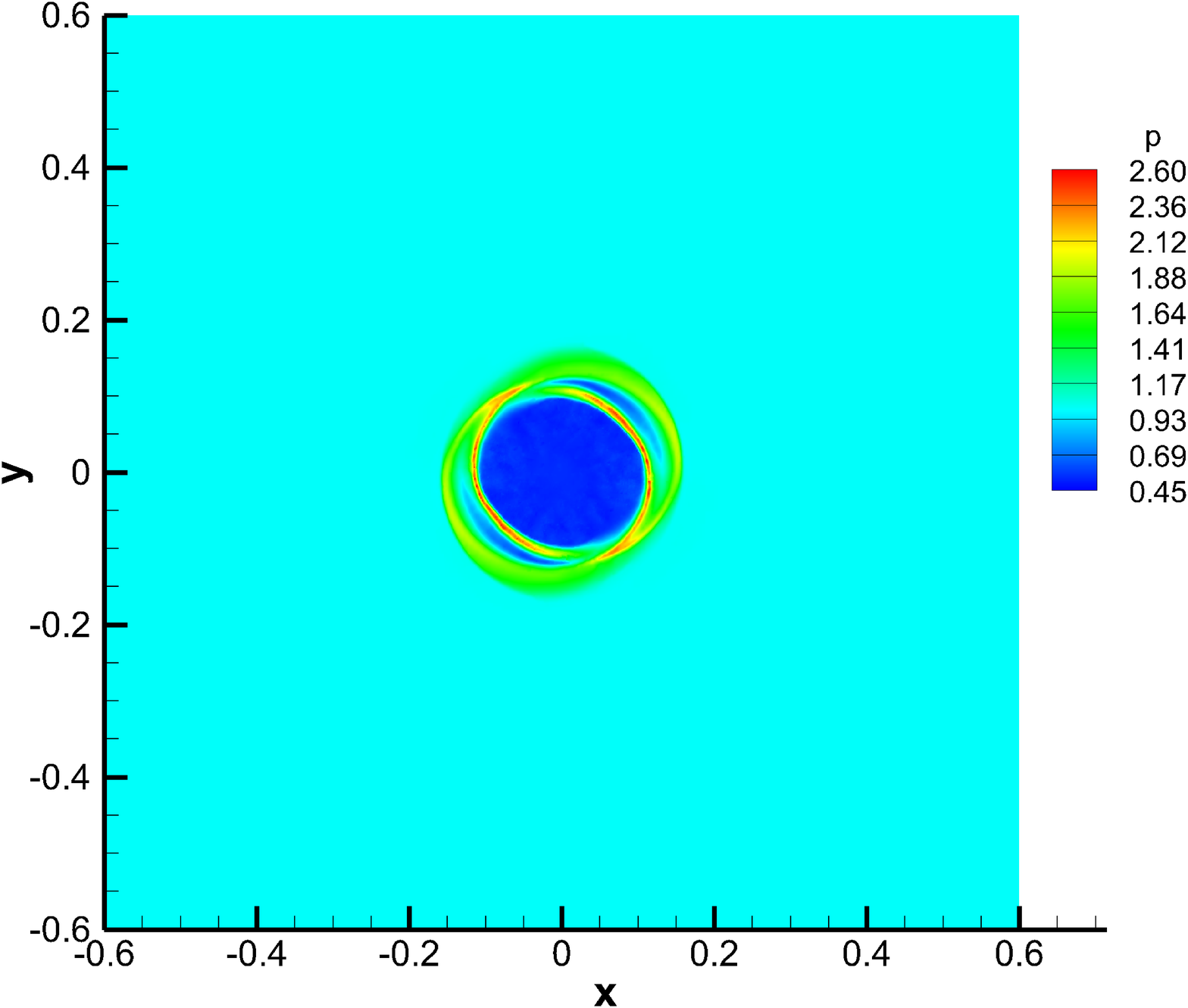}}
\end{center}
\hspace{1cm}
\begin{center}
{\includegraphics[angle=0,width=6.7cm,height=6.5cm]{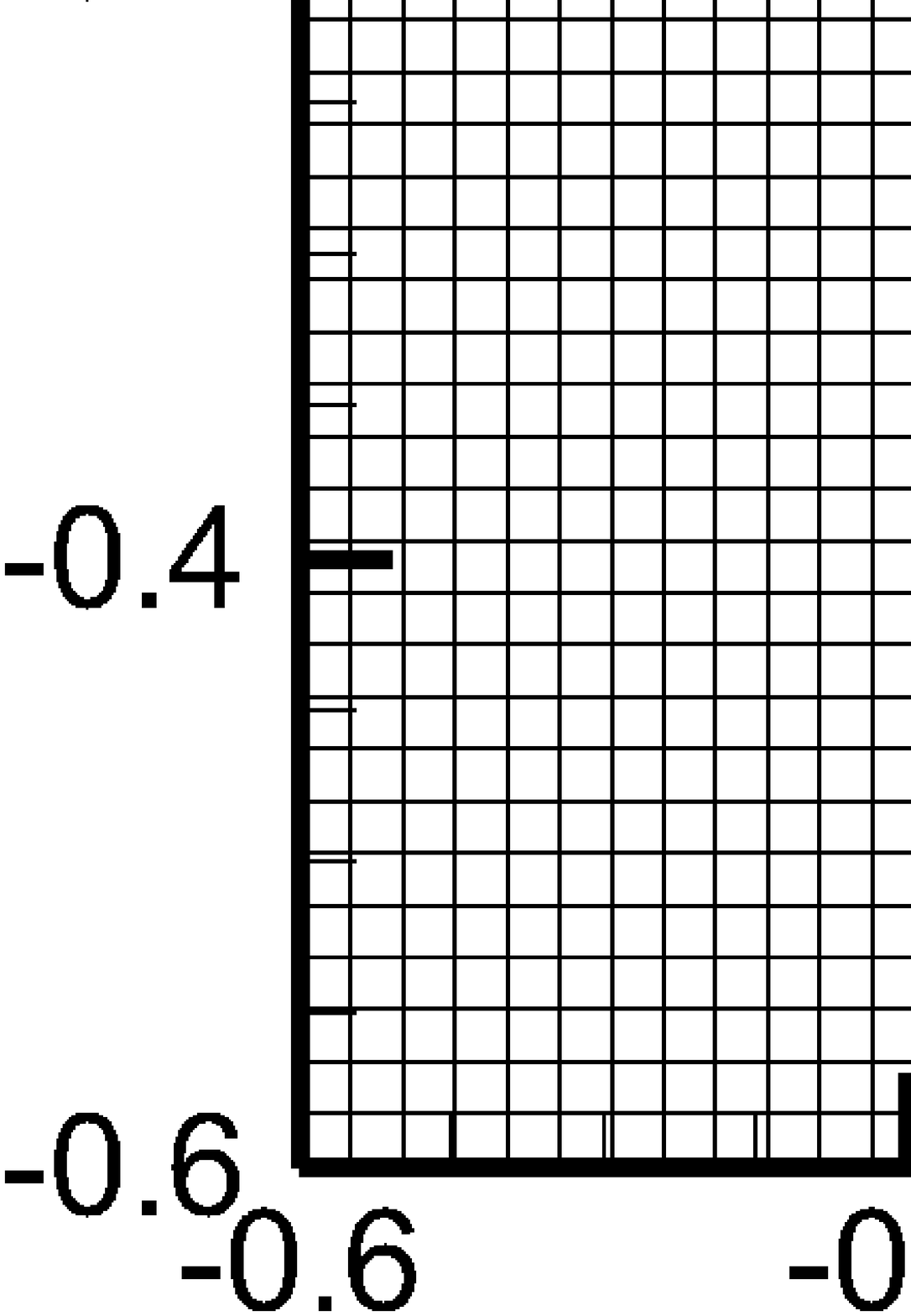}}
{\includegraphics[angle=0,width=6.7cm,height=6.5cm]{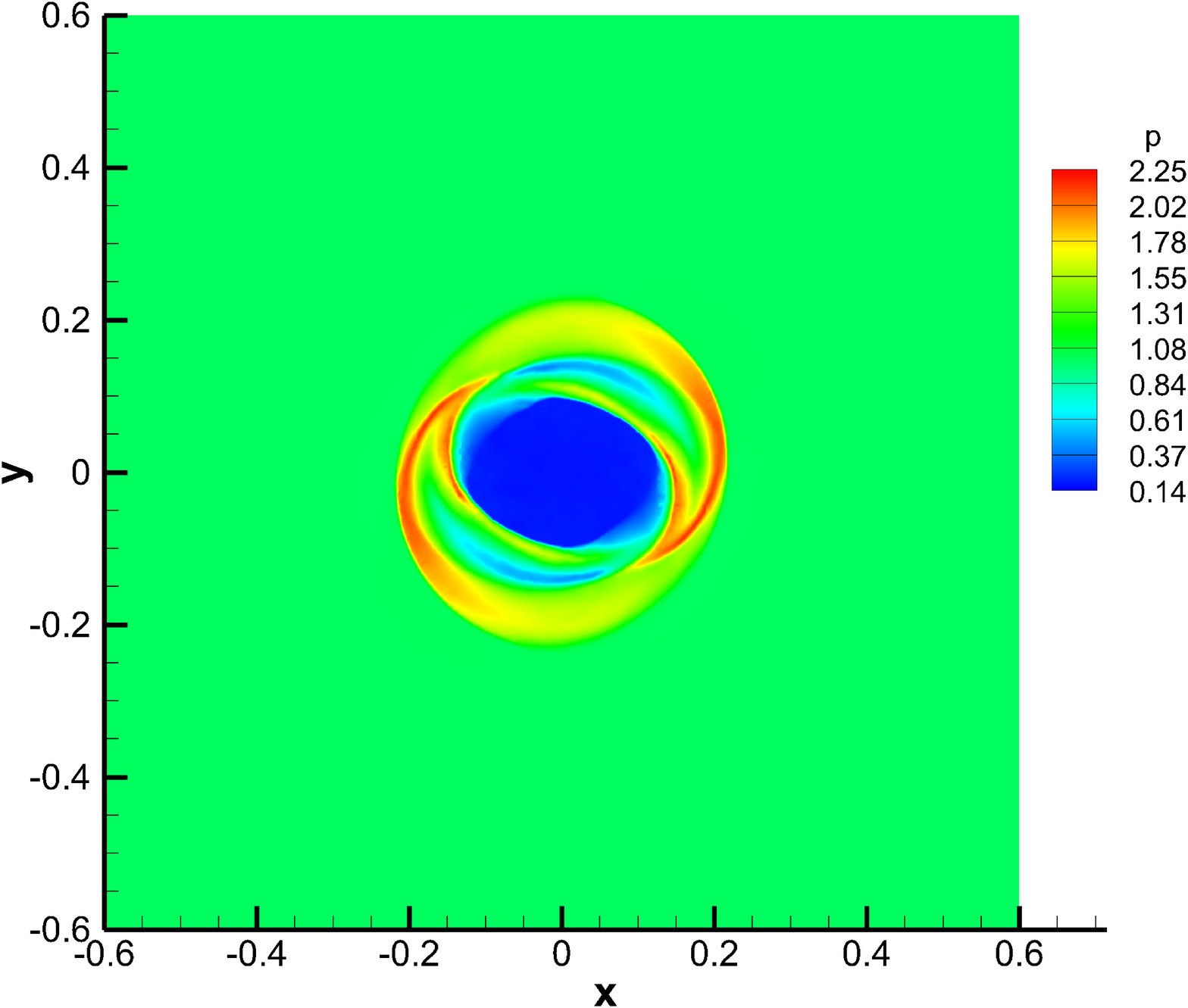}}
\end{center}
\begin{center}
{\includegraphics[angle=0,width=6.7cm,height=6.5cm]{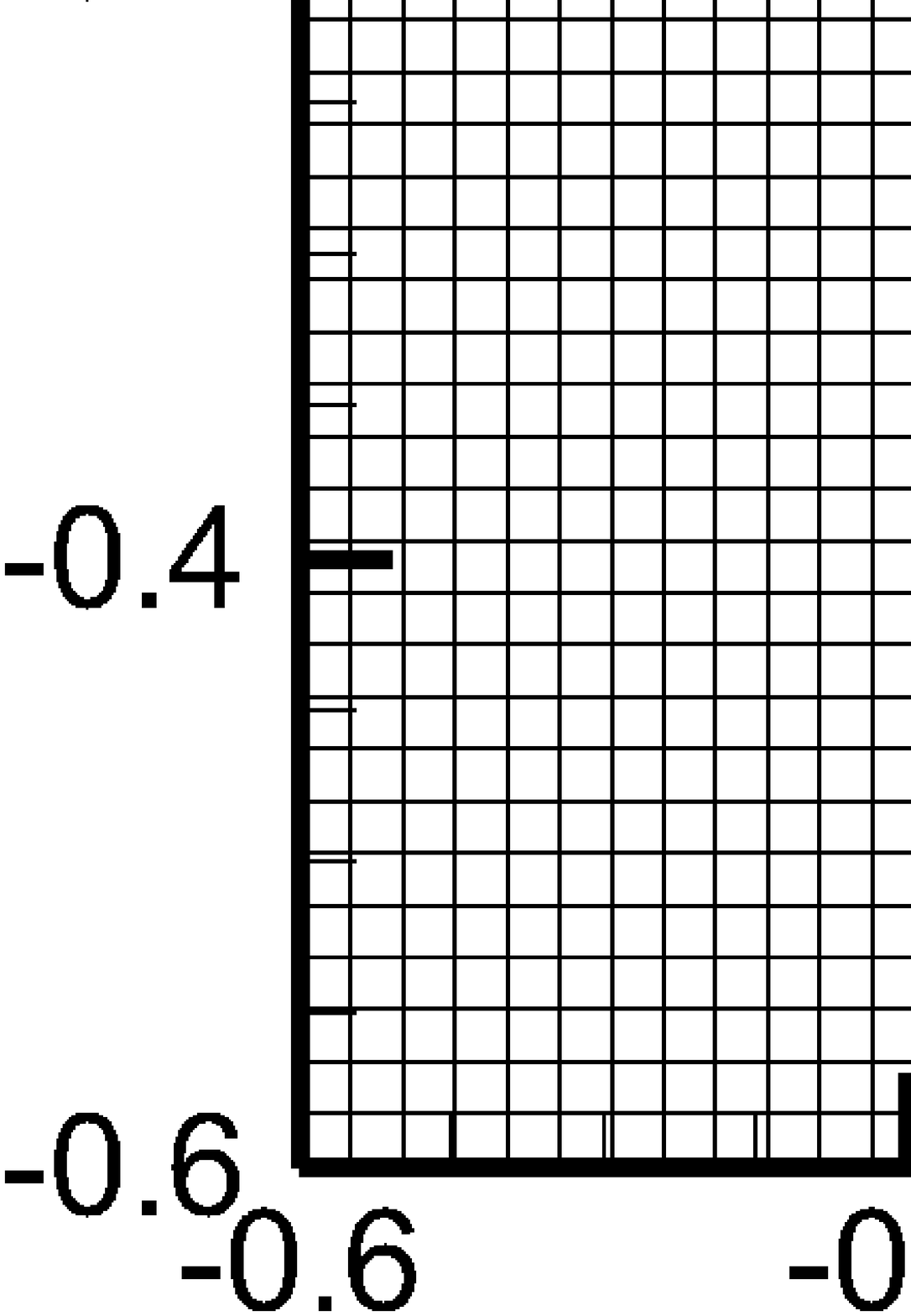}}
{\includegraphics[angle=0,width=6.7cm,height=6.5cm]{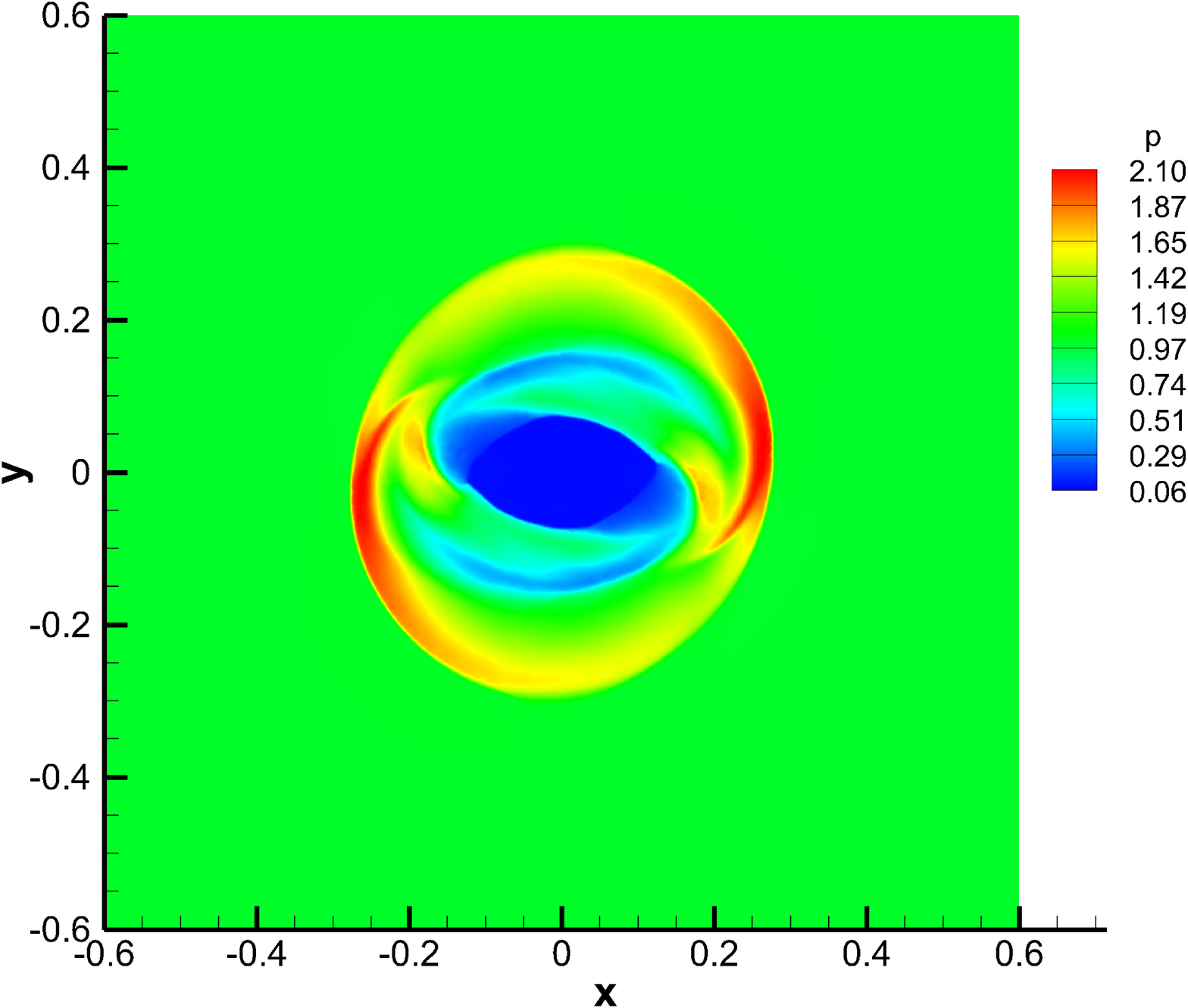}}
\end{center}
\caption{
Two-dimensional relativistic Rotor problem at time $t=0.1$, $t=0.2$ and $t=0.3$ (from top to bottom).
The left columns show the AMR grid while the right columns show the pressure field.
A fourth order ADER-WENO scheme with the HLL Riemann solver has been adopted. The initial grid has $60\times60$ cells, subsequently refined with $\ell_{\rm max}=2$ and $\mathfrak{r}=3$.
}
\label{fig.rmhdrotor}
\end{figure}

%-----------------------------------------------------------
%
In the MHD rotor problem proposed by \cite{Balsara99} strong torsional Alfven waves propagate from a spinning cylinder embedded in a light fluid.
The relativistic version of this test has been solved by~\citet{DelZanna2003a}
while resistivity effects were considered by \cite{Dumbser2009}.
The rotor has a radius $R_0=0.1$, it is placed at the center of the  
computational domain $\Omega=[-0.6;0.6]\times[-0.6;0.6]$ 
and it is rotating 
with an angular frequency $\omega_s = 8.5$.
The density is $\rho=10$ 
inside the rotor and $\rho=1$ in the ambient fluid at rest,
while the adiabatic index of the gas is $\gamma=4/3$.
The pressure $p=1$ is uniform everywhere, as well as the magnetic field, which, initially,
has only one non-vanishing component, namely $B_x=1$. The fact that $B_x$ is initially constant everywhere allows the transport of angular momentum 
from the rotor to the ambient through the torsional Alfven waves.  
 We have not applied any taper in the density of the rotor. Finally,
transmissive boundary conditions are applied at the
outer boundaries. We have solved this problem with the fourth order ADER-WENO scheme, starting with an initially uniform grid
with $70\times70$ elements. During the evolution, AMR is activated with 
relevant parameters given by $\mathfrak{r}=3$ and $\ell_{\rm max}=2$.
Fig.~\ref{fig.rmhdrotor} shows the AMR mesh (left panels) and the pressure field (right panels) at the three different  times $t=0.1$, $t=0.2$ and $t=0.3$ (from top to bottom).
Also in this test, but only in a limited number of troubled cells at the border between the rotor and the ambient medium, the accuracy of the solution has been intentionally 
reduced to first-order according to the MOOD-type approach of \citet{MOOD}.
%
%-----------------------------------------------------------
\subsection{RMHD Orszag--Tang}
\label{RMHD-OT}
\begin{figure}
\begin{center}
{\includegraphics[angle=0,width=6.7cm,height=6.5cm]{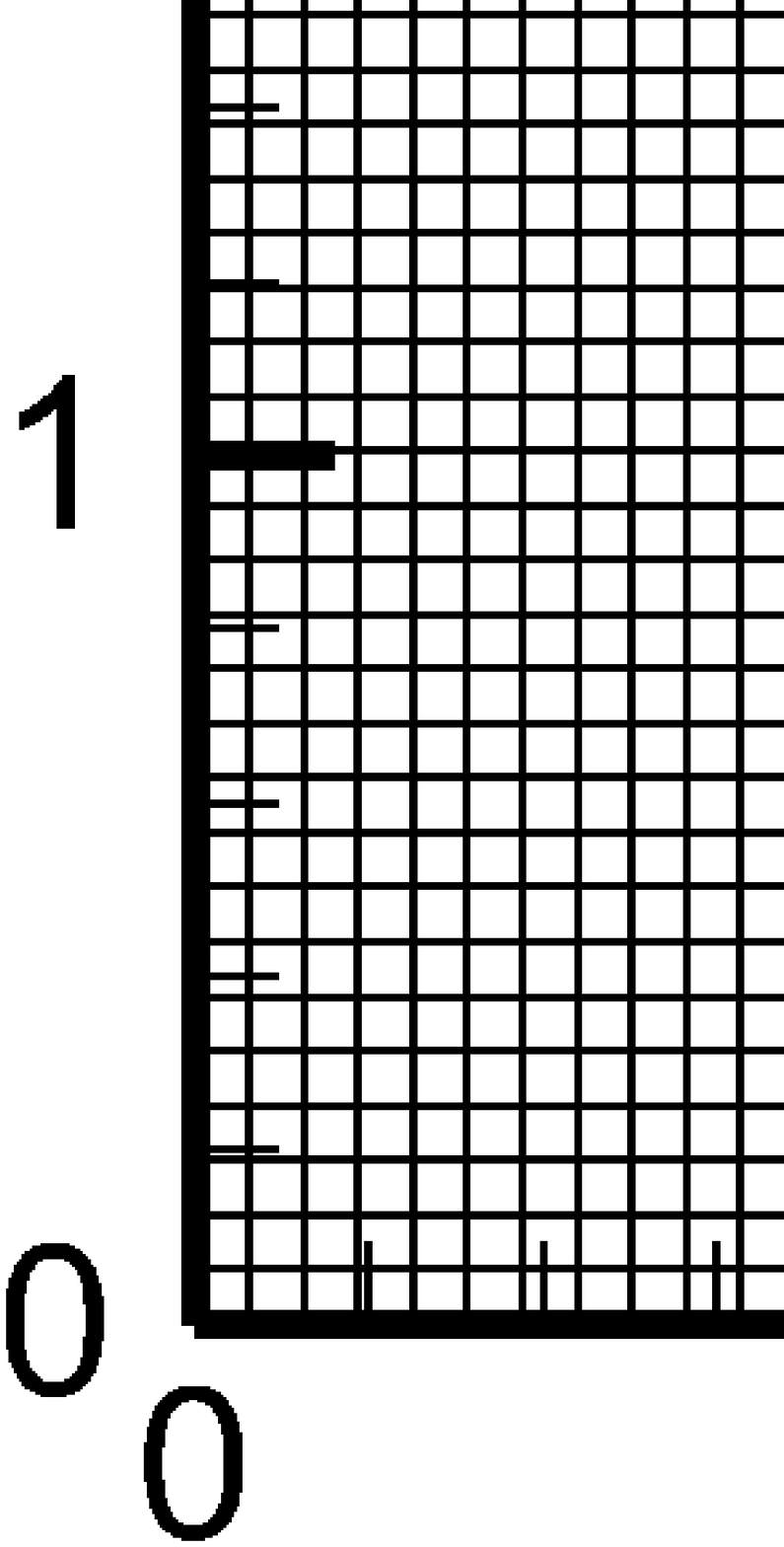}}
{\includegraphics[angle=0,width=6.7cm,height=6.5cm]{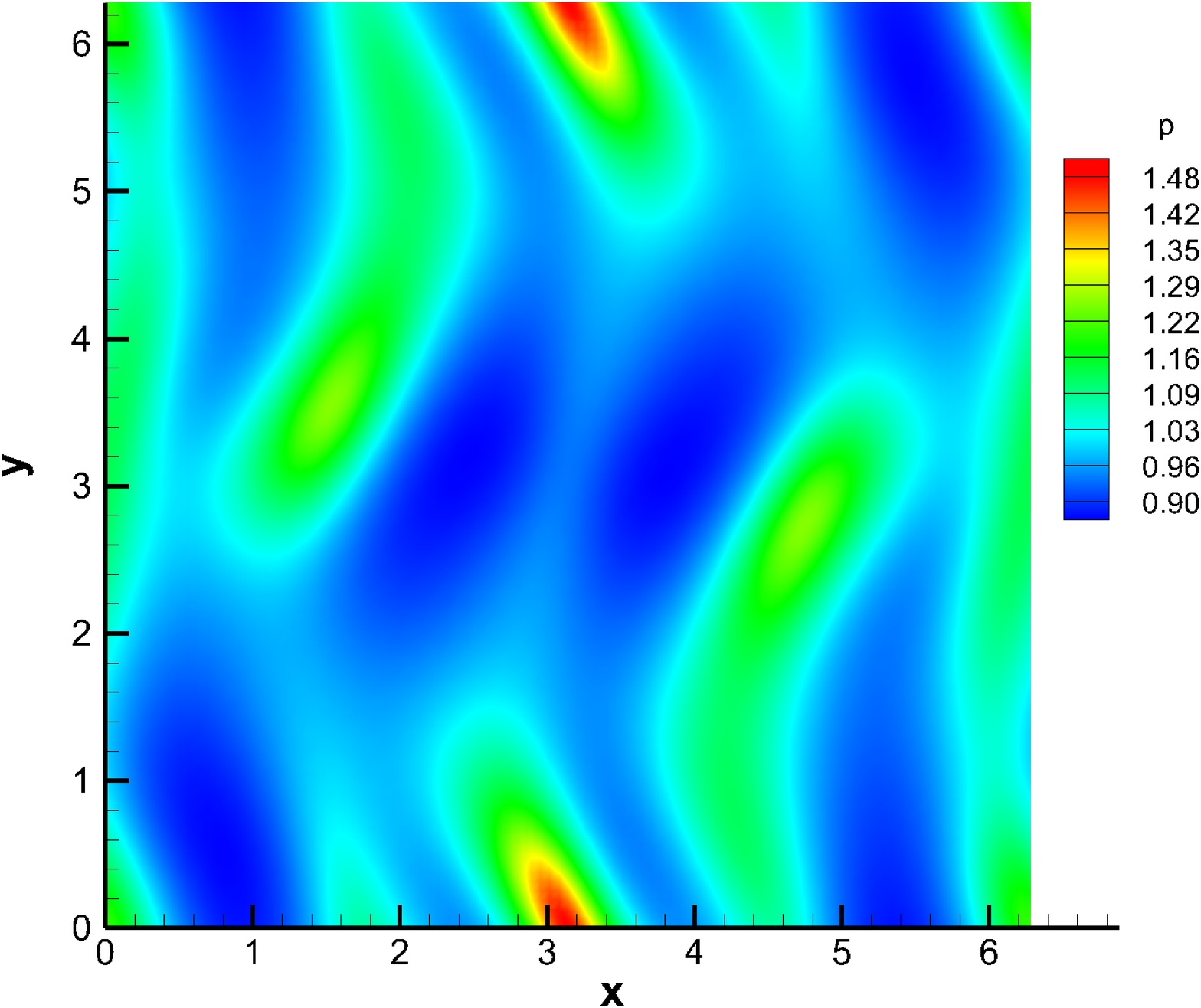}}
\end{center}
\hspace{1cm}
\begin{center}
{\includegraphics[angle=0,width=6.7cm,height=6.5cm]{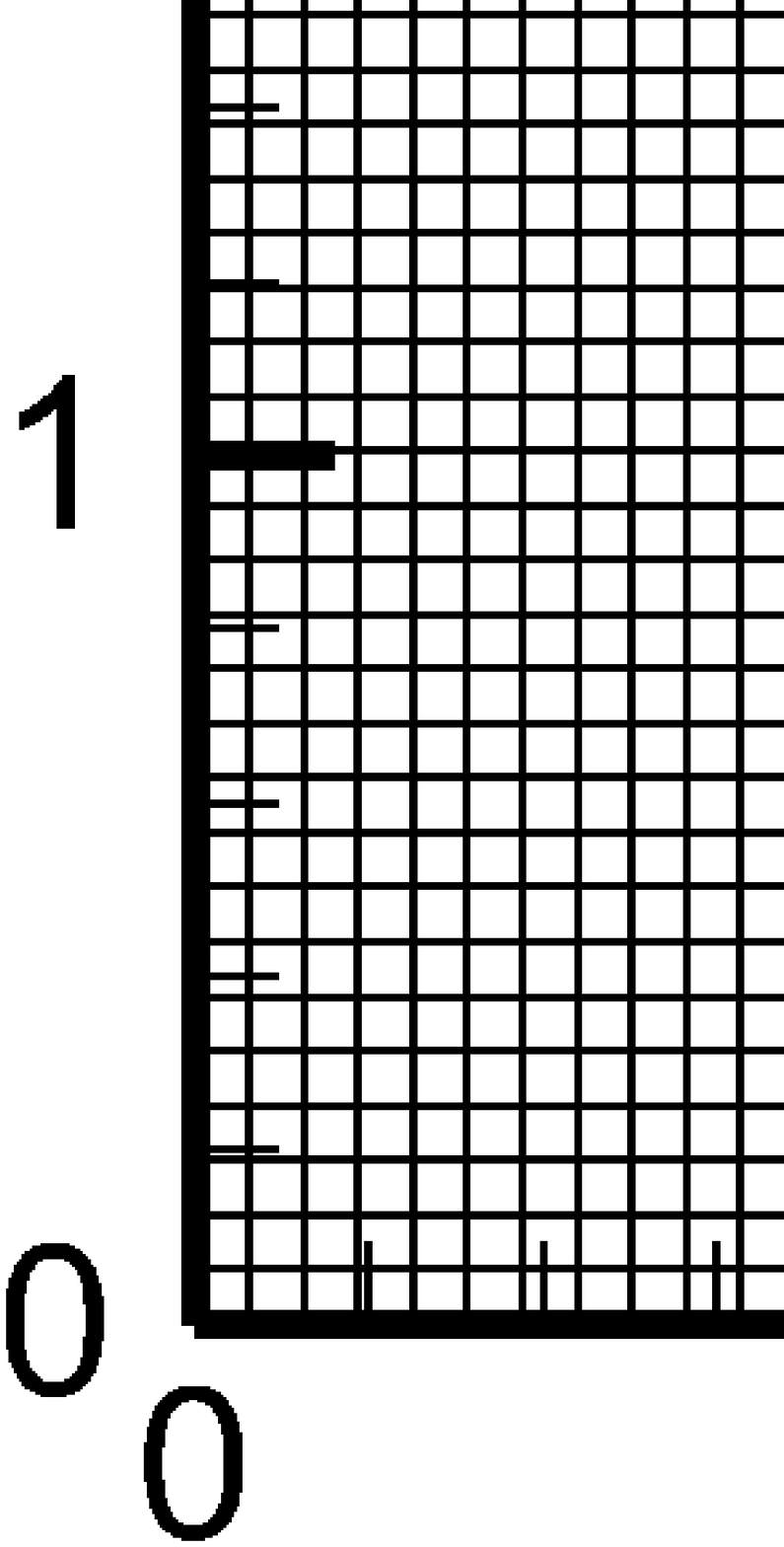}}
{\includegraphics[angle=0,width=6.7cm,height=6.5cm]{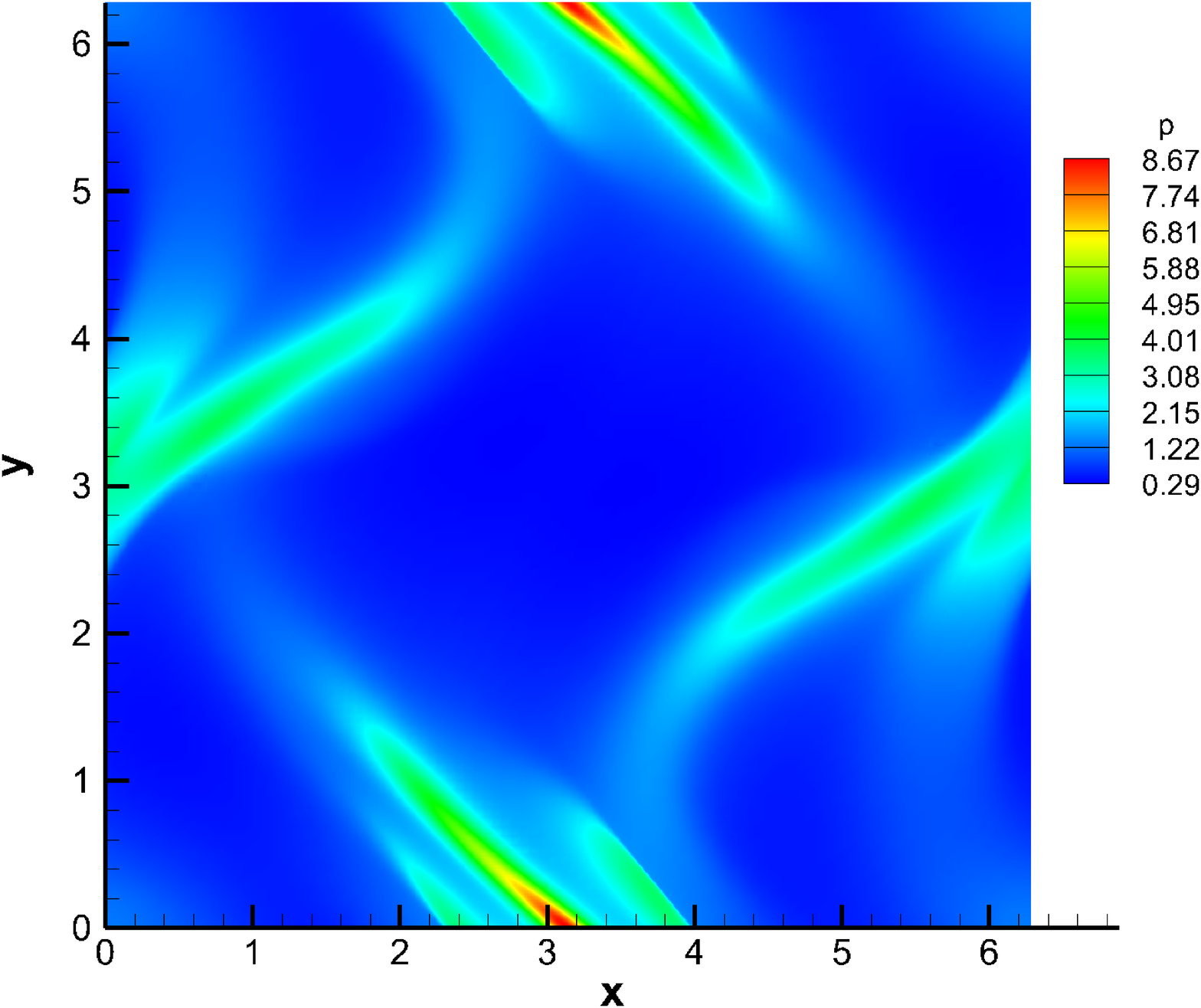}}
\end{center}
\begin{center}
{\includegraphics[angle=0,width=6.7cm,height=6.5cm]{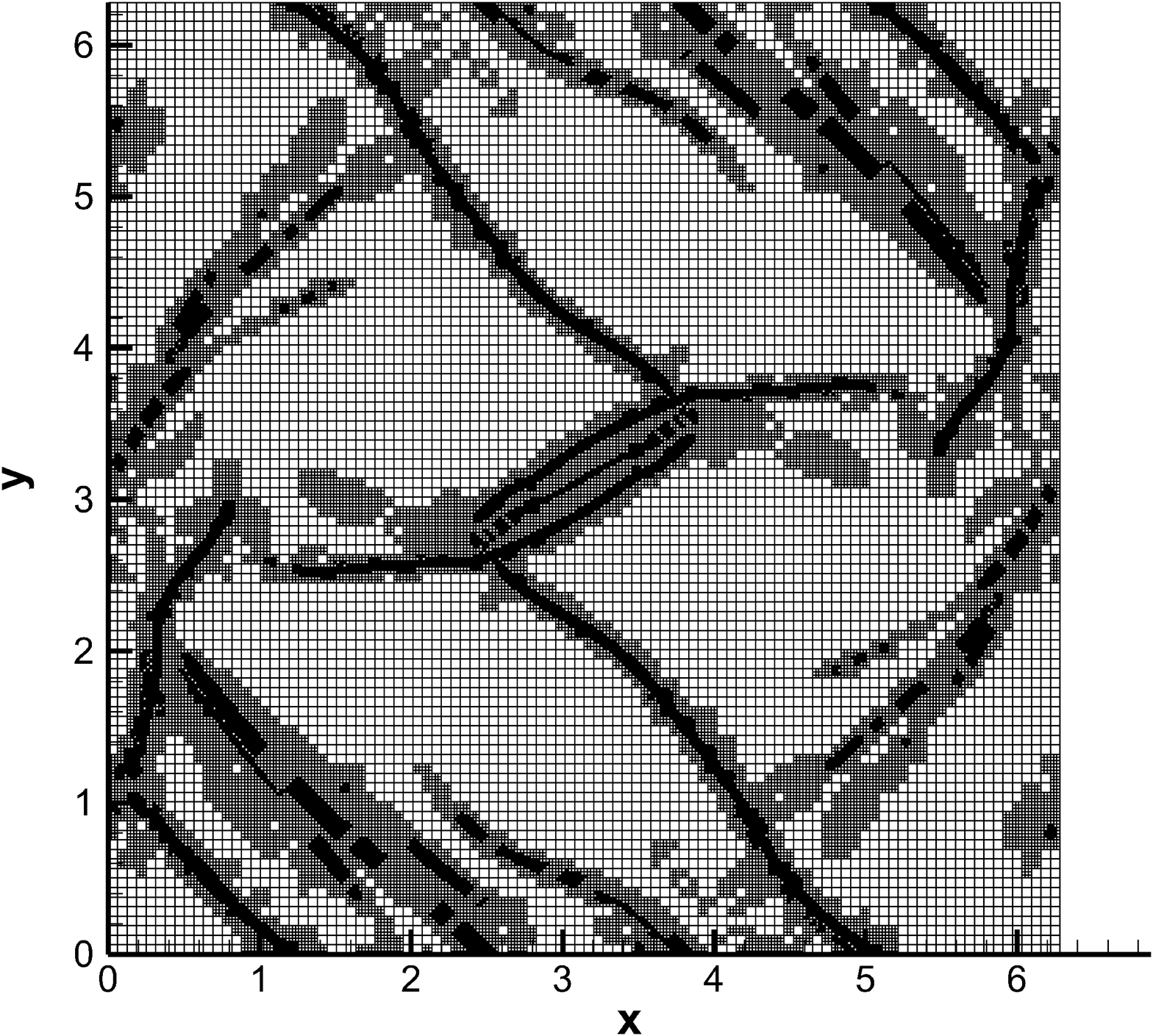}}
{\includegraphics[angle=0,width=6.7cm,height=6.5cm]{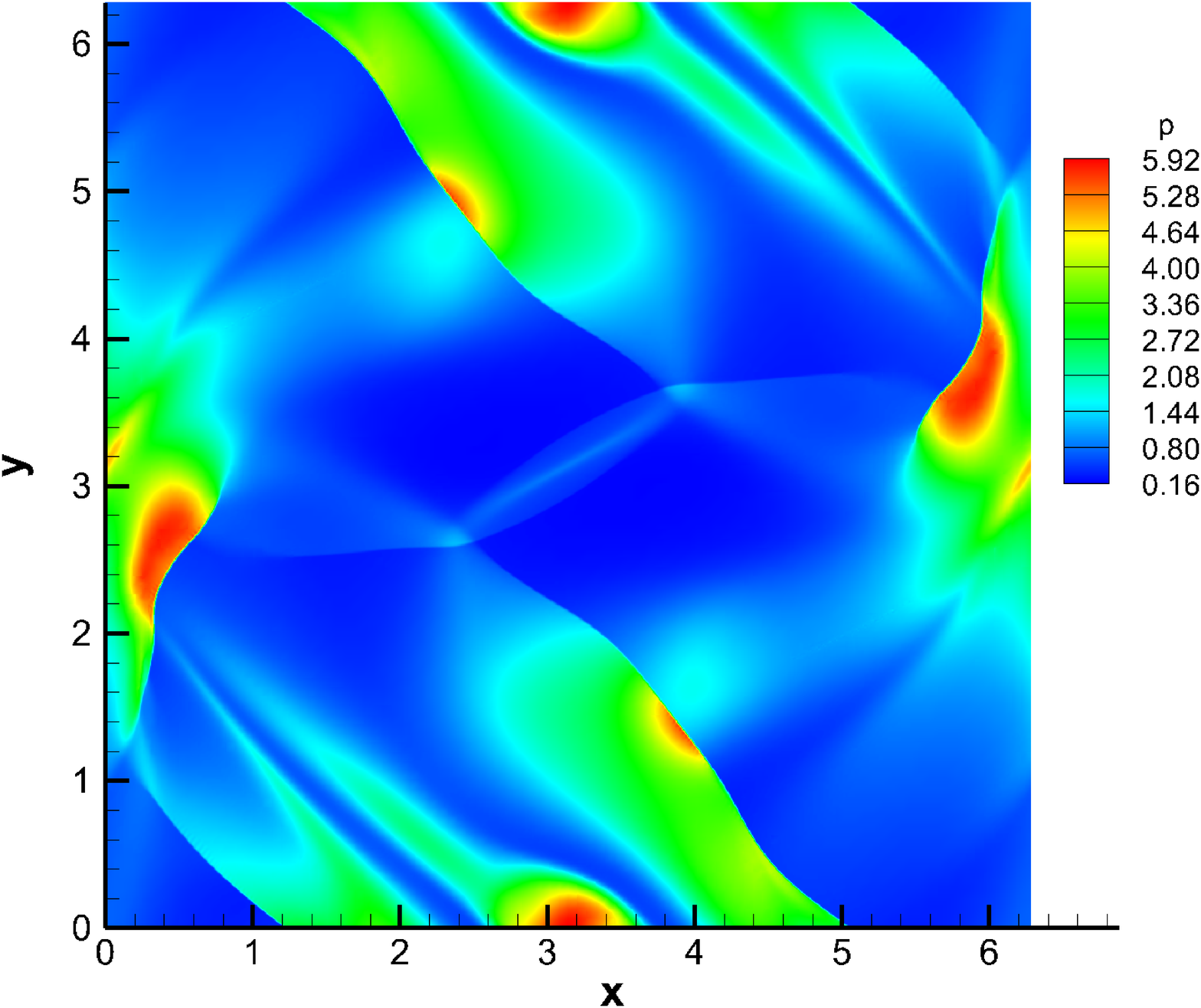}}
\end{center}
\caption{
Two-dimensional relativistic Orszag--Tang vortex problem at time $t=0.5$, $t=2.0$ and $t=4.0$ (from top to bottom).
The left columns show the AMR grid while the right columns show the pressure field.
A third order ADER-WENO scheme with the HLL Riemann solver has been adopted. The initial grid has $100\times100$ cells, subsequently refined with $\ell_{\rm max}=2$ and $\mathfrak{r}=3$.
}
\label{fig.otang}
\end{figure}

%-----------------------------------------------------------
As a final test for magnetized flows we have considered the 
relativistic version 
of the Orszag--Tang vortex problem, which was first introduced by \cite{OrszagTang} and later studied by \cite{DahlburgPicone} and \cite{JiangWu}.   
The initial conditions are given by  
\begin{equation}
  \left( \rho,v_x,v_y,v_z,p,B_x,B_y,B_z \right) = \left( 1, -\frac{v_0}{\sqrt 2}\sin(y), \frac{v_0}{\sqrt 2}\sin(x), 0, 1, - B_0 \sin(y), B_0 \sin(2x),0 \right)\,,
\end{equation}
where $v_0=0.75$, and $\gamma = 4/3$.
The problem is solved up to $t=4.0$ over the computational domain $\Omega = \left[0;2\pi\right]\times\left[0;2\pi\right]$. 
The initial grid at level zero  uses $100\times100$ cells, while during the evolution we have chosen 
$\mathfrak{r}=3$ and $\ell_{\rm max}=2$. The calculation has been performed at the third order of accuracy 
with the HLL Riemann solver.
The results are shown in the six panels of Fig. \ref{fig.otang} at times $t=0.5$, $t=2.0$ and $t=4.0$,
where the AMR grids (left column) and the thermal pressure (right column) can be seen.
They are in good agreement with the high conductivity limit of the same test by
\cite{Dumbser2009}.

\section{Conclusions}
\label{Conclusions}
Extending our recent work about classical gas dynamics~\citep{Dumbser2012b}, 
we have presented the first implementation of ADER methods in combination with Adaptive Mesh Refinement
for the solution of the special relativistic magnetohydrodynamics equations. The key idea of the ADER strategy, in 
the modern version proposed by \citet{DET2008}, is to build high order accurate one-step finite volume schemes 
which are based on a high order nonlinear WENO reconstruction operator and an element-local weak formulation of 
the governing partial differential equations in space-time. 
Under many respects, by treating space and time on an equal footing, the ADER scheme calls naturally for  an implementation to the relativistic regime.
Its property becomes also particularly advantageous in the presence of AMR, since, by avoiding the sub-steps typical 
of Runge--Kutta schemes, it substantially reduces the amount of necessary MPI communications and makes the implementation
of time-accurate local time stepping (LTS) very natural and simple. Our AMR implementation follows  
a 'cell-by-cell' refinement criterion through a tree-data structure, similarly to 
~\citet{Khokhlov1997}.

We have verified the ability of our new scheme
in coping with complex physical conditions, by considering
a wide set of relativistic hydrodynamic and magnetohydrodynamic tests, both for one and for multi-dimensional problems.
Our ADER-WENO AMR scheme represents the first step of a plan to incorporate more physical effects (viscous fluids, radiation hydrodynamics etc.) into a 
versatile tool suitable for studying high energy relativistic processes in astrophysics with high order of accuracy in both space and time. 

High order methods often become the only choice for problems characterized by a wide range of length-scales, and in particular when 
fluid instabilities are triggered. One of such cases is represented by 
the Richtmyer--Meshkov (RM) instability, which arises when a shock wave crosses the interface between two fluids having different densities. 
The RM instability is thought to play a significant role in several high energy physical systems, including inertial confinement fusion (ICF) 
and core-collapse supernovae explosions. Broadly speaking, it typically produces a pronounced mixing of the involved fluids. 
Relativistic effects on the RM instability have been considered only recently \citep{Inoue2012, Matsumoto2013, Mohseni2013}, showing, somewhat 
counter-intuitively, that the growth rate of the instability decreases as the Lorentz factor increases. In the near future we plan to shed light on 
these aspects by means of high order numerical simulations combined with space-time adaptive mesh refinement.

\section*{Acknowledgments}

We would like to thank Davide Radice and 
Francesco Fambri for useful discussions; Bruno Giacomazzo for providing the exact 
Riemann solver for the RMHD equations,
as well as an anonymous referee for various suggestions
that allowed us to improved the quality of this work.
The research conducted here has been financed in parts by the European Research Council under 
the European Union's Seventh Framework Programme (FP7/2007-2013) in the frame of the research 
project \textit{STiMulUs}, ERC Grant agreement no. 278267.

The numerical simulations presented in this paper were performed 
on the FERMI supercomputer at CINECA (Italy) within the IsB05-STiMulUs project
and on SuperMuc at  Leibniz-Rechenzentrum (Germany), within the PRACE 6-th CALL
(Project ID 2012071312).

\bibliographystyle{model1c-num-names}
\bibliography{references}

\begin{thebibliography}{87}
\expandafter\ifx\csname natexlab\endcsname\relax\def\natexlab#1{#1}\fi
\providecommand{\bibinfo}[2]{#2}
\ifx\xfnm\relax \def\xfnm[#1]{\unskip,\space#1}\fi
%Type = Article
\bibitem[{{Aloy} et~al.(1999){Aloy}, {Ib{\'a}{\~n}ez}, {Mart{\'{\i}}} and
  {M{\"u}ller}}]{Aloy1999c}
\bibinfo{author}{M.A. {Aloy}}, \bibinfo{author}{J.M. {Ib{\'a}{\~n}ez}},
  \bibinfo{author}{J.M. {Mart{\'{\i}}}}, \bibinfo{author}{E.~{M{\"u}ller}},
  \bibinfo{journal}{Astrohys. J. Suppl.} \bibinfo{volume}{122}
  (\bibinfo{year}{1999}) \bibinfo{pages}{151--166}.
%Type = Article
\bibitem[{Aloy and Rezzolla(2006)}]{Aloy:2006rd}
\bibinfo{author}{M.A. Aloy}, \bibinfo{author}{L.~Rezzolla},
  \bibinfo{journal}{Astrophys. J.} \bibinfo{volume}{640} (\bibinfo{year}{2006})
  \bibinfo{pages}{L115--L118}.
%Type = Article
\bibitem[{Anderson et~al.(2006)Anderson, Hirschmann, Liebling and
  Neilsen}]{Anderson2006a}
\bibinfo{author}{M.~Anderson}, \bibinfo{author}{E.~Hirschmann},
  \bibinfo{author}{S.L. Liebling}, \bibinfo{author}{D.~Neilsen},
  \bibinfo{journal}{Classical Quantum Gravity} \bibinfo{volume}{23}
  (\bibinfo{year}{2006}) \bibinfo{pages}{6503--6524}.
%Type = Book
\bibitem[{{Anile}(1990)}]{Anile_book}
\bibinfo{author}{A.M. {Anile}}, \bibinfo{title}{{Relativistic Fluids and
  Magneto-fluids}}, \bibinfo{publisher}{Cambridge University Press},
  \bibinfo{year}{1990}.
%Type = Article
\bibitem[{{Anile} et~al.(1983){Anile}, {Miller} and {Motta}}]{Anile1983}
\bibinfo{author}{A.M. {Anile}}, \bibinfo{author}{J.C. {Miller}},
  \bibinfo{author}{S.~{Motta}}, \bibinfo{journal}{Physics of Fluids}
  \bibinfo{volume}{26} (\bibinfo{year}{1983}) \bibinfo{pages}{1450--1460}.
%Type = Article
\bibitem[{{Ant{\'o}n} et~al.(2010){Ant{\'o}n}, {Miralles}, {Mart{\'{\i}}},
  {Ib{\'a}{\~n}ez}, {Aloy} and {Mimica}}]{Anton2010}
\bibinfo{author}{L.~{Ant{\'o}n}}, \bibinfo{author}{J.A. {Miralles}},
  \bibinfo{author}{J.M. {Mart{\'{\i}}}}, \bibinfo{author}{J.M.
  {Ib{\'a}{\~n}ez}}, \bibinfo{author}{M.A. {Aloy}},
  \bibinfo{author}{P.~{Mimica}}, \bibinfo{journal}{Astrophys. J. Suppl.}
  \bibinfo{volume}{188} (\bibinfo{year}{2010}) \bibinfo{pages}{1--31}.
%Type = Article
\bibitem[{Baeza et~al.(2012)Baeza, Mart\'inez-Gavara and Mulet}]{Mulet2}
\bibinfo{author}{A.~Baeza}, \bibinfo{author}{A.~Mart\'inez-Gavara},
  \bibinfo{author}{P.~Mulet}, \bibinfo{journal}{Applied Numerical Mathematics}
  \bibinfo{volume}{62} (\bibinfo{year}{2012}) \bibinfo{pages}{278--296}.
%Type = Article
\bibitem[{Baeza and Mulet(2006)}]{Mulet1}
\bibinfo{author}{A.~Baeza}, \bibinfo{author}{P.~Mulet},
  \bibinfo{journal}{International Journal for Numerical Methods in Fluids}
  \bibinfo{volume}{52} (\bibinfo{year}{2006}) \bibinfo{pages}{455--471}.
%Type = Phdthesis
\bibitem[{Baiotti(2004)}]{Baiotti2004thesis}
\bibinfo{author}{L.~Baiotti}, \bibinfo{title}{Numerical relativity simulations
  of non-vacuum spacetimes in three dimensions}, Ph.D. thesis, SISSA,
  International School for advanced studies, \bibinfo{year}{2004}.
%Type = Article
\bibitem[{{Balsara}(2001)}]{Balsara2001}
\bibinfo{author}{D.~{Balsara}}, \bibinfo{journal}{Astrophysical Journal Suppl.
  Series} \bibinfo{volume}{132} (\bibinfo{year}{2001})
  \bibinfo{pages}{83--101}.
%Type = Article
\bibitem[{{Balsara} and {Spicer}(1999)}]{Balsara99}
\bibinfo{author}{D.S. {Balsara}}, \bibinfo{author}{D.S. {Spicer}},
  \bibinfo{journal}{J. Comput. Phys.} \bibinfo{volume}{149}
  (\bibinfo{year}{1999}) \bibinfo{pages}{270--292}.
%Type = Article
\bibitem[{Beckwith and Stone(2011)}]{Beckwith2011}
\bibinfo{author}{K.~Beckwith}, \bibinfo{author}{J.M. Stone},
  \bibinfo{journal}{The Astrophysical Journal Supplement Series}
  \bibinfo{volume}{193} (\bibinfo{year}{2011}) \bibinfo{pages}{6}.
%Type = Article
\bibitem[{Berger(1986)}]{Berger86}
\bibinfo{author}{M.J. Berger}, \bibinfo{journal}{SIAM Journal of Scientific and
  Statistical Computing} \bibinfo{volume}{7} (\bibinfo{year}{1986})
  \bibinfo{pages}{904--916}.
%Type = Article
\bibitem[{Berger and Colella(1989)}]{Berger89}
\bibinfo{author}{M.J. Berger}, \bibinfo{author}{P.~Colella},
  \bibinfo{journal}{J. Comput. Phys.} \bibinfo{volume}{82}
  (\bibinfo{year}{1989}) \bibinfo{pages}{64--84}.
%Type = Article
\bibitem[{Berger and Oliger(1984)}]{Berger84}
\bibinfo{author}{M.J. Berger}, \bibinfo{author}{J.~Oliger},
  \bibinfo{journal}{J. Comput. Phys.} \bibinfo{volume}{53}
  (\bibinfo{year}{1984}) \bibinfo{pages}{484--512}.
%Type = Article
\bibitem[{{Brouillette}(2002)}]{Brouillette2002}
\bibinfo{author}{M.~{Brouillette}}, \bibinfo{journal}{Annual Review of Fluid
  Mechanics} \bibinfo{volume}{34} (\bibinfo{year}{2002})
  \bibinfo{pages}{445--468}.
%Type = Article
\bibitem[{{Bucciantini} and {Del Zanna}(2011)}]{Bucciantini2011}
\bibinfo{author}{N.~{Bucciantini}}, \bibinfo{author}{L.~{Del Zanna}},
  \bibinfo{journal}{Astron. Astrophys.} \bibinfo{volume}{528}
  (\bibinfo{year}{2011}) \bibinfo{pages}{A101}.
%Type = Incollection
\bibitem[{Choptuik(1989)}]{Choptuik89}
\bibinfo{author}{M.W. Choptuik}, in: \bibinfo{editor}{C.~Evans},
  \bibinfo{editor}{L.~Finn}, \bibinfo{editor}{D.~Hobill} (Eds.),
  \bibinfo{booktitle}{Frontiers in Numerical Relativity},
  \bibinfo{publisher}{Cambridge University Press}, \bibinfo{address}{Cambridge,
  England}, \bibinfo{year}{1989}, pp. \bibinfo{pages}{206--221}.
%Type = Article
\bibitem[{Clain et~al.(2011)Clain, Diot and Loub\`ere}]{MOOD}
\bibinfo{author}{S.~Clain}, \bibinfo{author}{S.~Diot},
  \bibinfo{author}{R.~Loub\`ere}, \bibinfo{journal}{Journal of Computational
  Physics} \bibinfo{volume}{230} (\bibinfo{year}{2011})
  \bibinfo{pages}{4028--4050}.
%Type = Article
\bibitem[{Dahlburg and Picone(1989)}]{DahlburgPicone}
\bibinfo{author}{R.B. Dahlburg}, \bibinfo{author}{J.M. Picone},
  \bibinfo{journal}{Phys. Fluids B} \bibinfo{volume}{1} (\bibinfo{year}{1989})
  \bibinfo{pages}{2153--2171}.
%Type = Article
\bibitem[{{De Colle} et~al.(2012){De Colle}, {Granot}, {L{\'o}pez-C{\'a}mara}
  and {Ramirez-Ruiz}}]{DeColle2011}
\bibinfo{author}{F.~{De Colle}}, \bibinfo{author}{J.~{Granot}},
  \bibinfo{author}{D.~{L{\'o}pez-C{\'a}mara}},
  \bibinfo{author}{E.~{Ramirez-Ruiz}}, \bibinfo{journal}{Astrophys. J.}
  \bibinfo{volume}{746} (\bibinfo{year}{2012}) \bibinfo{pages}{122}.
%Type = Article
\bibitem[{{Dedner} et~al.(2002){Dedner}, {Kemm}, {Kr\"oner}, {Munz},
  {Schnitzer} and {Wesenberg}}]{Dedner:2002}
\bibinfo{author}{A.~{Dedner}}, \bibinfo{author}{F.~{Kemm}},
  \bibinfo{author}{D.~{Kr\"oner}}, \bibinfo{author}{C.D. {Munz}},
  \bibinfo{author}{T.~{Schnitzer}}, \bibinfo{author}{M.~{Wesenberg}},
  \bibinfo{journal}{Journal of Computational Physics} \bibinfo{volume}{175}
  (\bibinfo{year}{2002}) \bibinfo{pages}{645--673}.
%Type = Article
\bibitem[{{Del Zanna} and {Bucciantini}(2002)}]{DelZanna2002}
\bibinfo{author}{L.~{Del Zanna}}, \bibinfo{author}{N.~{Bucciantini}},
  \bibinfo{journal}{Astron. Astrophys.} \bibinfo{volume}{390}
  (\bibinfo{year}{2002}) \bibinfo{pages}{1177--1186}.
%Type = Article
\bibitem[{{Del Zanna} et~al.(2003){Del Zanna}, {Bucciantini} and
  {Londrillo}}]{DelZanna2003a}
\bibinfo{author}{L.~{Del Zanna}}, \bibinfo{author}{N.~{Bucciantini}},
  \bibinfo{author}{P.~{Londrillo}}, \bibinfo{journal}{Astron. Astrophys.}
  \bibinfo{volume}{400} (\bibinfo{year}{2003}) \bibinfo{pages}{397--413}.
%Type = Article
\bibitem[{{Del Zanna} et~al.(2007){Del Zanna}, {Zanotti}, {Bucciantini} and
  {Londrillo}}]{DelZanna2007}
\bibinfo{author}{L.~{Del Zanna}}, \bibinfo{author}{O.~{Zanotti}},
  \bibinfo{author}{N.~{Bucciantini}}, \bibinfo{author}{P.~{Londrillo}},
  \bibinfo{journal}{Astron. Astrophys.} \bibinfo{volume}{473}
  (\bibinfo{year}{2007}) \bibinfo{pages}{11--30}.
%Type = Article
\bibitem[{{Dionysopoulou} et~al.(2013){Dionysopoulou}, {Alic}, {Palenzuela},
  {Rezzolla} and {Giacomazzo}}]{Dionysopoulou:2012pp}
\bibinfo{author}{K.~{Dionysopoulou}}, \bibinfo{author}{D.~{Alic}},
  \bibinfo{author}{C.~{Palenzuela}}, \bibinfo{author}{L.~{Rezzolla}},
  \bibinfo{author}{B.~{Giacomazzo}}, \bibinfo{journal}{Phys. Rev. D}
  \bibinfo{volume}{88} (\bibinfo{year}{2013}) \bibinfo{pages}{044020}.
%Type = Article
\bibitem[{{Dolezal}(1995)}]{Dolezal1995}
\bibinfo{author}{A.~{Dolezal}}, \bibinfo{journal}{Journal of Computational
  Physics} \bibinfo{volume}{120} (\bibinfo{year}{1995})
  \bibinfo{pages}{266--277}.
%Type = Article
\bibitem[{{D{\"o}nmez}(2004)}]{Donmez2004}
\bibinfo{author}{O.~{D{\"o}nmez}}, \bibinfo{journal}{Astrophys. Spac. Sci.}
  \bibinfo{volume}{293} (\bibinfo{year}{2004}) \bibinfo{pages}{323--354}.
%Type = Article
\bibitem[{{Dumbser} et~al.(2008{\natexlab{a}}){Dumbser}, {Balsara}, {Toro} and
  {Munz}}]{DBTM2008}
\bibinfo{author}{M.~{Dumbser}}, \bibinfo{author}{D.S. {Balsara}},
  \bibinfo{author}{E.F. {Toro}}, \bibinfo{author}{C.D. {Munz}},
  \bibinfo{journal}{Journal of Computational Physics} \bibinfo{volume}{227}
  (\bibinfo{year}{2008}{\natexlab{a}}) \bibinfo{pages}{8209--8253}.
%Type = Article
\bibitem[{{Dumbser} et~al.(2008{\natexlab{b}}){Dumbser}, {Enaux} and
  {Toro}}]{DET2008}
\bibinfo{author}{M.~{Dumbser}}, \bibinfo{author}{C.~{Enaux}},
  \bibinfo{author}{E.F. {Toro}}, \bibinfo{journal}{Journal of Computational
  Physics} \bibinfo{volume}{227} (\bibinfo{year}{2008}{\natexlab{b}})
  \bibinfo{pages}{3971--4001}.
%Type = Article
\bibitem[{{Dumbser} et~al.(2007){Dumbser}, {Kaeser}, {Titarev} and
  {Toro}}]{Dumbser2007}
\bibinfo{author}{M.~{Dumbser}}, \bibinfo{author}{M.~{Kaeser}},
  \bibinfo{author}{V.A. {Titarev}}, \bibinfo{author}{E.F. {Toro}},
  \bibinfo{journal}{Journal of Computational Physics} \bibinfo{volume}{226}
  (\bibinfo{year}{2007}) \bibinfo{pages}{204--243}.
%Type = Article
\bibitem[{Dumbser et~al.(2007)Dumbser, Käser and Toro}]{Dumbser2007c}
\bibinfo{author}{M.~Dumbser}, \bibinfo{author}{M.~Käser},
  \bibinfo{author}{E.F. Toro}, \bibinfo{journal}{Geophysical Journal
  International} \bibinfo{volume}{171} (\bibinfo{year}{2007})
  \bibinfo{pages}{695--717}.
%Type = Article
\bibitem[{Dumbser and Toro(2011)}]{OsherUniversal}
\bibinfo{author}{M.~Dumbser}, \bibinfo{author}{E.~Toro},
  \bibinfo{journal}{Communications in Computational Physics}
  \bibinfo{volume}{10} (\bibinfo{year}{2011}) \bibinfo{pages}{635--671}.
%Type = Article
\bibitem[{{Dumbser} and {Zanotti}(2009)}]{Dumbser2009}
\bibinfo{author}{M.~{Dumbser}}, \bibinfo{author}{O.~{Zanotti}},
  \bibinfo{journal}{Journal of Computational Physics} \bibinfo{volume}{228}
  (\bibinfo{year}{2009}) \bibinfo{pages}{6991--7006}.
%Type = Article
\bibitem[{{Dumbser} et~al.(2013){Dumbser}, {Zanotti}, {Hidalgo} and
  {Balsara}}]{Dumbser2012b}
\bibinfo{author}{M.~{Dumbser}}, \bibinfo{author}{O.~{Zanotti}},
  \bibinfo{author}{A.~{Hidalgo}}, \bibinfo{author}{D.S. {Balsara}},
  \bibinfo{journal}{Journal of Computational Physics} \bibinfo{volume}{248}
  (\bibinfo{year}{2013}) \bibinfo{pages}{257--286}.
%Type = Article
\bibitem[{{East} et~al.(2012){East}, {Pretorius} and {Stephens}}]{East2012b}
\bibinfo{author}{W.E. {East}}, \bibinfo{author}{F.~{Pretorius}},
  \bibinfo{author}{B.C. {Stephens}}, \bibinfo{journal}{Phys. Rev. D}
  \bibinfo{volume}{85} (\bibinfo{year}{2012}) \bibinfo{pages}{124010}.
%Type = Article
\bibitem[{Etienne et~al.(2010)Etienne, Liu and Shapiro}]{Etienne:2010ui}
\bibinfo{author}{Z.B. Etienne}, \bibinfo{author}{Y.T. Liu},
  \bibinfo{author}{S.L. Shapiro}, \bibinfo{journal}{Phys. Rev. D}
  \bibinfo{volume}{82} (\bibinfo{year}{2010}) \bibinfo{pages}{084031}.
%Type = Article
\bibitem[{Evans et~al.(2005)Evans, Iyer, Schnetter, Suen, Tao, Wolfmeyer and
  Zhang}]{Evans2005a}
\bibinfo{author}{E.~Evans}, \bibinfo{author}{S.~Iyer},
  \bibinfo{author}{E.~Schnetter}, \bibinfo{author}{W.M. Suen},
  \bibinfo{author}{J.~Tao}, \bibinfo{author}{R.~Wolfmeyer},
  \bibinfo{author}{H.M. Zhang}, \bibinfo{journal}{Phys. Rev. D}
  \bibinfo{volume}{71} (\bibinfo{year}{2005}) \bibinfo{pages}{081301(R)}.
%Type = Article
\bibitem[{Font(2008)}]{Font08}
\bibinfo{author}{J.A. Font}, \bibinfo{journal}{Living Rev. Relativ.}
  \bibinfo{volume}{6} (\bibinfo{year}{2008}) \bibinfo{pages}{4;
  http://www.livingreviews.org/lrr--2008--7}.
%Type = Article
\bibitem[{Gassner et~al.(2008)Gassner, L\"orcher and Munz}]{stedg2}
\bibinfo{author}{G.~Gassner}, \bibinfo{author}{F.~L\"orcher},
  \bibinfo{author}{C.D. Munz}, \bibinfo{journal}{Journal of Scientific
  Computing} \bibinfo{volume}{34} (\bibinfo{year}{2008})
  \bibinfo{pages}{260--286}.
%Type = Article
\bibitem[{Giacomazzo and Rezzolla(2006)}]{Giacomazzo:2005jy}
\bibinfo{author}{B.~Giacomazzo}, \bibinfo{author}{L.~Rezzolla},
  \bibinfo{journal}{Journal of Fluid Mechanics} \bibinfo{volume}{562}
  (\bibinfo{year}{2006}) \bibinfo{pages}{223--259}.
%Type = Article
\bibitem[{{Harten} et~al.(1987){Harten}, {Engquist}, {Osher} and
  {Chakravarthy}}]{eno}
\bibinfo{author}{A.~{Harten}}, \bibinfo{author}{B.~{Engquist}},
  \bibinfo{author}{S.~{Osher}}, \bibinfo{author}{S.R. {Chakravarthy}},
  \bibinfo{journal}{Journal of Computational Physics} \bibinfo{volume}{71}
  (\bibinfo{year}{1987}) \bibinfo{pages}{231--303}.
%Type = Article
\bibitem[{Harten et~al.(1983)Harten, Lax and van Leer}]{Harten83}
\bibinfo{author}{A.~Harten}, \bibinfo{author}{P.D. Lax},
  \bibinfo{author}{B.~van Leer}, \bibinfo{journal}{SIAM Rev.}
  \bibinfo{volume}{25} (\bibinfo{year}{1983}) \bibinfo{pages}{35}.
%Type = Article
\bibitem[{{Inoue}(2012)}]{Inoue2012}
\bibinfo{author}{T.~{Inoue}}, \bibinfo{journal}{Astrop. J.}
  \bibinfo{volume}{760} (\bibinfo{year}{2012}) \bibinfo{pages}{43}.
%Type = Article
\bibitem[{Jiang and Wu(1999)}]{JiangWu}
\bibinfo{author}{G.~Jiang}, \bibinfo{author}{C.~Wu}, \bibinfo{journal}{Journal
  of Computational Physics} \bibinfo{volume}{150} (\bibinfo{year}{1999})
  \bibinfo{pages}{561--594}.
%Type = Article
\bibitem[{Jiang and Shu(1996)}]{Jiang1996}
\bibinfo{author}{G.S. Jiang}, \bibinfo{author}{C.W. Shu}, \bibinfo{journal}{J.
  Comput. Phys} \bibinfo{volume}{126} (\bibinfo{year}{1996})
  \bibinfo{pages}{202--228}.
%Type = Article
\bibitem[{Khokhlov et~al.(1997)Khokhlov, Oran and Wheeler}]{Khokhlov1997}
\bibinfo{author}{A.M. Khokhlov}, \bibinfo{author}{E.S. Oran},
  \bibinfo{author}{J.C. Wheeler}, \bibinfo{journal}{Astrophys. J.}
  \bibinfo{volume}{478} (\bibinfo{year}{1997}) \bibinfo{pages}{678}.
%Type = Article
\bibitem[{{Komissarov}(1997)}]{Komissarov1997}
\bibinfo{author}{S.S. {Komissarov}}, \bibinfo{journal}{Physics Letters A}
  \bibinfo{volume}{232} (\bibinfo{year}{1997}) \bibinfo{pages}{435--442}.
%Type = Article
\bibitem[{{Komissarov}(1999)}]{Komissarov1999}
\bibinfo{author}{S.S. {Komissarov}}, \bibinfo{journal}{Mon. Not. R. Astron.
  Soc.} \bibinfo{volume}{303} (\bibinfo{year}{1999}) \bibinfo{pages}{343--366}.
%Type = Article
\bibitem[{{Lehner} et~al.(2012){Lehner}, {Palenzuela}, {Liebling}, {Thompson}
  and {Hanna}}]{Lehner2011}
\bibinfo{author}{L.~{Lehner}}, \bibinfo{author}{C.~{Palenzuela}},
  \bibinfo{author}{S.L. {Liebling}}, \bibinfo{author}{C.~{Thompson}},
  \bibinfo{author}{C.~{Hanna}}, \bibinfo{journal}{Phys. Rev. D}
  \bibinfo{volume}{86} (\bibinfo{year}{2012}) \bibinfo{pages}{104035}.
%Type = Article
\bibitem[{{Liebling} et~al.(2010){Liebling}, {Lehner}, {Neilsen} and
  {Palenzuela}}]{liebling_2010_emr}
\bibinfo{author}{S.L. {Liebling}}, \bibinfo{author}{L.~{Lehner}},
  \bibinfo{author}{D.~{Neilsen}}, \bibinfo{author}{C.~{Palenzuela}},
  \bibinfo{journal}{Phys. Rev. D} \bibinfo{volume}{81} (\bibinfo{year}{2010})
  \bibinfo{pages}{124023}.
%Type = Article
\bibitem[{{Liu} et~al.(1994){Liu}, {Osher} and {Chan}}]{Liu1994}
\bibinfo{author}{X.D. {Liu}}, \bibinfo{author}{S.~{Osher}},
  \bibinfo{author}{T.~{Chan}}, \bibinfo{journal}{Journal of Computational
  Physics} \bibinfo{volume}{115} (\bibinfo{year}{1994})
  \bibinfo{pages}{200--212}.
%Type = Article
\bibitem[{L\"ohner(1987)}]{Lohner1987}
\bibinfo{author}{R.~L\"ohner}, \bibinfo{journal}{Computer Methods in Applied
  Mechanics and Engineering} \bibinfo{volume}{61} (\bibinfo{year}{1987})
  \bibinfo{pages}{323--338}.
%Type = Article
\bibitem[{L\"orcher et~al.(2007)L\"orcher, Gassner and Munz}]{stedg1}
\bibinfo{author}{F.~L\"orcher}, \bibinfo{author}{G.~Gassner},
  \bibinfo{author}{C.D. Munz}, \bibinfo{journal}{Journal of Scientific
  Computing} \bibinfo{volume}{32} (\bibinfo{year}{2007})
  \bibinfo{pages}{175--199}.
%Type = Article
\bibitem[{Lucas-Serrano et~al.(2004)Lucas-Serrano, Font, Ibanez and
  Marti}]{Lucas-Serrano:2004aq}
\bibinfo{author}{A.~Lucas-Serrano}, \bibinfo{author}{J.A. Font},
  \bibinfo{author}{J.M. Ibanez}, \bibinfo{author}{J.M. Marti},
  \bibinfo{journal}{Astron. Astrophys.} \bibinfo{volume}{428}
  (\bibinfo{year}{2004}) \bibinfo{pages}{703--715}.
%Type = Article
\bibitem[{MacNeice et~al.(2000)MacNeice, Olson, Mobarry, de~Fainchtein and
  Packer}]{MacNeice00}
\bibinfo{author}{P.~MacNeice}, \bibinfo{author}{K.M. Olson},
  \bibinfo{author}{C.~Mobarry}, \bibinfo{author}{R.~de~Fainchtein},
  \bibinfo{author}{C.~Packer}, \bibinfo{journal}{Computer Physics
  Communications} \bibinfo{volume}{126} (\bibinfo{year}{2000})
  \bibinfo{pages}{330--354}.
%Type = Article
\bibitem[{Mart{\'\i} et~al.(1991)Mart{\'\i}, Ib{\'a}{\~n}ez and
  Miralles}]{Marti91}
\bibinfo{author}{J.M. Mart{\'\i}}, \bibinfo{author}{J.M. Ib{\'a}{\~n}ez},
  \bibinfo{author}{J.A. Miralles}, \bibinfo{journal}{Phys. Rev. D}
  \bibinfo{volume}{43} (\bibinfo{year}{1991}) \bibinfo{pages}{3794}.
%Type = Article
\bibitem[{Mart{\'\i} and M{\"{u}}ller(1994)}]{Marti94}
\bibinfo{author}{J.M. Mart{\'\i}}, \bibinfo{author}{E.~M{\"{u}}ller},
  \bibinfo{journal}{J. Fluid Mech.} \bibinfo{volume}{258}
  (\bibinfo{year}{1994}) \bibinfo{pages}{317--333}.
%Type = Article
\bibitem[{Mart{\'\i} and M{\"u}ller(2003)}]{Marti03}
\bibinfo{author}{J.M. Mart{\'\i}}, \bibinfo{author}{E.~M{\"u}ller},
  \bibinfo{journal}{Living Rev. Relativ.} \bibinfo{volume}{6}
  (\bibinfo{year}{2003}) \bibinfo{pages}{7;
  http://www.livingreviews.org/lrr--2003--7}.
%Type = Article
\bibitem[{{Matsumoto} and {Masada}(2013)}]{Matsumoto2013}
\bibinfo{author}{J.~{Matsumoto}}, \bibinfo{author}{Y.~{Masada}},
  \bibinfo{journal}{Astrop. J. Lett.} \bibinfo{volume}{772}
  (\bibinfo{year}{2013}) \bibinfo{pages}{L1}.
%Type = Article
\bibitem[{Meshkov(1968)}]{Meshkov1969}
\bibinfo{author}{E.~Meshkov}, \bibinfo{journal}{Fluid Dynamics}
  \bibinfo{volume}{43} (\bibinfo{year}{1968}) \bibinfo{pages}{101--104}.
%Type = Article
\bibitem[{{Mignone} and {Bodo}(2005)}]{Mignone2005}
\bibinfo{author}{A.~{Mignone}}, \bibinfo{author}{G.~{Bodo}},
  \bibinfo{journal}{Mon. Not. R. Astron. Soc.} \bibinfo{volume}{364}
  (\bibinfo{year}{2005}) \bibinfo{pages}{126--136}.
%Type = Article
\bibitem[{Mignone et~al.(2009)Mignone, Ugliano and Bodo}]{Mignone2009}
\bibinfo{author}{A.~Mignone}, \bibinfo{author}{M.~Ugliano},
  \bibinfo{author}{G.~Bodo}, \bibinfo{journal}{Monthly Notices of the Royal
  Astronomical Society} \bibinfo{volume}{393} (\bibinfo{year}{2009})
  \bibinfo{pages}{1141--1156}.
%Type = Article
\bibitem[{{Mignone} et~al.(2012){Mignone}, {Zanni}, {Tzeferacos}, {van
  Straalen}, {Colella} and {Bodo}}]{Mignone2012}
\bibinfo{author}{A.~{Mignone}}, \bibinfo{author}{C.~{Zanni}},
  \bibinfo{author}{P.~{Tzeferacos}}, \bibinfo{author}{B.~{van Straalen}},
  \bibinfo{author}{P.~{Colella}}, \bibinfo{author}{G.~{Bodo}},
  \bibinfo{journal}{Astrophys. J. Suppl. Ser.} \bibinfo{volume}{198}
  (\bibinfo{year}{2012}) \bibinfo{pages}{7}.
%Type = Article
\bibitem[{{Mohseni} et~al.(2013){Mohseni}, {Mendoza}, {Succi} and
  {Herrmann}}]{Mohseni2013}
\bibinfo{author}{F.~{Mohseni}}, \bibinfo{author}{M.~{Mendoza}},
  \bibinfo{author}{S.~{Succi}}, \bibinfo{author}{H.J. {Herrmann}},
  \bibinfo{journal}{ArXiv e-prints}  (\bibinfo{year}{2013}).
%Type = Article
\bibitem[{Orszag and Tang(1979)}]{OrszagTang}
\bibinfo{author}{S.A. Orszag}, \bibinfo{author}{C.M. Tang},
  \bibinfo{journal}{Journal of Fluid Mechanics} \bibinfo{volume}{90}
  (\bibinfo{year}{1979}) \bibinfo{pages}{129}.
%Type = Article
\bibitem[{{Palenzuela} et~al.(2009){Palenzuela}, {Lehner}, {Reula} and
  {Rezzolla}}]{Palenzuela:2008sf}
\bibinfo{author}{C.~{Palenzuela}}, \bibinfo{author}{L.~{Lehner}},
  \bibinfo{author}{O.~{Reula}}, \bibinfo{author}{L.~{Rezzolla}},
  \bibinfo{journal}{Mon. Not. R. Astron. Soc.} \bibinfo{volume}{394}
  (\bibinfo{year}{2009}) \bibinfo{pages}{1727--1740}.
%Type = Article
\bibitem[{{Radice} and {Rezzolla}(2011)}]{Radice2011}
\bibinfo{author}{D.~{Radice}}, \bibinfo{author}{L.~{Rezzolla}},
  \bibinfo{journal}{Phys. Rev. D} \bibinfo{volume}{84} (\bibinfo{year}{2011})
  \bibinfo{pages}{024010}.
%Type = Article
\bibitem[{{Radice} and {Rezzolla}(2012)}]{Radice2012a}
\bibinfo{author}{D.~{Radice}}, \bibinfo{author}{L.~{Rezzolla}},
  \bibinfo{journal}{Astron. Astrophys.} \bibinfo{volume}{547}
  (\bibinfo{year}{2012}) \bibinfo{pages}{A26}.
%Type = Article
\bibitem[{Rezzolla and Zanotti(2001)}]{Rezzolla01}
\bibinfo{author}{L.~Rezzolla}, \bibinfo{author}{O.~Zanotti},
  \bibinfo{journal}{Journ. of Fluid Mech.} \bibinfo{volume}{449}
  (\bibinfo{year}{2001}) \bibinfo{pages}{395}.
%Type = Article
\bibitem[{{Rezzolla} and {Zanotti}(2002)}]{Rezzolla02}
\bibinfo{author}{L.~{Rezzolla}}, \bibinfo{author}{O.~{Zanotti}},
  \bibinfo{journal}{Phys. Rev. Lett.} \bibinfo{volume}{89}
  (\bibinfo{year}{2002}) \bibinfo{pages}{114501}.
%Type = Book
\bibitem[{{Rezzolla} and {Zanotti}(2013)}]{Rezzolla_book:2013}
\bibinfo{author}{L.~{Rezzolla}}, \bibinfo{author}{O.~{Zanotti}},
  \bibinfo{title}{{Relativistic Hydrodynamics}}, \bibinfo{publisher}{Oxford
  University Press, Oxford UK}, \bibinfo{year}{2013}.
%Type = Article
\bibitem[{Rezzolla et~al.(2003)Rezzolla, Zanotti and Pons}]{Rezzolla03}
\bibinfo{author}{L.~Rezzolla}, \bibinfo{author}{O.~Zanotti},
  \bibinfo{author}{J.A. Pons}, \bibinfo{journal}{Journ. of Fluid Mech.}
  \bibinfo{volume}{479} (\bibinfo{year}{2003}) \bibinfo{pages}{199}.
%Type = Article
\bibitem[{Richtmyer(1960)}]{Richtmyer1960}
\bibinfo{author}{R.D. Richtmyer}, \bibinfo{journal}{Communications on Pure and
  Applied Mathematics} \bibinfo{volume}{13} (\bibinfo{year}{1960})
  \bibinfo{pages}{297--319}.
%Type = Article
\bibitem[{{Sod}(1978)}]{Sod1978}
\bibinfo{author}{G.A. {Sod}}, \bibinfo{journal}{Journal of Computational
  Physics} \bibinfo{volume}{27} (\bibinfo{year}{1978}) \bibinfo{pages}{1--31}.
%Type = Book
\bibitem[{Solin(2006)}]{Solin2006}
\bibinfo{author}{P.~Solin}, \bibinfo{title}{Partial Differential Equations And
  the Finite Element Method}, Pure and Applied Mathematics,
  \bibinfo{publisher}{Wiley-Interscience}, \bibinfo{year}{2006}.
%Type = Inproceedings
\bibitem[{{Tao} et~al.(2008){Tao}, {Suen}, {Wolfmeyer} and {Zhang}}]{Tao2008}
\bibinfo{author}{J.~{Tao}}, \bibinfo{author}{W.M. {Suen}},
  \bibinfo{author}{R.~{Wolfmeyer}}, \bibinfo{author}{H.M. {Zhang}}, in:
  \bibinfo{booktitle}{APS April Meeting Abstracts}, p. \bibinfo{pages}{K1012}.
%Type = Article
\bibitem[{{Taub}(1948)}]{Taub1948}
\bibinfo{author}{A.H. {Taub}}, \bibinfo{journal}{Phys. Rev.}
  \bibinfo{volume}{74} (\bibinfo{year}{1948}) \bibinfo{pages}{328--334}.
%Type = Article
\bibitem[{Taube et~al.(2009)Taube, Dumbser, Munz and Schneider}]{Taube2009}
\bibinfo{author}{A.~Taube}, \bibinfo{author}{M.~Dumbser}, \bibinfo{author}{C.D.
  Munz}, \bibinfo{author}{R.~Schneider}, \bibinfo{journal}{International
  Journal of Numerical Modelling: Electronic Networks, Devices and Fields}
  \bibinfo{volume}{22} (\bibinfo{year}{2009}) \bibinfo{pages}{77--103}.
  \bibinfo{note}{Cited By (since 1996):9}.
%Type = Article
\bibitem[{Tchekhovskoy et~al.(2007)Tchekhovskoy, McKinney and
  Narayan}]{Tchekhovskoy2007}
\bibinfo{author}{A.~Tchekhovskoy}, \bibinfo{author}{J.C. McKinney},
  \bibinfo{author}{R.~Narayan}, \bibinfo{journal}{Mon. Not. R. Astron. Soc.}
  \bibinfo{volume}{379} (\bibinfo{year}{2007}) \bibinfo{pages}{469--497}.
%Type = Book
\bibitem[{Toro(2009)}]{Toro09}
\bibinfo{author}{E.F. Toro}, \bibinfo{title}{Riemann Solvers and Numerical
  Methods for Fluid Dynamics}, \bibinfo{publisher}{Springer-Verlag},
  \bibinfo{year}{2009}.
%Type = Article
\bibitem[{Toro and Titarev(2002)}]{Toro2002}
\bibinfo{author}{E.F. Toro}, \bibinfo{author}{V.A. Titarev},
  \bibinfo{journal}{Proceedings of the Royal Society of London. Series A:
  Mathematical, Physical and Engineering Sciences} \bibinfo{volume}{458}
  (\bibinfo{year}{2002}) \bibinfo{pages}{271--281}.
%Type = Article
\bibitem[{{Toro} and {Titarev}(2005)}]{Toro2005}
\bibinfo{author}{E.F. {Toro}}, \bibinfo{author}{V.A. {Titarev}},
  \bibinfo{journal}{Journal of Computational Physics} \bibinfo{volume}{202}
  (\bibinfo{year}{2005}) \bibinfo{pages}{196--215}.
%Type = Article
\bibitem[{{van der Holst} et~al.(2008){van der Holst}, {Keppens} and
  {Meliani}}]{vanderHolst2008}
\bibinfo{author}{B.~{van der Holst}}, \bibinfo{author}{R.~{Keppens}},
  \bibinfo{author}{Z.~{Meliani}}, \bibinfo{journal}{Computer Physics
  Communications} \bibinfo{volume}{179} (\bibinfo{year}{2008})
  \bibinfo{pages}{617--627}.
%Type = Article
\bibitem[{Winkler et~al.(1984)Winkler, Norman and Mihalas}]{Winkler1984}
\bibinfo{author}{K.H.A. Winkler}, \bibinfo{author}{M.L. Norman},
  \bibinfo{author}{D.~Mihalas}, \bibinfo{journal}{Journal of Quantitative
  Spectroscopy and Radiative Transfer} \bibinfo{volume}{31}
  (\bibinfo{year}{1984}) \bibinfo{pages}{473 -- 478}.
%Type = Article
\bibitem[{{Zanotti} and {Dumbser}(2011)}]{Zanotti2011b}
\bibinfo{author}{O.~{Zanotti}}, \bibinfo{author}{M.~{Dumbser}},
  \bibinfo{journal}{Mon. Not. R. Astron. Soc.} \bibinfo{volume}{418}
  (\bibinfo{year}{2011}) \bibinfo{pages}{1004--1011}.
%Type = Article
\bibitem[{Zhang and MacFadyen(2006)}]{Zhang2006}
\bibinfo{author}{W.~Zhang}, \bibinfo{author}{A.~MacFadyen},
  \bibinfo{journal}{The Astrophysical Journal Supplement Series}
  \bibinfo{volume}{164} (\bibinfo{year}{2006}) \bibinfo{pages}{255}.

\end{thebibliography}

\appendix
\section{}

\begin{figure}
\begin{center}
\includegraphics[angle=0,width=0.5\textwidth]{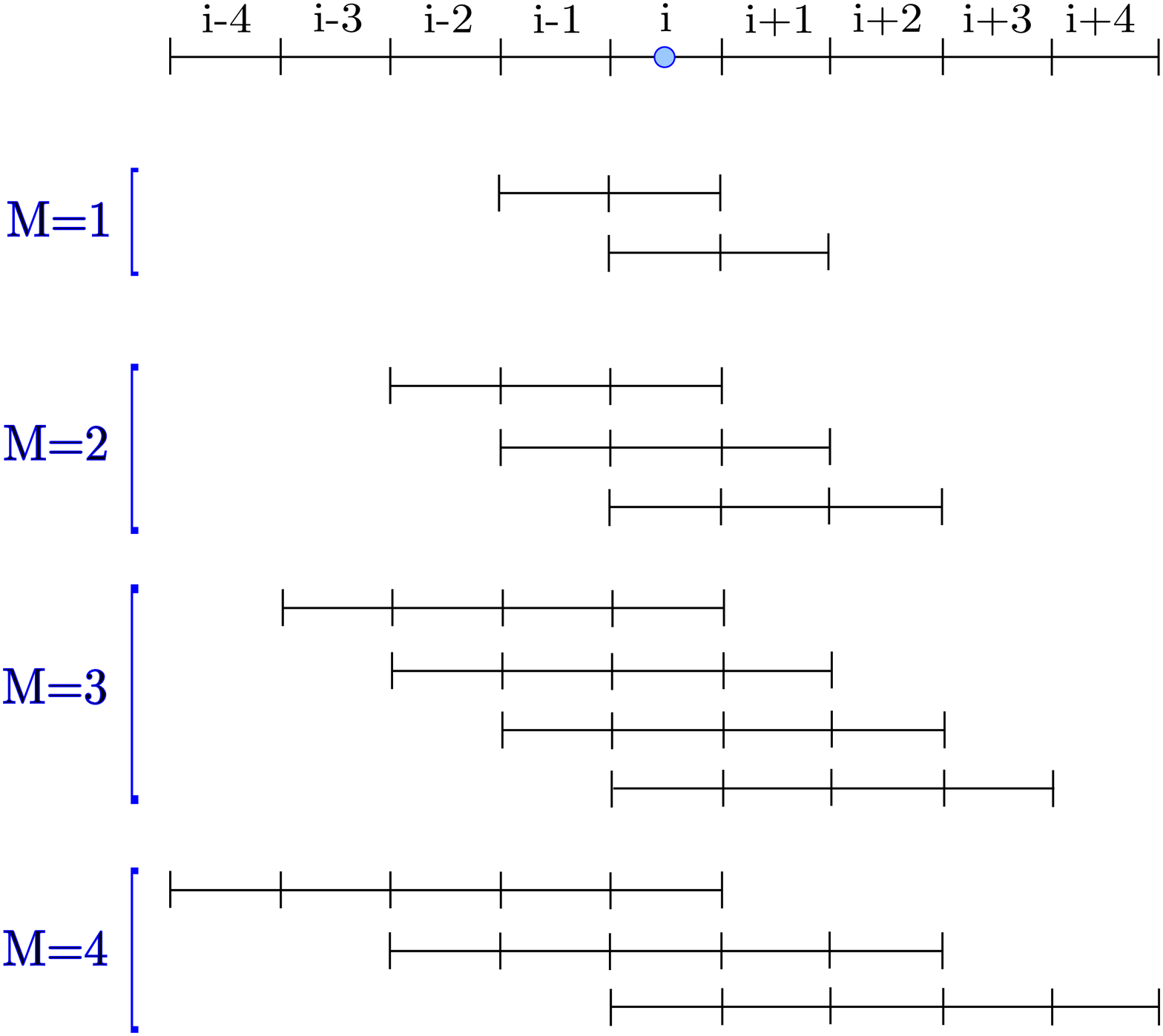} 
\caption{Representation of the one-dimensional stencils adopted up to $M=4$. 
Odd order schemes (even polynomials
of degree $M$) always use three stencils,
while 
even order schemes (odd polynomials of degree $M$) always
adopt four stencils, 
with the exception of the $M=1$ case, for which there are only two stencils.
}
\label{stencils}
\end{center}
\end{figure}

We list the most relevant parameters of the polynomials used in the WENO reconstruction that we have described 
in Sect.~\ref{sec:A dimension-by-dimension WENO reconstruction}. 
Table~\ref{tab-appendix} reports the coordinates $\lambda_k$ of the Gauss-Legendre quadrature nodes and the corresponding
nodal basis of polynomials $\psi_l$, for a few values of $M$ up to $M=4$. 
On the other hand, Fig.~(\ref{stencils}) shows the corresponding one-dimensional stencils that are used.
\begin{table}[!t]   
\caption{The table shows the coordinates of the Gauss--Legendre nodes and the corresponding nodal basis polynomials for a
few values of $M$.}
\begin{center} 
\renewcommand{\arraystretch}{1.0}
\begin{tabular}{lll} 
\hline
%\hline
%  $\ell_{\rm max}=1$ & $N_G\times N_G$  & $\epsilon_{L_2}$ & $\mathcal{O}(L_2)$ & & $N_G\times N_G$ & $\epsilon_{L_2}$ & $\mathcal{O}(L_2)$  \\ 
\hline
   & $\lambda_{k}$ & $\psi_l$    \\ 
  \hline
  $M=1$ &   &    \\ 
     & $\lambda_1=0.2113248654051$ & $\psi_1=1.366025403784 -1.732050807568 \xi $         \\ 
     & $\lambda_2=0.7886751345948$ & $\psi_2=-0.3660254037844 + 1.732050807568\xi $  \\
	\hline	
  $M=2$ &   &    \\ 
     & $\lambda_1=0.1127016653792$ & $\psi_1=1.478830557701 -4.624327782069 \xi $         \\ 
		 &                               & \hspace{0.7cm} $+ 3.333333333333 \xi^2$ \\        
     & $\lambda_2=0.5   $            & $\psi_2=-0.6666666666666 + 6.666666666666 \xi $  \\
		 &                               & \hspace{0.7cm} $-6.666666666666 \xi^2$ \\
		 & $\lambda_3=0.8872983346207$ & $\psi_3=0.1878361089654 -2.042338884597 \xi $ \\
		 &                               & \hspace{0.7cm} $+ 3.333333333333 \xi^2$ \\
 \hline
  $M=3$ &   &    \\ 
     & $\lambda_1=6.9431844202973\times10^{-2}$ & $\psi_1=1.526788125457 -8.546023607872 \xi$ \\
		 &                                  & \hspace{0.7cm} $ + 14.32585835417 \xi^2 -7.42054006803894 \xi^3$         \\ 
     & $\lambda_2=0.3300094782075   $ & $\psi_2=-0.8136324494869 + 13.80716692568 \xi $ \\
		 &                                  & \hspace{0.7cm} $-31.38822236344 \xi^2 + 18.79544940755 \xi^3$  \\ 
		 & $\lambda_3=0.6699905217924   $ & $\psi_3=0.4007615203116 -7.417070421462 \xi $ \\
		 &                                  & \hspace{0.7cm} $+24.99812585921 \xi^2 -18.79544940755 \xi^3$ \\ 
		 & $\lambda_4=0.9305681557970   $ & $\psi_4=-0.1139171962819 + 2.155927103645 \xi$ \\
		 &                                  & \hspace{0.7cm} $-7.935761849944 \xi^2 + 7.420540068038 \xi^3$ \\
\hline 
$M=4$ &   &    \\ 
     & $\lambda_1=4.6910077030668\times10^{-2}$ & $\psi_1=1.551408049094 -13.47028450119 \xi$ \\
		 &                                  & \hspace{0.7cm} $ + 38.64449905534 \xi^2 -44.98898505587 \xi^3$         \\ 
		 &                                  & \hspace{0.7cm} $ + 18.33972111443 \xi^4$ \\
     & $\lambda_2=0.2307653449471   $ & $\psi_2=-0.8931583920000 + 22.92433355572 \xi $ \\
		 &                                  & \hspace{0.7cm} $-88.22281082816 \xi^2 + 117.8634151266 \xi^3$  \\ 
		 &                                  & \hspace{0.7cm} $-51.93972111443 \xi^4$ \\
		 & $\lambda_3=0.5                 $ & $\psi_3=0.5333333333333 -14.93333333333 \xi $ \\
		 &                                  & \hspace{0.7cm} $+82.13333333333 \xi^2 -134.4000000000 \xi^3$ \\ 
		 &                                  & \hspace{0.7cm} $+67.20000000000 \xi^4$ \\
		 & $\lambda_4=0.7692346550528   $ & $\psi_4=-0.2679416522233 + 7.689927178385 \xi$ \\
		 &                                  & \hspace{0.7cm} $-46.27089213480 \xi^2 + 89.89546933107 \xi^3$ \\
		 &                                  & \hspace{0.7cm} $-51.93972111443 \xi^4$ \\
		 & $\lambda_5=0.9530899229693   $ & $\psi_5= 7.635866179581\times10^{-2} -2.210642899581 \xi$ \\
		 &                                  & \hspace{0.7cm} $+13.71587057429 \xi^2 -28.36989940184 \xi^3 $ \\
		 &                                  & \hspace{0.7cm} $+18.33972111443 \xi^4 $\\
\hline		
\end{tabular} 
\end{center}
\label{tab-appendix}
\end{table} 




%% \bibitem must have one of the following forms:
%%   \bibitem[Jones et al.(1990)]{key}...
%%   \bibitem[Jones et al.(1990)Jones, Baker, and Williams]{key}...
%%   \bibitem[Jones et al., 1990]{key}...
%%   \bibitem[\protect\citeauthoryear{Jones, Baker, and Williams}{Jones
%%       et al.}{1990}]{key}...
%%   \bibitem[\protect\citeauthoryear{Jones et al.}{1990}]{key}...
%%   \bibitem[\protect\astroncite{Jones et al.}{1990}]{key}...
%%   \bibitem[\protect\citename{Jones et al., }1990]{key}...
%%   \harvarditem[Jones et al.]{Jones, Baker, and Williams}{1990}{key}...
%%


% \end{thebibliography}

\end{document}